\documentclass[a4paper, 11pt, oneside]{Thesis}  
\graphicspath{Figures/}  

\usepackage{epigraph}
\usepackage{color}
\usepackage{csquotes}
\usepackage[square, numbers, comma, sort&compress]{natbib}  
\usepackage{verbatim}  
\usepackage{vector}  
\hypersetup{urlcolor=blue, colorlinks=true}  

\begin{document}
\frontmatter      

\title  {Cosmological Implications of Affine Gravity}
\authors  {\textcolor{black}{Hemza AZRI}
            }
\addresses  {\groupname\\\deptname\\\univname}  
\date       {April 2018}
\subject    {}

\maketitle

\setstretch{1.3}  

\acknowledgements{
\addtocontents{toc}{\vspace{1em}}  

Thanks to everyone for the support throughout the PhD journey...
}
\clearpage  

\setstretch{1.3}  

\fancyhead{}  
\rhead{\thepage}  
\lhead{}  

\pagestyle{fancy}  




\addtotoc{Abstract}  
\abstract{
\addtocontents{toc}{\vspace{1em}}  

The main aim of this thesis is to reveal some interesting aspects of the purely affine theory of gravity and its cosmological
implication. A particular attention will be devoted to its consequences when applied to cosmological inflation. Primarily, affine spacetime, composed of geodesics with no notion of length and angle, accommodates gravity but not matter. The thesis study is expected to reveal salient properties of matter dynamics in affine spacetime and may reveal an intimate connection between vacuum state and metrical gravity. An interesting application of the framework is the inflationary regime, where it is shown that affine gravity prefers only a unique metric tensor such that the transition from
nonminimal to minimal coupling of the inflaton is performed only via redefinition of the latter. This allows us to avoid the use of the so called \textit{conformal frames}. In fact, unlike metric gravity, the metric tensor in affine gravity is generated and not postulated
\textit{a priori}, thus this tensor is absent in the actions and conformal transformation does not make sense. Last but not least, we try to show how metric gravity can be induced through a simple structure that contains only affine connection and scalar fields. General relativity arises classically only at the vacuum, and this view of gravity may be considered as a new way to inducing metric elasticity of space, not through quantum corrections as in standard induced gravity, but only classically. The thesis is concluded by analyzing affine gravity in a particular higher-dimensional manifold (product of two spaces) in an attempt to understand both, the cosmological constant and matter dynamically.

}

\clearpage  
\Declaration{

\addtocontents{toc}{\vspace{1em}}  

This thesis work has been conducted in the Department of Physics in IZTECH, and it is mainly based on the results of the author's own work, in addition to a collaboration with his supervisor. 
The main chapters of this thesis are based on the following papers: 

\begin{enumerate}
\item Hemza Azri, \textbf{Are there really conformal frames? Uniqueness of affine inflation} (2018) arXiv:1802.01247 [gr-qc], to appear in Int. J. Mod. Phys. D.

\item Hemza Azri and Durmu\c{s} Demir, \textbf{Induced affine inflation}, Phys. Rev. D 97, no. 4, 044025 (2018) doi:10.1103/PhysRevD.97.044025 [arXiv:1802.00590 [gr-qc]].

\item Hemza Azri and Durmu\c{s} Demir, \textbf{ Affine inflation}, Phys. Rev. D 95, no. 12, 124007 (2017), doi:10.1103/PhysRevD.95.124007 [arXiv:1705.05822 [gr-qc]]

\item Hemza Azri, \textbf{Separate Einstein-Eddington spaces and the cosmological constant}, Annalen Phys. 528, 404 (2016), doi:10.1002/andp.201500270, [arXiv:1511.06600 [gr-qc]]

\item Hemza Azri, \textbf{Eddington's Gravity in Immersed Spacetime}, Class. Quant. Grav. 32, no. 6, 065009 (2015), doi:10.1088/0264-9381/32/6/065009, [arXiv:1501.06177 [gr-qc]]
\end{enumerate}

 
}
\clearpage  


\pagestyle{fancy}  

\lhead{\emph{Contents}}  
\tableofcontents  

\lhead{\emph{List of Figures}}  
\listoffigures  

\lhead{\emph{List of Tables}}  
\listoftables  

{

}

{

}

{
}

\setstretch{1.3}  

\pagestyle{empty}  

\null\vfill
\textit{``Si nous attribuons les ph\'{e}nom\`{e}nes inexpliqu\'{e}s au hasard, ce n'est que par des lacunes de notre connaissance.''}

\begin{flushright}
Pierre Simon de Laplace
\end{flushright}

\vfill\vfill\vfill\vfill\vfill\vfill\null
\clearpage  

\addtocontents{toc}{\vspace{2em}}  

\mainmatter	  
\pagestyle{fancy}  

\clearpage  
\lhead{\emph{Chapter 1}}  
\chapter{Introductory notes and motivations}
\label{intro}
\vspace{-0.5 cm}
\epigraph{\textit{If I have ever made any valuable discoveries, it has been owing more to patient attention, than to any other talent.\\}\, --- Sir Isaac Newton}{}

\section{Why does gravity matter?}
Newton formulated the laws that govern the gravitational attractions between macroscopic bodies. He has shown that these laws are not only applied to small objects on Earth, as it has been tested for the first time, but it holds for the whole universe. Gravity in this sense is universal. Later in 1916, Einstein provided a new description\footnote{The new theory is called Einstein's general theory of relativity and it will be discussed in some details in chapter three.} to this interaction based on new concepts (at the time) of space and time. As observed by Galileo, all freely falling bodies accelerate in the same rate in a given gravitational field. This simple remark has led Einstein to describe gravity by an accelerating frame of reference. Roughly speaking, the gravitational interaction became one aspect of the curvature of spacetime, and every type of energy in the universe responds to these manifestations, which finally appear as gravitational effects. On the other hand, spacetime gains a dynamical character due to the presence of every kind of energy which distort it.        

Universality of spacetime, and gravity as its curvature, lead us to think about the universe itself as a \textit{physical object}, and Einstein himself tried to study the evolution of the universe based on his description of gravity. In fact, in studying the evolution of the universe at large scales, the only long range force that acts everywhere is gravity. 

Below we summarize some of the interesting \textit{puzzling} physical  cases where gravity was always the main force behind them:       

\begin{itemize}
\item \textbf{Dark side of the universe}:

As it is usually stated, \textit{all the known stuff only adds up to 5\% of the content of the observable universe}. The known stuff here includes all the baryonic matter which are formed by protons and neutrons, as well as radiations (photons, neutrinos...etc). In other words, everything which is formed by (or includes) the standard model elementary particles. 

Here, gravity played an important role in the indication of the 95\% missing matter and energy. Assuming that gravity is described by Newton's theory in leading approximation, it has been shown that clumps of a non baryonic matter, which does not interact through any non gravitational force, are present in the outer galaxy halos. This non-luminous \textbf{dark matter} which forms almost 28\% the contents of the universe, acts only through its gravitational effects \cite{rotation-curves1,rotation-curves2}. 

The remaining 67\% is supposed to be \textbf{vacuum} or \textit{dark energy}. It is the cosmological fluid that makes the expansion of the universe speeding up rather than slowing down if only matter is considered \cite{supernovae Ia}. Again, the assumption of the existence of this energy is based on the theory of gravity at hand; Einstein's general relativity where not only density but pressure plays also an important role. Rather than attracting in a standard way, gravity in this sense stretches space apart due to the negative pressure of vacuum energy. The nature of dark or vacuum energy is one of the mysteries in cosmology where its signature came only through gravity.         

\item \textbf{Early universe and black holes}:

Our understanding of the big bang model, the most successful model of the universe, is mainly based on the expanding universe which is a direct consequence of relativistic gravity (general relativity). However, extrapolating further the history of the universe in the standard big bang model, one encounters an \textbf{initial singularity}. If general relativity is taken as the theory of gravity at this phase, an infinite energy density would lead to an unacceptable infinite curvature of space. This phenomenon is not much more different than the center of a \textbf{black hole}. Understanding the nature of these singularities necessitates a better understanding of gravity itself. 

\item \textbf{Unifying endeavor}:

Besides gravity, the physical world runs along three other fundamental interactions. The first of these is the Electromagnetic force described by Maxwell's classical electrodynamics and its successful quantum version; Quantum Electro-Dynamics (QED). At shorter ranges, nuclear particles such as protons and neutrons obey the weak and strong nuclear forces. A unified picture of electromagnetic and weak forces, namely Electroweak interaction, is successfully understood in the context of gauge field theory, and finally a standard model of particle physics is set up as a successful description of the three interactions.

\textit{Now, what about gravity?} If it is \enquote{fundamental} too, then the first aim would be its possible unification with the mentioned forces. One way towards a unified theory that includes gravity is to write the other forces in a geometric form like general relativity. The first request of this kind of unification (by Einstein) was to describe electrodynamics in terms of geometry in order to put it in the same geometric framework as gravity. This early view which has also led to postulating a fifth spatial dimension turned out to be misleading. The same thing happens when we try to geometrize the other forces. This is simply impossible because the geometric description of gravity is mainly based on the equivalence between the latter and the accelerating reference frames, which is not the case for the other forces.

The other schema of unification stands on the quantum description of gravity. The main approach to quantum gravity is to extend and apply the techniques of quantum field theory to general relativity. Despite the similarity between electromagnetic and the gravitational interactions (both are long range forces), it turned out that the latter suffer from infinities. While QED is successfully renormalizable \cite{qed}, quantum gravity is not \cite{qg}.  

The failure of the unification has led people to think about gravity from different directions. Some of these directions run through the possibility that \textit{gravity is not a fundamental force !}. Gravity in this case is considered as an \textit{induced} or \textit{emergent} phenomenon. In the former, gravity may arise from elementary particles through one loop corrections to particle fields, whereas the latter suggests that gravity may gain an emergent character from black holes thermodynamics \cite{sakharov0,jacobson}. The problem of gravity at very small scales has not been settled down yet.     
\end{itemize}

\section{If it is all about curvature, then which geometry is viable?}

Gravity is believed to be one aspect of the curvature of spacetime and then the correct geometric view is essential in any theory of gravity. In Einstein's general relativity, it is assumed that metric geometry is relevant to the theory. The spacetime then is supposed to be pseudo-Riemannian, i.e, a space endowed with a metric tensor which describes the intervals (distances and times) and angles between different events in the curved background.

The concept of the metric tensor which is postulated \textit{a priori} is certainly fundamental in the large scale structure where notions such as lengths and angles are present. However, it is worth noting that the very existence of spacetime may not accommodate these concepts, but rather, they may arise from a more fundamental requirements.

The concept of the curvature of spacetime does not require a metric field. In fact, that is the rule of parallel displacement which provides a measure of the curvature of the manifolds. This rule is incorporated in the so called affine connection through covariant derivatives. Theories of gravity which are based on this affine connections as fundamental fields are called purely affine theories. Interestingly, what is known as metric tensor in general relativity appears in affine gravity \textit{a posteriori} as a solution of the equations of motion. In this sense, metrical structure is generated \textit{dynamically}.

Affine spacetime involves only trajectories generated by connections with no notion of length and angle. In this spacetime there are no invariants; even a constant energy density is \enquote{difficult} to define. The only meaningful structure is determinants of tensors. In this vision, possible gravity actions are constructed from determinants of the tensors at hand, among these tensors, we have the pure geometrical tensors; the Riemann tensor, the Ricci tensor as well as the torsion tensor. These are defined only in terms of the affine connection and no other entities (such as metric tensor) are required. Besides these tensorial quantities, matter fields may also enter the actions in a tensorial form. In fact, from scalar fields $\phi$, in addition to its potential energy, a symmetric tensor $\nabla_{\mu}\phi\nabla_{\nu}\phi$ can also be formed \cite{affine inflation}.

An interesting feature of affine spaces, is that they accommodate scalar fields only for nonzero potential energy, a property which is at the heart of the inflationary cosmology. Nonzero potential energy means at least nonzero vacuum energy, thus metrical structure which is generated \textit{a posteriori} may gain in this sense an \textit{induced} character from vacuum energy \cite{induced affine inflation}. 

Studying purely affine gravity from its different aspects will be the primary objective of our thesis which will be organized as in the following section.    

\section{Plan of this thesis}

In this thesis we will provide a detailed study of purely affine gravity. Our aim here is to show the viability of this theory in both theoretical and observational sides. However, to make it more pedagogical, the thesis will contain some introductions on general relativity, and relativistic cosmology including inflation. This will make it easy for the reader to see the differences between general relativity and affine gravity and extract the new features of the latter.    

We organize the thesis as follows: Chapter two will be devoted to the geometrical framework where we introduce the concepts of metric tensors and affine connections. This part is important and it shows how those concepts, though both fundamental, are completely independent. It is also shown how curvature which is at the heart of any relativistic theory of gravity, is related to the affine connection without introducing any metrical structure. 

In Chapter three, we present the general theory of relativity. We show how Einstein was able to come out with his interesting description of gravity in terms of curvature of spacetime, based only on a simple remark of free fall. We give a detailed derivation of Einstein's field equations using the variational principle where the fundamental field that plays the role of the gravitational field is the metric tensor.  

Purely affine gravity, the objective of the thesis, will be addressed in the fourth Chapter. A particular and the simplest affine gravity is Eddington's gravity, which is based solely on the Ricci tensor and a nonzero cosmological constant. We show how this theory, free of any matter fields, becomes equivalent to Einstein's gravity after generating the metric tensor. We proceed and extend Eddington's theory by adding scalar fields. Here, two important cases are studied separately; minimal and nonminimal couplings. We will show that affine gravity is different than metric gravity for the nonminimal coupling case. The differences appear in both the gravitational equations and the scalar field equation. This, as we will see, is the consequence of the first order (linear) affine action. 

An interesting part in this Chapter is about induced gravity in the affine context. It is shown that in the affine picture, both the scale of gravity and the metric tensor gain an induced character. Gravity arises from vacuum expectation value of heavy scalar fields, while the metric tensor arises from the vacuum energy \cite{induced affine inflation}.       

Chapter five will be devoted to standard cosmology and inflation, and in Chapter six we tackle the problem of inflation in the context of affine gravity. We apply the affine formalism presented in Chapter five to inflation, and show that the inflationary regime arises naturally for slowly rolling fields with predictions compatible with the recent data. Induced affine gravity however shows slight deviations from observation. We will also discuss the question of different frames, namely Jordan and Einstein frames in metric gravity, and the frame ambiguities in inflation. We show that these frames do not make sense in affine gravity where there is a unique metric generated from nonzero vacuum energy. The transformation from nonminimal to minimal coupling can be obtained only through field redefinition, but not metric transformation.
 
In Chapter seven, we study Eddington-like gravity in a particular higher dimensional spaces, namely the \textit{product spaces}, as an attempt to give a dynamical nature to matter from high dimensions. Finally, we summarize and conclude our thesis in Chapter eight.   
 
\clearpage  
\lhead{\emph{Chapter 2}}  
  \chapter{Spacetime: Metricity and affinity }
\label{chapter 2}
\vspace{-0.5 cm}
\epigraph{\textit{The theory of relativity brought the insight that space and time are not merely the stage on which the piece is produced, but are themselves actors playing an essential part in the plot.}\, ---Willem de Sitter}{}

\section{When metric tensor is necessary?}
\label{sec:when metric tensor is necessary?}
\subsection{Minkowski spacetime}
We are usually interested in the local character of the physical world. The local properties require physical measurements associated with clock ticks and spacial distances, or \textit{spacetime} intervals in relativistic view. This imposes the concepts of a \textit{distance} and \textit{angle} in the physical world. In fact, in our four dimensional spacetime, the infinitesimal interval which generalizes the three dimensional Euclidean distance is given by
\begin{eqnarray}
\label{minkowski line element1}
(d s)^{2}=-(c dt)^{2}+ d\vec{\textbf{x}}.d\vec{\textbf{x}},
\end{eqnarray}   
where $dt$ and $d\vec{\textbf{x}}$ are time and space coordinate differentials connecting two close points, and $c$ is the speed of light.

The quantity (\ref{minkowski line element1}) describes the length of a spacetime coordinate differential
\begin{eqnarray}
d x^{a}=\left(c dt, d\vec{\textbf{x}} \right),
\end{eqnarray} 
and it is written in the following form
\begin{eqnarray}
\label{minkowski line element2}
(d s)^{2}=\eta_{ab} dx^{a} dx^{b}.
\end{eqnarray}
where clearly the quantity $\eta$ is nothing but the $4\times 4$ matrix with constant coefficients
\[
\eta_{ab}=
  \begin{pmatrix}
    -1 & 0 & 0 & 0 \\
     0 & +1 & 0 & 0 \\
     0 & 0 & +1 & 0 \\
     0 & 0 & 0 & +1
  \end{pmatrix}.
\]
The above writings allow us to associate to every \textit{contravariant} vector $A^{a}$, the square of its \textit{length}, given by
\begin{eqnarray}
\label{flat norm of a vector}
A^{2}\equiv \eta_{ab} A^{a}A^{b}.
\end{eqnarray}
The square of the length (\ref{flat norm of a vector}) is a particular case of the scalar product of two vectors of components $A^{a}$ and $B^{b}$ which is given by 
\begin{eqnarray}
\label{flat scalar product}
A.B=\eta_{ab}A^{a}B^{b}
\end{eqnarray}
The above structure is inherited from the spacetime \textit{symmetries}, where quantities like (\ref{minkowski line element1}) are invariant under the so called Lorentz transformation. It is defined by a linear transformation of the form
\begin{eqnarray}
\label{lorentz transformation}
x^{a} \rightarrow \hat{x}^{a}= \Lambda ^{a}_{\,\,\, b} x^{b}
\end{eqnarray}  
that leaves the line element (\ref{minkowski line element1}) \textit{invariant}.
\newpage
\begin{figure}[h]
\centering
    \includegraphics[width=0.7\textwidth]{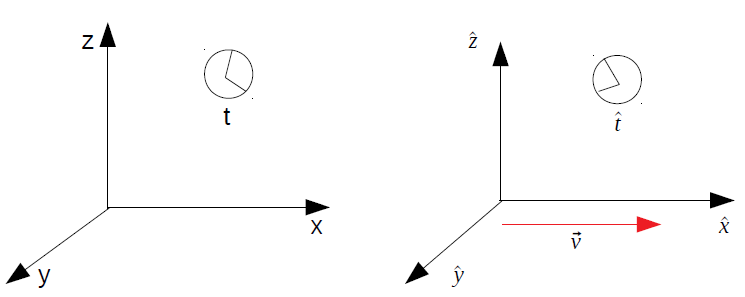}
\caption{Two inertial frames in relative motion along the $x$ direction. A coordinate system is attached to each frame and they are related to each other by the Lorentz transformation (\ref{lorentz transformation}). }
\label{fig:relative-frames}
\end{figure}
This is applied to the norms of vectors (\ref{flat norm of a vector}) and scalar products (\ref{flat scalar product}), and consequently, the quantity $\eta$ must satisfy
\begin{eqnarray}
\eta_{ab}=\Lambda^{c}_{\,\,\,a}\Lambda^{d}_{\,\,\,b}\, \eta_{cd}.
\end{eqnarray}
The $4\times 4$ matrix elements $\Lambda^{a}_{\,\,\,b}$ describe rotations and boost transformations and they relate the spacetime coordinates of the same event recorded in two inertial frames. 

\begin{figure}[h]
\centering
    \includegraphics[width=0.5\textwidth]{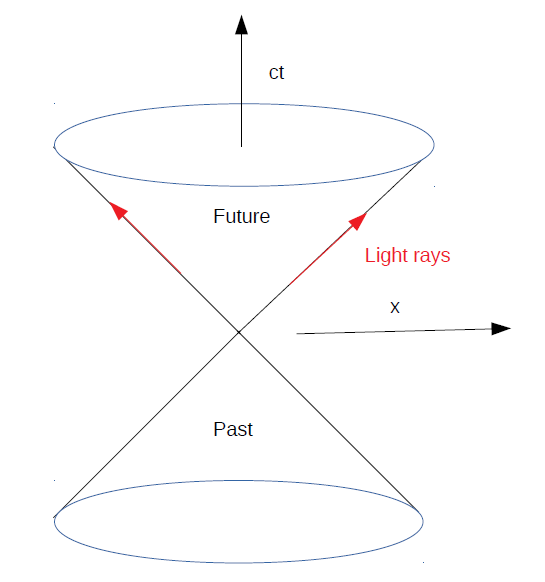}
\caption{Light rays travel in straight lines (generators of the cone) between any two points separated by a null interval $ds^{2}=0$. Physical (massive) particles move along lines with $ds^{2} < 0$.}
\label{fig:cone}
\end{figure}

Along the $x$ direction where the two inertial frames are in uniform relative motion with a velocity $v$ (figure~\ref{fig:relative-frames}), the Lorentz transformations are parametrized by  
\[
\Lambda^{\nu}_{\,\,\,\mu}=
  \begin{pmatrix}
    \cosh \phi & \sinh \phi & 0 & 0 \\
     \sinh \phi & \cosh \phi & 0 & 0 \\
     0 & 0 & 1 & 0 \\
     0 & 0 & 0 & 1
  \end{pmatrix},
\]
where
\begin{eqnarray}
\cosh \phi=\frac{1}{\sqrt{1-\frac{v^{2}}{c^{2}}}}, \quad \quad \quad
\sinh \phi =\frac{v/c}{\sqrt{1-\frac{v^{2}}{c^{2}}}}
\end{eqnarray}
Like rotations in Euclidean space, the Lorentz transformations form a group $O(3,1)$. This Lorentz group differs from the Euclidean group $O(4)$ by the non positive spacetime invariants. In fact, according to (\ref{flat norm of a vector}), vectors may take positive, negative or zero norms. To that end, a vector $A$ is said to be \textit{time like}, \textit{space like} or \textit{light like} if $A^{2}$ is negative, positive, or zero respectively. The speed of light c provides a limit to particle velocities, and then the regions which are causally\footnote{The regions must be separated by a timelike interval, $ds^{2}<0$.} connected, are described only by points inside the \textit{light cone} (see figure~\ref{fig:cone}).  

Next, we will discover the structure of distances and line elements in curved spacetime and introduce the notion of the \textbf{metric tensor}, where the invariance are manifested by general coordinates rather than Lorentz transformations.
\subsection{General coordinates and metrical structure}

As we will see later in this Chapter, the gravitational phenomena \textit{necessitate}, the use of arbitrary reference frames. The spacetime geometry must then be described by general (curvilinear) coordinates rather than the special coordinates used in the last subsection. Under general coordinate transformations $\hat{x}^{\mu}(x^{\lambda})$, the coordinate differentials transform as
\begin{eqnarray}
\label{differentials transformation}
d x^{\mu}=\frac{\partial x^{\mu}}{\partial \hat{x}^{\nu}} d\hat{x}^{\nu},
\end{eqnarray}
where we have used Greek indices to refer to general coordinates.

Quantities $A^{\mu}$ that transform like (\ref{differentials transformation}) under coordinate transformations are called \textit{contravariant} four vectors
\begin{eqnarray}
\label{vector transformation}
A^{\mu}=\frac{\partial x^{\mu}}{\partial \hat{x}^{\nu}} \hat{A}^{\nu}.
\end{eqnarray}

Like coordinate differentials (\ref{differentials transformation}), it is also convenient to write the transformation rule of the partial derivative of a scalar $\phi$. This is simply given by
\begin{eqnarray}
\label{partial transformation}
\partial_{\mu}\phi=\frac{\partial \hat{x}^{\nu} }{\partial x^{\mu}}
\partial_{\hat{\nu}}\phi.
\end{eqnarray} 
Quantities $A_{\mu}$ which transform as (\ref{partial transformation}) are called \textit{covariant} four vectors
\begin{eqnarray}
A_{\mu}=\frac{\partial \hat{x}^{\nu} }{\partial x^{\mu}}
\hat{A}_{\nu}.
\end{eqnarray}
Based on the above rules of transformations, one may simply call a covariant tensor of rank two, an element $T_{\mu\nu}$ of sixteen quantities which transform like the product of the components of two covariant vectors, thus  
\begin{eqnarray}
T_{\mu\nu}=
\frac{\partial \hat{x}^{\alpha} }{\partial x^{\mu}}
\frac{\partial \hat{x}^{\beta} }{\partial x^{\nu}}
\hat{T}_{\alpha\beta}.
\end{eqnarray}
In the same manner, contravariant and mixed rank two tensors $T^{\mu\nu}$ and $T^{\mu}_{\,\,\,\nu}$ are introduced respectively.

Now, the line element that describes the distance between two infinitesimally close points with coordinates $x^{\mu}$ and $x^{\mu}+dx^{\mu}$ is given by
\begin{eqnarray}
\label{general line element}
(ds)^{2}=g_{\mu\nu}(x)dx^{\mu}dx^{\nu},
\end{eqnarray} 
where $g_{\mu\nu}$ are the components of a covariant rank two tensor which coincides with the Minkowski metric tensor in a local reference frame (at a point $P$)
\begin{eqnarray}
g_{\mu\nu}(P)=\eta_{\mu\nu}.
\end{eqnarray}
The tensor $g_{\mu\nu}$ is symmetric, i.e, $g_{\nu\mu}=g_{\mu\nu}$ and it is called the \textit{metric tensor}. In Riemannian spaces, this tensor is called the Riemannian metric. However, since its signature is not positive (negative) definite, the metric then is called Lorentzian.   

The metric tensor plays an important role in defining two quantities; \textit{norm} of vectors and \textit{angles}. The norm of a vector of components $A^{\mu}$ is given by
\begin{eqnarray}
|A|= \sqrt{g_{\mu\nu} A^{\mu}A^{\nu}}.
\end{eqnarray}
This clearly coincides with (\ref{flat norm of a vector}) in a local reference frame.

The angle $\theta$ between two vectors of components $A^{\mu}$ and $B^{\nu}$ is defined by 
\begin{eqnarray}
\cos \theta =\frac{g_{\mu\nu}A^{\mu}B^{\nu}}{|A|.|B|},
\end{eqnarray}
where the numerator is nothing but the scalar product of the two vectors as written in (\ref{flat scalar product}) in local reference frames. 

Then like flat (Minkowski) spacetime, this (pseudo) Riemannian spacetime is an arena which is endowed with a property for measuring distances and angles. This property is encoded in the metric tensor. Later we will see that this \textbf{metrical structure} is not necessary for describing the essential properties of spacetime geometry.

\section{Affinity on spacetime continuum}
\subsection{Notion of affine connection}

Primary requirements in space geometry is how to compare vectors at different points in space. This is trivial in Euclidean (or Lorentzian) geometry where equality of the components of two vectors implies the equality of the vector themselves. The previous assumption may not be correct if the space has non Euclidean geometry.     

The reason of this can be understood from the fundamental transformation rule (\ref{differentials transformation}). The vector $A^{\mu}$ in (\ref{vector transformation}) which transforms exactly like (\ref{differentials transformation}) can be considered as a displacement vector from a point P of coordinates $x^{\mu}$ to a close point Q of coordinates $x^{\mu}+dx^{\mu}$. Since the coefficients $\partial \hat{x}^{\mu}/\partial x^{\nu}$ depend on the coordinates (change from point to point), then it becomes impossible to compare directly the same vector at the points P and Q as in Euclidean space, even if these points are very close to each other.   

Comparison of the same vector at two neighboring points can be related to the way we \textit{parallel} transfer this vector between these two points, and then, to how to make derivative of the vector along a given curve \cite{wald,hawking-ellis}. This process is not trivial in curved spaces because of the different tangent spaces at different points of the space. This requires a new machinery that allows the \textit{connection} between these tangent spaces.

Here we suppose that the vector $A^{\mu}$ at the point $P$ takes the form $A^{\mu}+\delta A^{\mu}$ which infinitesimally differs from $A^{\mu}$. The change in the vector $A^{\mu}$ is given in terms of both, the infinitesimal displacement $dx^{\mu}$ and the vector $A^{\mu}$ itself, as follows
\begin{eqnarray}
\label{affine connection}
\delta A^{\mu}= -\Gamma^{\mu}_{\,\,\alpha\beta} A^{\alpha}dx^{\beta},
\end{eqnarray}   
where $\Gamma^{\mu}_{\,\,\alpha\beta}$ are arbitrary functions of $x^{\mu}$, and they are called the coefficients of an \textit{affine connection}. 

In four dimensional space, the affine connection is determined by its 64 components. Since the spacetime acquires a pseudo Euclidean geometry locally, where vectors are transported from two neighbor points with no changes in the vector, i.e, $\delta A^{\mu}=0$, this implies that the affine connection locally vanishes
\begin{eqnarray}
\Gamma^{\mu}_{\,\,\alpha\beta}(P)=0.
\end{eqnarray}
Relation (\ref{affine connection}) describes then the \textit{parallel displacement} of the vector $A^{\alpha}$ along $dx^{\beta}$. This property, in non-Euclidean spaces, is at the heart of the concept of curvature. We call the new derivative (of a vector) which is based on parallel transfer, the \textit{covariant derivative} and it is defined as
\begin{eqnarray}
\label{covariant derivative}
\nabla_{\mu} A^{\alpha}=\partial_{\mu} A^{\alpha}
+\Gamma^{\alpha}_{\,\,\lambda\mu} A^{\lambda}.
\end{eqnarray}
This rule coincides with the ordinary derivative locally when the space is considered flat $(\Gamma(P)=0)$.

The presence of the affine connection in (\ref{covariant derivative}) can be understood from the tensorial character of the derivatives of vectors. One may easily show that under general coordinate transformation where the vector $A^{\mu}$ transforms as (\ref{vector transformation}), the derivative of $A^{\mu}$ transforms as
\begin{eqnarray}
\frac{\partial \hat{A}^{\alpha}}{\partial \hat{x}^{\mu}}=
\frac{\partial \hat{x}^{\alpha}}{\partial x^{\lambda}}
\frac{\partial x^{\sigma}}{\partial \hat{x}^{\mu}}
\frac{\partial A^{\lambda}}{\partial x^{\sigma}}
+\frac{\partial^{2}\hat{x}^{\alpha}}{\partial x^{\lambda}\partial x^{\sigma}}\frac{\partial x^{\sigma}}{\partial \hat{x}^{\mu}} A^{\lambda}.
\end{eqnarray}
The last term of this expression spoils the tensorial character of the ordinary derivative $\partial_{\mu} A^{\alpha}$. Thus, the ordinary derivative is an object that depends on coordinate systems. 

Based on this, one may show that the tensorial character of the total term (\ref{covariant derivative}) implies the following transformation rule of the affine connection
\begin{eqnarray}
\label{transformation rule of a connection}
\hat{\Gamma}^{\alpha}_{\,\,\lambda\mu}=
\frac{\partial \hat{x}^{\alpha}}{\partial x^{\beta}}
\frac{\partial x^{\sigma}}{\partial \hat{x}^{\lambda}}
\frac{\partial x^{\tau}}{\partial \hat{x}^{\mu}}
\Gamma^{\beta}_{\,\,\sigma \tau}
+\frac{\partial^{2} x^{\beta}}{\partial \hat{x}^{\lambda}\partial \hat{x}^{\mu}}\frac{\partial \hat{x}^{\alpha}}{\partial x^{\beta}}.
\end{eqnarray}   
From the transformation rule of this non tensorial object, we extract the following interesting properties:
\begin{itemize}
\item
First, the second term of the right hand side of equation (\ref{transformation rule of a connection}) does not depend on the connection but rather it is a coordinate system dependent quantity. Thus, if the connection tends to be zero at one reference frame, then the presence of this term forbids it to vanish everywhere.
\item 
However, if one envisages two connections $\Gamma^{\alpha}_{\,\,\mu \nu}$ and $\bar{\Gamma}^{\alpha}_{\,\,\mu \nu}$ in the same space, then the difference $\Gamma^{\alpha}_{\,\,\mu \nu}-\bar{\Gamma}^{\alpha}_{\,\,\mu \nu}$ forms the components of a tensor. A particular and interesting case is the \textit{torsion} tensor which we will discover later. The same for the infinitesimal variation $\delta \Gamma^{\alpha}_{\,\,\mu \nu}$ which is a tensor. In other words, one may always write any affine connection as a sum of a second connection and a tensor.   
\end{itemize}

\subsection{Curvature of space}
Up to now, we have referred to flatness only locally where the connection vanishes. This is what observers realize in Galilean inertial frames. As we have seen, the affine connection does not follow a tensorial transformation, thus, it may not provide us with the real character of space, and to that end, one has to explore a new quantity which has to be a tensor and characterizes the intrinsic form of space.

\begin{figure}[h]
\centering
    \includegraphics[width=0.7\textwidth]{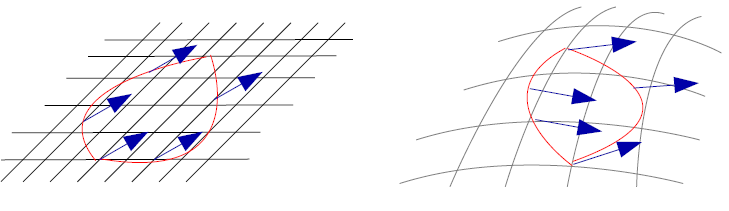}
\caption{In flat space (left), any vector is displaced infinitesimally along a closed curve without losing its initial direction and magnitude. However, in a curved space (right), parallel transfer affects the form of the vector due to the curvature of space.}
\label{fig:parallel}
\end{figure}

The parallel displacement which has led us to the concept of a connection is also the key point towards the concept of \textit{curvature}. In flat space, like the familiar two dimensional plan, a vector at an initial point can be parallel transported along a closed curve and it returns back to its original form. However, if the space is curved, the vector returns back to the initial point but with a different direction. The inequality of a vector (after parallel displacement) and its original form (before the parallel displacement) originates from the curvature of space \cite{wald,hawking-ellis,lovelock}. This fact is schematically illustrated in Figure~\ref{fig:parallel}.   

To proceed to the definition of the curvature tensor, it is important to mention that parallel transport of a vector field may be described by successive derivation of the vector itself. The success of a vector to return to its original initial value in flat space can be related to the fact that in this space, the derivatives commute   
\begin{eqnarray}
\left[\partial_{\mu},\partial_{\nu}  \right]A^{\alpha}
=\partial_{\mu} \partial_{\nu}A^{\alpha}
-\partial_{\nu} \partial_{\mu}A^{\alpha}
=0.
\end{eqnarray}
The curvature of space will be then the origin of the failure of this commutativity which is now written in its general covariant form as 
\begin{eqnarray}
[\nabla_{\mu},\nabla_{\nu}]A^{\alpha}=
\nabla_{\mu} \nabla_{\nu}A^{\alpha}
-\nabla_{\nu} \nabla_{\mu}A^{\alpha}
\neq 0.
\end{eqnarray}
This new commutation of the covariant derivative takes the following form
\begin{eqnarray}
[\nabla_{\mu},\nabla_{\nu}]A^{\alpha}&&=
\partial_{\mu}\partial_{\nu} A^{\alpha}
+\partial_{\mu}\Gamma^{\alpha}_{\,\,\nu\lambda}A^{\lambda}
+\Gamma^{\alpha}_{\,\,\nu\lambda}\partial_{\mu}A^{\lambda}
+\partial_{\nu}A^{\lambda}\Gamma^{\alpha}_{\,\,\mu\lambda}
+\Gamma^{\alpha}_{\,\,\mu\lambda}\Gamma^{\lambda}_{\,\,\nu\sigma}A^{\sigma} \nonumber \\
&&-\partial_{\lambda}A^{\alpha}\Gamma^{\lambda}_{\,\,\mu\nu}
-\Gamma^{\lambda}_{\,\,\mu\nu}\Gamma^{\alpha}_{\,\,\lambda\sigma}A^{\alpha}
-(\nu \mu),
\end{eqnarray}
where we have used the definition of the covariant derivatives (\ref{covariant derivative}), and the term $(\nu\mu)$ refers to the same expression with $\mu\leftrightarrow \nu$.

If we write the last term of the right hand side explicitly, and simplify the expression, we finally get
\begin{eqnarray}
\label{covariant derivative commutator}
[\nabla_{\mu},\nabla_{\nu}]A^{\alpha}&&=
\left(\partial_{\mu}\Gamma^{\alpha}_{\,\,\nu\lambda}
-\partial_{\nu}\Gamma^{\alpha}_{\,\,\mu\lambda}
+\Gamma^{\alpha}_{\,\,\mu\sigma}\Gamma^{\sigma}_{\,\,\nu\lambda}
-\Gamma^{\alpha}_{\,\,\nu\sigma}\Gamma^{\sigma}_{\,\,\mu\lambda}  \right)A^{\lambda} \nonumber \\
&&-\left(\Gamma^{\lambda}_{\,\,\mu\nu}-\Gamma^{\lambda}_{\,\,\nu\mu} \right)
\nabla_{\lambda}A^{\alpha}.
\end{eqnarray}
Two important quantities appear in this expression, namely, the \textit{curvature tensor}
\begin{eqnarray}
\label{riemann tensor}
R^{\alpha}_{\,\,\,\lambda\mu\nu}\left[\Gamma \right]=
\partial_{\mu}\Gamma^{\alpha}_{\,\,\nu\lambda}
-\partial_{\nu}\Gamma^{\alpha}_{\,\,\mu\lambda}
+\Gamma^{\alpha}_{\,\,\mu\sigma}\Gamma^{\sigma}_{\,\,\nu\lambda}
-\Gamma^{\alpha}_{\,\,\nu\sigma}\Gamma^{\sigma}_{\,\,\mu\lambda},
\end{eqnarray}
and the \textit{torsion tensor}
\begin{eqnarray}
S^{\lambda}_{\,\,\mu\nu}=\Gamma^{\lambda}_{\,\,[\mu\nu]}=
\frac{1}{2}\left(\Gamma^{\lambda}_{\,\,\mu\nu}-\Gamma^{\lambda}_{\,\,\nu\mu} \right),
\end{eqnarray}
and then, the commutator (\ref{covariant derivative commutator}) takes the form
\begin{eqnarray}
[\nabla_{\mu},\nabla_{\nu}]A^{\alpha}=
R^{\alpha}_{\,\,\,\lambda\mu\nu}  A^{\lambda}
-2S^{\lambda}_{\,\,\mu\nu} \nabla_{\lambda}A^{\alpha}.
\end{eqnarray} 
These two quantities have some interesting properties that worth pointing out here:
\begin{enumerate}
\item
\underline{\textbf{Curvature tensor}}:

Although it is given in terms of the connection coefficients which are not tensors, the quantities $R^{\alpha}_{\,\,\,\lambda\mu\nu}$ are the components of a true tensor. Since the commutator on the left hand side is antisymmetric in the indices $\mu$ and $\nu$, so the curvature tensor. An important identity satisfied by the curvature tensor is the \textit{Bianchi} identity which will be very useful later. This is written as 
\begin{eqnarray}
\label{bianchi0}
\nabla_{\alpha}R^{\lambda}_{\,\,\,\mu\kappa\nu}
+\nabla_{\nu}R^{\lambda}_{\,\,\,\mu\alpha\kappa}
+\nabla_{\kappa}R^{\lambda}_{\,\,\,\mu\nu\alpha}=0.
\end{eqnarray}
In addition to this property, a useful rank-two tensor, namely the Ricci tensor, can be extracted from the curvature tensor as
\begin{eqnarray}
\label{ricci-general}
R_{\mu\nu} \left[\Gamma \right]
= R^{\lambda}_{\,\,\,\mu\lambda\nu} \left[\Gamma \right].
\end{eqnarray} 
\item
\underline{\textbf{Torsion tensor}}: 

The torsion tensor $S^{\lambda}_{\,\,\mu\nu}$ is nothing but the antisymmetric part of the affine connection, and thus, it is antisymmetric in the indices $\mu$ and $\nu$. This means that an affine connection is symmetric if it is \textit{torsionless}. As we will see later, a symmetric connection is important when Einstein's equivalent principle is applied \cite{landau}.  
\end{enumerate}
In constructing these important quantities, we were dealing with only connections. The latter is introduced in (\ref{affine connection}) based on parallel transfers of tensors, but without referring to any concepts of distances and angles. It is only from this simple spacetime structure that one may use to describe a theory of gravity. This will be the aim of this thesis and it will be started from Chapter 4.

\subsection{Geodesics and geodesic deviation}    

Like Euclidean space or the flat spacetime of special relativity, the general curved spaces have the properties of straight lines. We are familiar with the fact that these straight lines are the shortest paths connecting two points. However, this is the case only if the concept of distance is taken into account, and then the space is endowed with a metric tensor. Nevertheless, the straightest possible lines or \textit{geodesics} are completely independent of distances. They are the paths whose tangent vectors are parallel transported along themselves. If a geodesic is parametrized by an affine parameter $\tau$ as $x^{\alpha}(\tau)$, its tangent vector $u^{\alpha}$ must satisfy the parallel transport condition
\begin{eqnarray}
\label{parallel tangent vector}
u^{\mu}\nabla_{\mu}u^{\alpha}=0,
\end{eqnarray}  
where $\nabla$ is the covariant derivative operator with respect to the affine connection $\Gamma$.

Since the tangent vector is given as $u^{\alpha}=dx^{\alpha}/d\tau$, then the condition (\ref{parallel tangent vector}) is written explicitly as
\begin{eqnarray}
\label{affine geodesic equation}
\frac{d^{2}x^{\alpha}}{d\tau^{2}}+
\Gamma^{\alpha}_{\,\,\mu\nu}\frac{dx^{\mu}}{d\tau}\frac{dx^{\nu}}{d\tau}=0.
\end{eqnarray}
This represents a system of second order differential equations, and then a unique solution is guaranteed by providing an initial conditions for $x^{\alpha}(\tau)$ and $dx^{\alpha}/d\tau$. This unique solution describes a curve which we have called a geodesic. A point particle moving through a geodesic has a vector velocity (tangent vector) that keeps the same direction along this path. 

\begin{figure}[h]
\centering
    \includegraphics[width=0.5\textwidth]{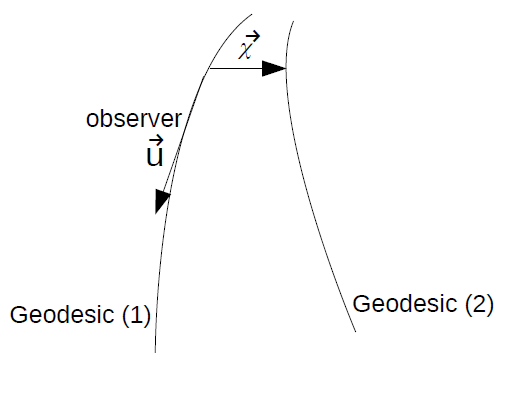}
\caption{In a curved space, two geodesics (1) and (2) which start parallel, do not remain parallel but they deviate from each other. This deviation is the effect of the curvature of space as written in equation (\ref{deviation from curvature}). }
\label{fig:geodesic-deviation}
\end{figure}

The concept of curvature that we have met in the last section can be addressed here from the notion of \textit{geodesic deviation}. Suppose, we are given a geodesic curve with tangent vector $u^{\alpha}$, and its separation from a nearby curve is denoted by the deviation vector $\chi^{\alpha}$ as in Figure~\ref{fig:geodesic-deviation}. Then the rate of change of this deviation, or the relative velocity of the neighboring geodesics, is given by the vector $u^{\mu}\nabla_{\mu}\chi^{\alpha}$. The relative acceleration of the infinitesimally close geodesics reads \cite{wald} 
\begin{eqnarray}
\mathfrak{a}^{\mu}=
u^{\lambda}\nabla_{\lambda}(u^{\nu}\nabla_{\nu}\chi^{\mu}).
\end{eqnarray}
By writing the covariant derivatives explicitly in terms of the connection, one may show that
\begin{eqnarray}
\label{geodesic deviation equation1}
\mathfrak{a}^{\mu}=
-R_{\,\,\,\nu\lambda\rho}^{\mu}\,\,u^{\nu}\chi^{\lambda}u^{\rho},
\end{eqnarray}
where $R_{\,\,\,\nu\lambda\rho}^{\mu}$ is the curvature or the Riemann tensor (\ref{riemann tensor}).

This equation shows that, in curved space, two lines which are initially parallel will not remain parallel, and their \enquote{deviation} is caused by the curvature of space. To leading order, the equation of the geodesic deviation (\ref{geodesic deviation equation1}) takes the following form \cite{hartle}
\begin{eqnarray}
\label{deviation from curvature}
\mathfrak{a}^{i}\equiv
\frac{d^{2}\chi^{i}}{dt^{2}}=
-R^{i}_{\,\,\,0j0}\chi^{j}.
\end{eqnarray}

An analogy with gravity can be illustrated as follows. In a gravitational potential $\phi(x)$, a freely falling particle is governed by Newton's equation of motion which is written in an inertial frame as
\begin{eqnarray}
\frac{d^{2}x^{i}}{dt^{2}}=
-\frac{\partial \phi(x^{k})}{\partial x^{j}} \delta^{i}_{j}.
\end{eqnarray}
If the position of a second (freely falling) particle is $x^{i}(t)+\chi^{i}(t)$, where $\chi^{i}(t)$ measures the separation of the two particles, then the equation of motion of the second particle reads
\begin{eqnarray}
\frac{d^{2}(x^{i}+\chi^{i})}{dt^{2}}=
-\frac{\partial \phi(x^{k}+\chi^{k})}{\partial x^{j}}\delta^{i}_{j}.
\end{eqnarray}
To linear order of (small) $\chi^{j}$, one finds the time evolution of the separation vector as 
\begin{eqnarray}
\frac{d^{2}\chi^{i}}{dt^{2}}=
-\left(\frac{\partial^{2}\phi}{\partial x^{j}\partial x^{k}}\right)\delta^{i}_{j}
\chi^{k}.
\end{eqnarray}
This equation is at the heart of the so called \textit{tidal gravitational forces}, and it provides a measure of the relative accelerations of two (test) freely falling particles. When compared to equation (\ref{deviation from curvature}), one may easily find 
\begin{eqnarray}
\left(\frac{\partial^{2}\phi}{\partial x^{j}\partial x^{k}}
\right)\delta^{i}_{j} \propto
R^{i}_{\,\,\,0k0}.
\end{eqnarray}
This shows that the effects of these forces are nothing but the manifestation of the curvature of spacetime. In the next Chapter, we will show how gravity is related to the curvature of spacetime in the general theory of relativity. From the equivalence principle, Einstein was able to describe the gravitational phenomena as the manifestation of the curvature of spacetime. Near a gravitational object, two freely falling particles which move through two infinitely close geodesics, accelerate towards one another due to tidal effect. This effect is directly connected to the curvature of spacetime.

%

\clearpage  
\lhead{\emph{Chapter 3}}  
  \chapter{Metric gravity and Palatini formalism}
\label{chapter 3}
\vspace{-0.5 cm}
\epigraph{\textit{I was sitting in a chair in the patent office at Bern when all of a sudden a thought occurred to me: \enquote{If a person falls freely he will not feel his own weight.} I was startled. This simple thought made a deep impression on me. It impelled me towards a theory of gravitation.}\, ---Albert Einstein}{}

\section{General theory of relativity}
\label{sec:gr}

Newtonian theory of gravity has been successful in describing the motion of bodies  under the influence of the force of gravity, not only of the Earth but also of all the planets of the Solar System. This has extremely determined the correct trajectories of these planets around the Sun. However, it has been known for a long time that Newtonian gravity fails in explaining the shifts observed in the orbit of planet Mercury, the closest planet to the Sun. It turns out that this planet follows only an approximate ellipse where its closest approach to the Sun, called perihelion, moves slightly around the Sun. This has been known as the \textit{anomalous precession} of the perihelion of Mercury, and its explanation by general relativity, as we will see in this section, is considered as one of the accurate tests of the later.     

Besides the anomalous precession, Newton's theory of gravity is simply based on the classical concepts of absolute space and time. These concepts have been rejected after the new study of classical electrodynamics made in 1905, which has led to the formulation of special relativity. People believed then that a successful theory of gravity has to be based on the new \enquote{spacetime} concept.

\subsection{Equivalence principle and Einstein's field equations}

A remarkable property of gravity which has been noticed by Galileo and then by Newton himself is that under the same initial conditions, all bodies fall in the same manner in a gravitational field and they gain a unique acceleration independently of their masses. This observable fact which is restricted to gravitational forces has been accurately tested on earth \cite{etvos}. Einstein has realized that this property can be viewed by observers located in a \textit{noninertial} reference frames in the absence of gravity. In fact, bodies which are at rest or in a uniform motion relative to an inertial frame are seen to be in a freely falling state by a noninertial observer. Einstein then states his \textit{equivalence principle} \cite{landau,misner}:
\begin{center}
\textit{Noninertial reference frames are equivalent to some gravitational fields.}
\end{center}    
This means that one may eliminate a given gravitational field in a region of space by referring to a noninertial reference. However, since gravity becomes negligible and vanishes at farther points (from the source), it is then impossible to find a noninertial frame that is able to eliminate the gravitational field at \textit{infinity}.     

The physics of special relativity briefly described in chapter 1 ignores completely any gravitational field, and since this later becomes equivalent to a certain accelerated (noninertial) reference frame, one may directly think about the form of the physical laws under \textit{general coordinate transformations}. In effect, a uniform accelerated reference frame is described by general coordinate system.

One of the important \enquote{geometrical} quantities that depends on the general coordinates is the metric tensor $g_{\mu\nu}$ which gives the invariant line element (\ref{general line element}). Eliminating a gravitational field then, means finding a (general) coordinate system which reduces the tensor $g_{\mu\nu}$ to Minkowsi metric tensor $\eta_{ab}$
\begin{eqnarray}
\label{constant components of the metric}
g_{00}=-1, \quad \quad g_{11}=1, \quad \quad g_{22}=1, \quad \quad g_{33}=1.
\end{eqnarray}
This turns out to be impossible, since the gravitational field does not vanish everywhere. The failure of finding such a transformation where the metric has the components (\ref{constant components of the metric}) means that the spacetime is \enquote{curved}. This remarkable fact means that the metric tensor responsible for distances and angles is not geometrically fixed as in (\ref{constant components of the metric}) but it describes a physical phenomenon; \textit{the gravitational field}. 

For an inertial observer, a free particle is known to be at rest or in a uniform motion, and its path is described in terms of the coordinates $x^{\mu}$ by 
\begin{eqnarray}
\label{free particle}
\frac{d^{2}x^{\mu}}{d\tau^{2}}=0,
\end{eqnarray}
where the parameter $\tau$ refers to the proper\footnote{Proper time is the duration measured in an inertial frame in which two events are simultaneous.} time.

This equation represents the shortest path (straight line) the particle can take between two points $A$ and $B$. Here, the concept of distance is essential, and then, equation (\ref{free particle}) is nothing but an extrema of the line element
\begin{eqnarray}
\label{exterma of distance}
s=\int_{A}^{B} \sqrt{g_{\mu\nu}dx^{\mu}dx^{\nu}},
\end{eqnarray}
where $g_{\mu\nu}$ has the constant components (\ref{constant components of the metric}).

Since gravitational phenomenon is equivalent to a certain noninertial reference frame, the equations of motion of a freely falling particle is derived then from the same line element (\ref{exterma of distance}) but for general coordinates where $g_{\mu\nu}$ is no longer constant. 

Now, we suppose that a particle follows a curve which is parametrized by a parameter $\lambda$ as $x^{\mu}(\lambda)$. The quantity (\ref{exterma of distance}) is a parameter independent, and we may easily apply the variational principle on the following invariant 
\begin{eqnarray}
\int_{\lambda_{1}}^{\lambda_{2}} d\lambda L\left( x^{\mu},\frac{dx^{\mu}}{d\lambda}\right),
\end{eqnarray}
where the Lagrangian function is given by
\begin{eqnarray}
\label{lagrangian-length}
L\left(x^{\mu},\frac{dx^{\mu}}{d\lambda}\right)=
\sqrt{g_{\mu\nu}\frac{dx^{\mu}}{d\lambda}\frac{dx^{\nu}}{d\lambda}}.
\end{eqnarray}
As in special relativity, the particle here follows a time-like worldline between the two points A and B in spacetime. This world line simply extremise the \textit{proper time}; the time measured by the clock along this world line. Now, if we fix the end points such that for any deviation from the curve, noted $\delta x^{\mu}$, satisfies $\delta x^{\mu}(\lambda_{1})=\delta x^{\mu}(\lambda_{2})=0$, the variational principle implies the Euler-Lagrange equations 
\begin{eqnarray}
\label{euler-lagrange}
\frac{\partial L}{\partial x^{\mu}}=
\frac{d}{d\lambda}\left(\frac{\partial L}{\partial (dx^{\mu}/d\lambda) } \right).
\end{eqnarray}
Using the explicit form of the Lagrangian (\ref{lagrangian-length}), the Euler-Lagrange equation (\ref{euler-lagrange}) leads to the following equation 
\begin{eqnarray}
\frac{d^{2}x^{\mu}}{d\lambda^{2}}+
\frac{1}{2}g^{\mu\sigma}\left( \partial_{\alpha} g_{\sigma\beta}
+\partial_{\beta} g_{\alpha\sigma}-\partial_{\sigma} g_{\alpha\beta} \right)\frac{d x^{\alpha}}{d \lambda}
\frac{d x^{\beta}}{d \lambda}=0.
\end{eqnarray}
This can be written equivalently in terms of the proper time $\tau$ as
\begin{eqnarray}
\label{geodesic equation-metric}
\frac{d^{2}x^{\mu}}{d\tau^{2}}
=-\Gamma^{\mu}_{\,\, \alpha \beta} (g) \frac{d x^{\alpha}}{d \tau}
\frac{d x^{\beta}}{d \tau},
\end{eqnarray}
where we have put
\begin{eqnarray}
\label{metric connection}
\Gamma^{\mu}_{\,\, \alpha \beta} (g)=
\frac{1}{2} g^{\mu \lambda} 
\left(\partial_{\alpha} g_{\lambda \beta}
+\partial_{\beta} g_{\lambda \alpha}
-\partial_{\lambda} g_{\alpha \beta}
 \right).
\end{eqnarray}
Straightforward comparison with equation (\ref{free particle}) of the free particle, shows that the latter is simply the limit of (\ref{geodesic equation-metric}) when the metric tensor takes the constant components (\ref{constant components of the metric}). This is valid only locally at a point P, where an inertial observer uses $g_{\mu\nu}|_{P}\equiv \eta_{\mu\nu}$ to label distances and time intervals. Thus, the quantity (\ref{metric connection}) vanishes at this frame, $\Gamma_{\alpha\beta}^{\mu}(\eta)=0$. This quantity defines a \textit{particular} connection, called the \textit{Levi-Civita connection}, and it is a unique function of the metric and its first derivatives.

Equation (\ref{geodesic equation-metric}) is also called the \textit{geodesic} equation. But unlike the general form (\ref{affine geodesic equation}), it is written in terms of the metric connection and not the affine connection. The geodesics, the world lines described by the geodesic equation (\ref{geodesic equation-metric}), are the shortest paths in spacetime and they generalize the concept of straight lines in Euclidean (flat space) geometry. Interestingly, along these special world lines, the metric which is responsible for distances and angles, is parallel transported, or 
\begin{eqnarray}
\label{metric compatibility}
\nabla_{\mu} g_{\alpha\beta}=0,
\end{eqnarray}
where in this case, the covariant derivative is with respect to the metric connection (\ref{metric connection}).

The last equation is called the \textit{compatibility} condition, and if it is solved with respect to an arbitrary connection, it leads to the unique Levi-civita connection (\ref{metric connection}).

The general theory of relativity, which we are exploring here, is essentially based on this metrical structure, where the metric tensor plays a crucial role in defining the gravitational field. In fact, the gravitational effects on the freely falling particle appear on the right hand side of equation (\ref{geodesic equation-metric}) as first derivatives of the metric. One may see this fact clearly if the test particle is non-relativistic and plunged in a weak gravitational field. This case is exactly the Newtonian description of gravity, and it is attained from a very weak deviation from the flatness of spacetime. Since our fundamental field is the metric tensor, then this later has to be expanded around the Minkowski metric as 
\begin{eqnarray}
\label{weak metric expansion}
g_{\mu\nu}\simeq \eta_{\mu\nu}+h_{\mu\nu},
\end{eqnarray}
where the tensor $h_{\mu\nu}$ is very small, $|h_{\mu\nu}| \ll 1$.

In this limit, the proper time and the velocities that characterize the motion of the particle must satisfy
\begin{eqnarray}
\tau \rightarrow t, \quad \quad \frac{dx^{0}}{d\tau} \sim c \quad \quad
\frac{dx^{i}}{d\tau} \ll c,
\end{eqnarray}
where the last term represents the components of the three vector velocity which has to be negligible.

Taking these together, one may easily check that the $i$ component of the geodesic equation (\ref{geodesic equation-metric}) takes the form 
\begin{eqnarray}
\frac{d^{2}x^{i}}{dt^{2}} \simeq
-c^{2}\Gamma^{i}_{\,\, 00}.
\end{eqnarray}
The last term can be calculated in terms of the components of the metric by using equation (\ref{metric connection}) where the inverse of the metric is given by
\begin{eqnarray}
g^{\mu\nu} \simeq \eta^{\mu\nu}-h^{\mu\nu},
\end{eqnarray}
then
\begin{eqnarray}
\Gamma^{i}_{\,\, 00} \simeq
-\frac{1}{2}\eta^{ii}\partial_{i} h_{00}.
\end{eqnarray}
Finally, the Newtonian limit of the geodesic equation reads 
\begin{eqnarray}
\frac{d^{2}x^{i}}{dt^{2}} \simeq
\frac{c^{2}}{2}\nabla_{i} h_{00}.
\end{eqnarray}
Thus, we recover Newton's second law, where the inertial force is given by the gradient of the gravitational potential $\phi$ of a certain gravitational source. An interesting conclusion of all this, is that the component $h_{00}$ of the (weak) metric tensor is nothing but the gravitational potential itself 
\begin{eqnarray}
\label{h00}
h_{00}=-\frac{2\phi}{c^{2}}.
\end{eqnarray}
This result shows the crucial relation between the spacetime geometry (metric tensor) and the gravitational forces.

What is left now is to discover the general evolution of this metric tensor which must be described by some fundamental differential equations that provide us with all the components of the metric. These equations are called Einstein's field equations, and we are now in a position to explore them. 

The Newtonian limit which has led us to the metric component (\ref{h00}) describes only the evolution of test particles not the field itself. The later is described by Laplace equation where the gravitational field propagates in vacuum, or the general Poisson's equations in the presence of the gravitational source itself. The gravitational potential in this case satisfies  
\begin{eqnarray}
\label{poisson equation}
\vec{\nabla}^{2} \phi = 4 \pi G_{N} \rho_{\text{m}},
\end{eqnarray}
where $G_{N}$ being the Newton's constant, and $\rho_{\text{m}}$ is the energy density of the matter distributions that generate the gravitational field.

Again, this equation has to be realized for weak field limits of a general equations of motion that take into account the general coordinate systems (noninertial frames). First, the left hand side of equation (\ref{poisson equation}) contains a second order partial differential operators that acts on the gravitational potential. The later, as we have seen, is one component of the metric tensor in the weak field limit. Thus, a generalization of this side may be realized using second order derivatives of the metric tensor $g_{\mu\nu}$. This directly implies that the right hand side must be represented by a covariant tensor. This implication is clearly understood, since the energy density is not a covariant quantity and it could not enter a relativistic equation of motion. 

The tensorial character of matter is generally represented by the so called Energy-momentum or Stress-energy tensor, noted $T_{\alpha\beta}$. It describes the flux of the four-momentum $p_{\alpha}$ of the particle (or a fluid) across a surface of constant $x^{\beta}$ \cite{schutz}. This definition provides a meaningful energy density which is then the flux of the zero momentum across a surface of constant $t$. A good example of the energy-momentum tensor is Maxwell's stress-energy tensor that describes the electromagnetic field \cite{landau}. 

To that end, Einstein's field equations (tensorial equations) provide a connection between the spacetime geometry (metric and its derivatives) and a source which describes all kind of matter and energy except gravity. These equations are generically written as
\begin{eqnarray}
\label{formal einstein equations}
\left\lbrace g, \partial g, \partial^{2} g \right\rbrace_{\alpha\beta}=
\kappa T_{\alpha\beta},
\end{eqnarray}
where $\kappa$ is a dimensionful parameter which must be related to Newton's constant in order to realize the weak field limit given in (\ref{poisson equation}).

In constructing the object of the left hand side of (\ref{formal einstein equations}), one has to take into account the following necessary properties: 
\begin{enumerate}
\item Like the energy-momentum tensor, this object has to be a rank-two symmetric tensor.
\item It involves up to second derivative of the metric tensor.
\item An important condition that has to be satisfied, is the covariant conservation law. This law must be written in a covariant form (general coordinates) which generalizes the conservation law written in an inertial frame in flat spacetime, thus
\begin{eqnarray}
\nabla^{\alpha} T_{\alpha\beta}=0.
\end{eqnarray} 
The new tensor then has to satisfy the same relation.
\end{enumerate}
It might be difficult to extract the form of this new tensor that satisfies the above conditions, from all the geometric quantities that depend on the metric tensor. Nevertheless, the only symmetric, rank-two tensors that we have at hand are, the metric tensor itself, and the Ricci tensor (\ref{ricci-general}) of the Levi-Civita connection. Indeed, these tensors do not have all the properties given above. However, though not be trivial, the \enquote{requested} tensor can be constructed from these two quantities, and it has the following form
\begin{eqnarray}
\label{einstein tensor plus lambda}
R_{\alpha\beta}-\frac{1}{2}g_{\alpha\beta}g^{\mu\nu}R_{\mu\nu}
+\Lambda g_{\alpha\beta}.
\end{eqnarray}
The first two terms of this quantity that include the Ricci tensor $R_{\alpha\beta}$, form the so called Einstein tensor
\begin{eqnarray}
\label{einstein tensor}
G_{\alpha\beta}=
R_{\alpha\beta}-\frac{1}{2}g_{\alpha\beta}R,
\end{eqnarray}
where the last term on the right hand side is called the Ricci scalar which is the contracted Ricci tensor
\begin{eqnarray}
R= g^{\mu\nu}R_{\mu\nu}.
\end{eqnarray}
The last term in (\ref{einstein tensor plus lambda}) which includes a constant $\Lambda$, is introduced here since it satisfies condition 3. However, in the following chapters, we will explore the physical necessity of this term. Einstein tensor is written in terms of the Levi-Civita connection (\ref{metric connection}) and its first derivative, and it can be constructed easily from the Ricci tensor (\ref{ricci-general}), which in this case is given by
\begin{eqnarray}
&&R_{\alpha\beta}\left[\Gamma(g)\right]=
R^{\mu}_{\,\,\,\alpha\mu\beta}\left[\Gamma(g)\right] \nonumber \\
&&=
\partial_{\mu}\Gamma^{\mu}_{\,\,\alpha\beta}(g)
-\partial_{\beta}\Gamma^{\mu}_{\,\,\alpha\mu}(g)
+\Gamma^{\mu}_{\,\,\sigma\mu}(g)\Gamma^{\sigma}_{\,\,\alpha\beta}(g)
-\Gamma^{\mu}_{\,\,\sigma\beta}(g)\Gamma^{\sigma}_{\,\,\alpha\mu}(g).
\label{ricci tensor terms of metric}
\end{eqnarray}
Here, we briefly show that the combination (\ref{einstein tensor plus lambda}) satisfies the third condition stated above. The main point is that the curvature tensor, which is given by its general form (\ref{riemann tensor}) satisfies the \textit{Bianchi identity} (\ref{bianchi0}) 
\begin{eqnarray}
\label{bianchi-general}
\nabla_{\alpha}R^{\lambda}_{\,\,\,\mu\kappa\nu}
+\nabla_{\nu}R^{\lambda}_{\,\,\,\mu\alpha\kappa}
+\nabla_{\kappa}R^{\lambda}_{\,\,\,\mu\nu\alpha}=0.
\end{eqnarray}
This is a tensorial equation, and in order to prove its validity, we simply check that it holds in a local reference frame at a point $P$, where the connection coefficients $\Gamma^{\lambda}_{\,\,\mu\nu}(P)=0$. Using equation (\ref{riemann tensor}), we have in this frame 
\begin{eqnarray}
\partial_{\alpha}R^{\lambda}_{\,\,\,\mu\kappa\nu}=
\partial_{\alpha}\partial_{\kappa}\Gamma^{\lambda}_{\,\,\mu\nu}
-\partial_{\alpha}\partial_{\nu}\Gamma^{\lambda}_{\,\,\mu\kappa}.
\end{eqnarray}  
With an index permutation, we easily get
\begin{eqnarray}
\partial_{\alpha}R^{\lambda}_{\,\,\,\mu\kappa\nu}
+\partial_{\nu}R^{\lambda}_{\,\,\,\mu\alpha\kappa}
+\partial_{\kappa}R^{\lambda}_{\,\,\,\mu\nu\alpha}=0,
\end{eqnarray}
which is exactly the identity (\ref{bianchi-general}) in the chosen local frame. 

Thus, the Bianchi identity (\ref{bianchi-general}) holds for every coordinate system. Now, what we need is to contract this identity two times, the first contraction $(\lambda \leftrightarrow \kappa)$ leads to
\begin{eqnarray}
\nabla_{\alpha}R_{\mu\nu}+\nabla_{\nu}R^{\lambda}_{\,\,\,\mu\alpha\lambda}
+\nabla_{\lambda}R^{\lambda}_{\,\,\,\mu\nu\alpha}=0,
\end{eqnarray}
and the second contraction (multiplying by $g^{\mu\alpha}$) gives
\begin{eqnarray}
2\nabla^{\mu}R_{\mu\nu}-\nabla_{\nu}R=0.
\end{eqnarray}
This equation gives us the required identity (condition three above)
\begin{eqnarray}
\nabla^{\mu} \left(R_{\mu\nu}-\frac{1}{2}g_{\mu\nu}R \right)=0,
\end{eqnarray}
where the last term in equation (\ref{einstein tensor plus lambda}) can be trivially added here due to the compatibility condition (\ref{metric compatibility}).

Finally, the gravitational field equations, or Einstein equations, are a system of second order differential equations of the metric sourced by energy-momentum that generates the curvature of spacetime, and are written as
\begin{eqnarray}
\label{ad hoc einstein's equations}
G_{\alpha\beta}(g)+\Lambda g_{\alpha\beta}=\kappa T_{\alpha\beta}.
\end{eqnarray}
Before discussing the properties of these equations and determining the constant $\kappa$, we will first give a detailed derivation of these equations based on a principle of least action.

\subsection{Variational principle: Einstein-Hilbert action}
The equation of motion of a test particle (\ref{geodesic equation-metric}) has been derived from the extrema of the invariant length (\ref{exterma of distance}). This procedure is called the principle of least action, and it is well known in analytical mechanics of point particles and its generalization to field theories. In a physical theory, the system in question is described by fundamental fields that satisfy (field) equations that govern their behavior. These equations are equivalent to the Euler-Lagrange equations derived from a certain action through the calculus of variation \cite{lovelock}. To proceed then, one needs a covariant integral, an integral which is invariant under general coordinate transformations.

In special relativity, integrals contain functions which are Lorentz invariant. This simply works well since the four dimensional integrals in flat space are invariant too. In fact, under Lorentz transformation (\ref{lorentz transformation}), the volume element transforms as
\begin{eqnarray}
\label{flat volume element}
d^{4}\hat{x}= d^{4}x ||\Lambda || = d^{4}x, \quad \text{since}\quad
||\Lambda ||=1,
\end{eqnarray}
where $||.||$ refers to the determinant sign.

Thus, the \textit{volume element} itself is invariant under Lorentz transformations (or in flat spacetime). However, the volume element (\ref{flat volume element}) is not generally covariant due to the Jacobian term that appears under general coordinate transformations $\hat{x}(x)$
\begin{eqnarray}
\label{general volume element}
d^{4}\hat{x}= d^{4}x \left|\left| \frac{\partial \hat{x}}{\partial x }\right| \right|,
\end{eqnarray}
where the right hand side includes the trivial determinant of the Jacobian matrix $J=\partial \hat{x}^{\alpha}/\partial x^{\beta}$ of the diffeomorphism $x^{\beta} \rightarrow \hat{x}^{\alpha}(x)$.

Nevertheless, one may simply notice that the volume elements might be generalized by introducing a quantity which eliminates the Jacobian under general coordinate transformations. Although there are some different way out to this problem, we concentrate here on the general concept of volume itself. Volumes are like areas and lengths, and their measurements have to be frame (coordinate) independent. A relevant covariant quantity is the metric tensor $g_{\mu\nu}$. While it provides a prescription for measuring volumes, the metric is a rank-two tensor, and then it is the quantity that one needs to cancel out the unwanted (Jacobian) term in the volume element (\ref{general volume element}). In fact, under general coordinate transformations, it transforms as
\begin{eqnarray}
\hat{g}_{\alpha\beta}(\hat{x})=
\frac{\partial x^{\mu}}{\partial \hat{x}^{\alpha}}
\frac{\partial x^{\nu}}{\partial \hat{x}^{\beta}}
g_{\mu\nu}(x).
\end{eqnarray}
The quantity that can be constructed from the metric tensor and which includes the determinant of the Jacobian matrix, is then the square root of its determinant 
\begin{eqnarray}
\label{determinant transformation}
\sqrt{||\hat{g}||}=
\left|\left| \frac{\partial \hat{x}}{\partial x }\right| \right|^{-1}
\sqrt{||g||}.
\end{eqnarray}
This quantity transforms as a scalar density of \textit{weight} $w=+1$ \cite{lovelock}.

Taking the product of the relations (\ref{general volume element}) and
(\ref{determinant transformation}) we get the correct covariant four-dimensional volume element 
\begin{eqnarray}
\label{metric covariant volume element}
d^{4}\hat{x} \sqrt{||\hat{g}||}=
d^{4}x \sqrt{||g||}.
\end{eqnarray}
This defines and generalize the volume measure of flat space, and  theory which is a coordinate free can then be derived by integrating scalars in the spacetime manifold. Additionally, all fields that form the required scalars will automatically be coupled to geometry due to the presence of the metric tensor in the volume itself. This leads to what is called minimal coupling to gravity.  

To derive Einstein's field equations (\ref{ad hoc einstein's equations}), we need to construct an action based on the integral of the volume element (\ref{metric covariant volume element}) and contains the following quantities:
\begin{enumerate}
\item 
A geometric scalar that includes the metric tensor and its derivatives up to second order. This part is essential in obtaining the geometric part of the field equations which is given by Einstein's tensor (\ref{einstein tensor}). The relevant quantity is the Ricci scalar $R(g)$.
\item
A second geometric term that gives the covariant part $\Lambda g_{\alpha\beta}$ in the field equations (\ref{ad hoc einstein's equations}). This can be obtained simply by adding a constant $\Lambda$. 
\item
A scalar that depends on the matter fields that fill the space and generate the gravitational field. To come out with a correct scalars, this part may include the metric tensor itself, and in this case, we say that matter is coupled minimally to gravity (the metric). However, in what follows, we will assume that this part can generally be described by a scalar $L_{\text{m}}(g)$ (a Lagrangian of matter).
\end{enumerate}
The generally invariant action which has the above properties takes the following form \cite{misner,blau} 
\begin{eqnarray}
\label{e-h action plus matter}
\text{S}[g]=  
\frac{1}{2\kappa} \int d^{4}x 
\sqrt{||g||} \left(
 R(g)
 -2\Lambda
\right) + \int d^{4}x \sqrt{||g||}\, L_{\text{m}}.
\end{eqnarray}
The first part of the integral which includes only the Ricci scalar is called Einstein-Hilbert action, and the second term $\Lambda$ is called the cosmological constant. The last part of the action describes the general form of matter interacting with gravity. 

The action will be varied with respect to the fundamental field; the metric tensor. To that end, it is useful to show briefly how to make the variation of the determinant, the important quantity in all parts of the integral. With the help of the matrix properties, the determinant of any matrix $\textbf{M}$ can be written as \cite{blau} 
\begin{eqnarray}
||\textbf{M}||=e^{\text{Tr} \log \textbf{M}}.
\end{eqnarray}
This leads to
\begin{eqnarray}
\label{matrix variation}
\delta ||\textbf{M}|| =||\textbf{M}||\,
\text{Tr}\left[\textbf{M}^{-1}\delta \textbf{M}\right].
\end{eqnarray}
This is easily applied to the metric tensor and leads to the important formula
\begin{eqnarray}
\label{delta g}
\delta \sqrt{||g||}=
\frac{1}{2}\sqrt{||g||}\,g^{\alpha\beta}\delta g_{\alpha\beta}.
\end{eqnarray}
One can write this differently, by noticing that
\begin{eqnarray}
g^{\alpha\lambda}g_{\lambda\beta}=\delta^{\alpha}_{\beta},\quad \quad \text{then} \quad \quad g^{\alpha\beta}\delta g_{\alpha\beta}=
-g_{\alpha\beta}\delta g^{\alpha\beta}.
\end{eqnarray}
Finally
\begin{eqnarray}
\label{variation of the metric}
\delta \sqrt{||g||}&&=-
\frac{1}{2}\sqrt{||g||}\, g_{\alpha\beta}\delta g^{\alpha\beta}.
\end{eqnarray}
Variation of the Ricci scalar scalar reads
\begin{eqnarray}
\delta R(g)=R_{\alpha\beta}\delta g^{\alpha\beta}
+g^{\alpha\beta}\delta R_{\alpha\beta}. 
\end{eqnarray}
The last part will be evaluated using the explicit form of the Ricci tensor (\ref{ricci tensor terms of metric}). Briefly, the part $\delta R_{\alpha\beta}$ includes terms such as $\partial(\delta \Gamma)$ and $\Gamma \delta \Gamma$, where $\Gamma$ is again the Levi-Civita connection (indices are implicitly included). In chapter 2, we have seen that, unlike $\Gamma$, the term $\delta \Gamma$ is a true tensor. One may then apply the covariant derivatives on this tensor as
\begin{eqnarray}
\label{covariant derivative of delta gamma}
\nabla_{\mu}(\delta \Gamma^{\lambda}_{\,\,\alpha\beta})=
\partial_{\mu}(\delta \Gamma^{\lambda}_{\,\,\alpha\beta})
+\Gamma^{\lambda}_{\,\,\kappa\mu}\delta \Gamma^{\kappa}_{\,\,\alpha\beta}-\Gamma^{\kappa}_{\,\,\mu\alpha}\delta \Gamma^{\lambda}_{\,\,\kappa\beta}-\Gamma^{\kappa}_{\,\,\mu\beta} \delta \Gamma^{\lambda}_{\,\,\alpha\kappa}.
\end{eqnarray}
Now, by using the expressions (\ref{ricci tensor terms of metric}) and (\ref{covariant derivative of delta gamma}) we can show the following important relation
\begin{eqnarray}
\label{delta ricci tensor}
\delta R_{\alpha\beta}=
\nabla_{\lambda}\left(\delta \Gamma^{\lambda}_{\,\,\alpha\beta} \right)
-\nabla_{\beta}\left(\delta\Gamma^{\lambda}_{\,\,\alpha\lambda} \right).
\end{eqnarray}
Although we have referred to the Levi-Civita connection, the last relation is completely independent on the metric tensor and it holds for an arbitrary (symmetric) affine connection. Expression (\ref{delta ricci tensor}) will be very important in the affine dynamics in the next chapter. 

Having the two expressions (\ref{variation of the metric}) and (\ref{delta ricci tensor}) at hand, variation of action (\ref{e-h action plus matter}) takes the following form
\begin{eqnarray}
\label{variation of e-h action}
\delta \text{S}=&&
\frac{1}{2\kappa}
\int d^{4}x \sqrt{||g||}
\left(R_{\alpha\beta}-\frac{1}{2}g_{\alpha\beta}R 
+\Lambda g_{\alpha\beta} \right)
\delta g^{\alpha \beta} \nonumber \\
&&+\frac{1}{2\kappa} \int d^{4}x \sqrt{||g||}\left[ g^{\alpha\beta}\left(\nabla_{\lambda}(\delta \Gamma^{\lambda}_{\,\,\alpha\beta})-\nabla_{\beta}(\delta \Gamma^{\lambda}_{\,\,\alpha\lambda}) \right)\right]
\nonumber \\
&&+\int d^{4}x \sqrt{||g||}
\left(\frac{\delta L_{\text{m}}}{\delta g^{\alpha\beta}} 
-\frac{1}{2}g_{\alpha\beta}L_{\text{m}}
\right)\delta g^{\alpha\beta}.
\end{eqnarray}
Now since the covariant derivative of the metric tensor (and its inverse) vanishes, then the second term in (\ref{variation of e-h action}) takes the form
\begin{eqnarray}
\label{surface term gr}
\int d^{4}x \sqrt{||g||}
\left[\nabla_{\lambda}(g^{\alpha\beta} \delta \Gamma^{\lambda}_{\,\,\alpha\beta})-\nabla_{\beta}(g^{\alpha\beta} \delta \Gamma^{\lambda}_{\,\,\alpha\lambda}) \right]=
\int d^{4}x \sqrt{||g||}\, \nabla_{\lambda} V^{\lambda},
\end{eqnarray}
where we have defined the vector $V^{\mu}$ as
\begin{eqnarray}
V^{\lambda}=
g^{\alpha\beta} \delta \Gamma^{\lambda}_{\,\,\alpha\beta}-g^{\alpha\lambda} \delta \Gamma^{\kappa}_{\,\,\alpha\kappa}
\end{eqnarray}
Explicitly, its covariant divergence is written as
\begin{eqnarray}
\label{trace of covariant derivative}
\nabla_{\lambda}V^{\lambda}=
\partial_{\lambda}V^{\lambda}+\Gamma^{\lambda}_{\,\,\kappa\lambda}\,V^{\kappa},
\end{eqnarray}
where the \enquote{trace} of the connection on the last term can be easily calculated in terms of the metric from (\ref{metric connection}) as
\begin{eqnarray}
\Gamma^{\lambda}_{\,\,\kappa\lambda}=\frac{1}{2}g^{\mu\nu}\partial_{\kappa} g_{\mu\nu}.
\end{eqnarray}
As we have done for the variation of the metric tensor to derive the expression (\ref{delta g}), similarly, the right had side of the last equation can be written in terms of the determinant of the metric tensor. Thus
\begin{eqnarray}
\Gamma^{\lambda}_{\,\,\kappa\lambda}=
\frac{1}{2}\frac{\partial_{\kappa}||g||}{||g||} \equiv
\frac{\partial_{\kappa}\sqrt{||g||}}{\sqrt{||g||}}.
\end{eqnarray}
The last expression is substituted into the covariant derivative (\ref{trace of covariant derivative}) leading to the useful expression of the covariant divergence
\begin{eqnarray}
\nabla_{\lambda} V^{\lambda}=
\frac{1}{\sqrt{||g||}}\partial_{\lambda}\left(\sqrt{||g||}V^{\lambda} \right).
\end{eqnarray}
What is remarkable here is that with this expression, the term (\ref{surface term gr}) is nothing but a total ordinary divergence. Finally, the variation of the gravitational action takes the form
\begin{eqnarray}
\label{delta s final}
\delta \text{S}=
\int d^{4}x \sqrt{||g||}
\left\lbrace \frac{1}{2\kappa}\left( G_{\alpha\beta}(g)+\Lambda g_{\alpha\beta}\right)-\frac{1}{2}T^{\text{m}}_{\alpha\beta} 
\right\rbrace \delta g^{\alpha\beta}+\int d^{4}x \partial_{\lambda}V^{\lambda},
\end{eqnarray}
where the last term is called the energy-momentum tensor of matter
\begin{eqnarray}
\label{energy momentum tensor of matter}
T^{\text{m}}_{\alpha\beta}= -\frac{2\delta L_{\text{m}}}{\delta g^{\alpha\beta}} 
+g_{\alpha\beta}L_{\text{m}}.
\end{eqnarray}
It is important to shed light on the nature of the last term of (\ref{delta s final}). At first glance one may think that this term is a total divergence and then by applying the Stokes's theorem, this integral vanishes. However, this is not trivial as one may think. The Stokes's theorem would lead to a surface term which vanishes at infinity if the metric variation vanishes at infinity. The essential problem is that the integral at hand depends on the derivative of the metric variation too, and then it will clearly contribute to the boundary term. This is a consequence of the nonlinearity of action (\ref{e-h action plus matter}) in the metric, which requires an additional term in order to cancel the unwanted boundary contribution \cite{blau}.

Finally, \enquote{assuming} that the last term in (\ref{delta s final}) does not contribute to the total variation, we get the gravitational field equations, namely Einstein's equations with matter
\begin{eqnarray}
\frac{1}{2\kappa}\left( G_{\alpha\beta}(g)+\Lambda g_{\alpha\beta}\right)-\frac{1}{2}T^{\text{m}}_{\alpha\beta}=0
\end{eqnarray}
or in a standard form
\begin{equation}
\label{final einstein equations}
G_{\alpha\beta}(g)+\Lambda g_{\alpha\beta}=
\kappa T_{\alpha\beta}^{\text{m}}.
\end{equation}

The simplest case is Einstein's field equations in vacuum, this is the case where space is free of matter and any source of gravity, then $T_{\alpha\beta}^{\text{m}}=0$ and $\Lambda=0$. One solution of Einstein's equations in this case is clearly the flat spacetime metric (\ref{constant components of the metric}). In the other cases, the solutions are curved spacetime metrics which are described by components of the energy-momentum tensor. These can be a mass, an energy density and pressure...etc. Another simple but important case, is the spacetime with only a cosmological constant, $T_{\alpha\beta}^{\text{m}}=0$ but $\Lambda \neq 0$. The spacetime curvature in this case is caused by a source of constant energy and pressure and it could be described by an energy-momentum tensor of the form  
\begin{eqnarray}
\label{energy-momentum tensor of lambda}
T_{\alpha\beta}^{\Lambda}=-\frac{\Lambda}{\kappa}\, g_{\alpha\beta}.
\end{eqnarray}
In general relativity, this tensor can be simply postulated to be zero if the cosmological constant $\Lambda$ vanishes. This can be realized from the beginning in the action (\ref{e-h action plus matter}). However, for some reasons which will be clear in the next chapters, this cosmological constant may receive nonzero contributions from different sources, the case that makes it impossible to vanish. We will call the tensor (\ref{energy-momentum tensor of lambda}), the energy-momentum tensor of \textit{vacuum}. But here let us take $\Lambda=0$ for simplicity

To determine the constant $\kappa$, one has to make the weak field limit where Newtonian gravity is valid. To simplify the calculation, it is useful to write the left hand side of Einstein's equations (\ref{final einstein equations}) in terms of only the Ricci tensor. By contracting both sides of equation (\ref{final einstein equations}), we get 
\begin{eqnarray}
-R=\kappa T^{\text{m}}, \quad \quad \text{where}\quad T^{\text{m}}=g^{\alpha\beta}T_{\alpha\beta}^{\text{m}}.
\end{eqnarray}
Now we substitute this in equation (\ref{final einstein equations}) and finally we get 
\begin{eqnarray}
\label{einstein equations different writing }
R_{\alpha\beta}=\kappa \left(T_{\alpha\beta}^{\text{m}}
-\frac{1}{2} g_{\alpha\beta}T^{\text{m}}\right),
\end{eqnarray} 
which are Einstein's field equations written differently.

Under the weak field approximation (\ref{weak metric expansion}), the Ricci tensor (\ref{ricci tensor terms of metric}) takes the form 
\begin{eqnarray}
\label{weak ricci}
R_{\alpha\beta}=
\frac{1}{2}\eta^{\kappa\sigma}\left( 
\partial_{\alpha}\partial_{\kappa}h_{\sigma\beta}
+\partial_{\sigma}\partial_{\beta}h_{\alpha\kappa}
-\partial_{\sigma}\partial_{\kappa}h_{\alpha\beta}
-\partial_{\alpha}\partial_{\beta}h_{\sigma\kappa}
\right).
\end{eqnarray}
Then, the $00$ component reads
\begin{eqnarray}
\label{weak poisson equation}
R_{00}=-\frac{1}{2}\eta^{\kappa\sigma} \partial_{\sigma}\partial_{\kappa}h_{00}
=-\frac{1}{2}\vec{\nabla}^{2}h_{00},
\end{eqnarray}
where we have taken a static field, $\partial_{0}h_{00}=0$.

If matter is taken pressureless, which is a good approximate case, the $00$ component of the right hand side of equation (\ref{einstein equations different writing }) will be given then by the energy density as \cite{schutz}
\begin{eqnarray}
T_{00}^{\text{m}}
-\frac{1}{2} g_{00}T^{\text{m}}=\frac{1}{2}\rho.
\end{eqnarray}
Now remember that the component of the perturbed metric $h_{00}$ is given in terms of the gravitational potential $\phi$ as in (\ref{h00}), and then the weak field limit of the gravitational equations (\ref{weak poisson equation}) reads 
\begin{eqnarray}
\frac{1}{c^{2}}\vec{\nabla}^{2}\phi=\frac{c^{2}\kappa}{2}\rho.
\end{eqnarray}
This equation must coincide with the Poisson equation (\ref{poisson equation}), and finally the constant $\kappa$ becomes  
\begin{eqnarray}
\kappa = \frac{8\pi G_{N}}{c^{4}}.
\end{eqnarray}
In fundamental units, this defines an inverse of a square of a mass, called the \textit{Planck} mass $M_{Pl}$. In fact, it is believed that in addition to Newton's gravitational constant, the fundamental constants $c$ (speed of light) and the Planck constant $\hbar$ when combined together lead to a fundamental scale at which gravity is \enquote{supposed} to gain a quantum description ! This scale is embodied in the following quantities which define a mass, length and a time respectively   
\begin{eqnarray}
M_{Pl}=\sqrt{\frac{\hbar c}{G_{N}}}\simeq 10^{19}\,\text{GeV} \\ 
l_{Pl}=\sqrt{\frac{\hbar G_{N}}{c^{3}}}\simeq 10^{-33}\,\text{cm}\\
t_{Pl}=\sqrt{\frac{\hbar G_{N}}{c^{5}}}\simeq 10^{-44}\,\text{sec}.
\end{eqnarray}
In the following chapters, we will use Planck mass rather than Newton's constant, and we will take (with $\hbar =c=1$)
\begin{eqnarray}
M^{2}_{Pl}\simeq (8\pi G_{N})^{-1}.
\end{eqnarray}

\subsection{Testable predictions of general relativity}

Like every physics theory, Einstein's general relativity had to be confronted with experiment. In the following, we present the tests of Einstein's gravity which are made in the Solar System, some of these tests are proposed by Einstein himself when he formulated his theory. The details behind understanding the proposed tests are based on particular (spherically symmetric) solution of Einstein equations (\ref{final einstein equations}) called \textit{Schwarzschild} solution. Solutions of Einstein's field equations are not the aim of this thesis, however one may find more details in the references given below.
  
\begin{itemize}
\item \textbf{Bending of light (Gravitational lensing)}:

One of the phenomena that has not been known or detected before the appearance of general relativity is the \textit{deflection} or bending of light by a gravitating mass. Einstein proposed this fact as a first possible test of his theory. In flat spacetime, light travels in straight lines, i.e, the geodesics of the flat geometry. In the presence of a massive object, like the Sun, light rays will follow the geodesics of the new curved background around this object, leading to a deflection of its path as shown in Figure \ref{fig:deflection}. 
\begin{figure}[h]
\centering
    \includegraphics[width=0.4\textwidth]{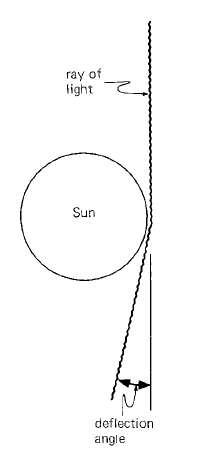}
\caption{Light rays coming from a star are deflected by the gravitational field of the Sun (curvature of spacetime around it) \cite{harrison}.}
\label{fig:deflection}
\end{figure}
Detailed calculation based on \textit{Schwarzschild} solution, show that the angle of deflection is given in terms of the mass of the object $M$ as follows        
\begin{eqnarray}
\delta \phi = \frac{4G_{N}M}{bc^{2}},
\end{eqnarray}
where $b$ is the \textit{apparent impact parameter} \cite{wald}.

This predicts a bending of $1.75$ second of arc, which has been confirmed by Eddington in 1919 during a Solar eclipse \cite{weinberg, eclipse}.

\item \textbf{Precession of planet Mercury}:

Newtonian theory of gravity predicts closed paths of the planets in the Solar System in agreement with elliptical orbits observed by Kepler. Neglecting any possible gravitational perturbation from other objects, the angle swept out by any planet during one revolution is indeed $\Delta \varphi= 2\pi$. However, in the case of planet Mercury, an \textit{anomalous perihelion} shift (from $2\pi$) given by \cite{weinberg,blau}  
\begin{eqnarray}
\delta \varphi =43.11^{''} \pm 0.45^{''} \quad \text{per century}
\end{eqnarray} 
remained unexplained in the context of Newton's theory even by considering the gravitational perturbations of nearby planets.  

The \textit{Schwarzschild} solution provides a deviation from the closed elliptical orbits. The deviation is translated by the following predicted perihelion shift \cite{weinberg,blau} 
\begin{eqnarray}
\delta \varphi =
\frac{6\pi G_{N}}{c^{2}}\left(\frac{M}{a(1-e^{2})} \right),
\end{eqnarray}
where $M$ is the mass of the Sun, $a$ is the semi-major axis of the orbit and $e$ is the so called eccentricity.

This has led to a value $\delta \varphi = 43.04^{''}$ per century, which is precisely the well known precession of the perihelion (see Figure~\ref{fig:mercury} below).  

Planet Mercury is the closest planet to the Sun where the (strong) gravitational effects need relativistic corrections.   

\begin{figure}[h]
\centering
    \includegraphics[width=0.4\textwidth]{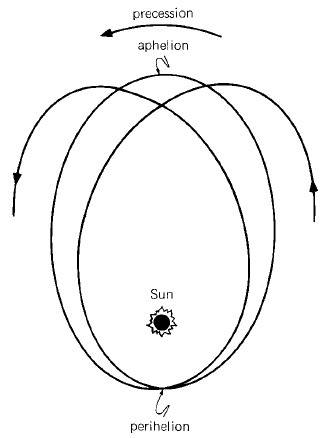}
\caption{Diagram illustrating the precession of the perihelion of Planet Mercury \cite{harrison}.}
\label{fig:mercury}
\end{figure}
\newpage
\item \textbf{Gravitational redshift}:

For a distant observer, an atomic light spectrum emitted near a gravitational field is redshifted. In fact, at a fixed point in space, i.e, $dx^{i}=0$, the passage of time is given by the proper time, $cd\tau=\sqrt{g_{00}}dt$, which is written near a gravitational field as
\begin{eqnarray}
d\tau = \sqrt{1-\frac{2\phi}{c^{2}}}\, dt
\simeq \left(1-\frac{\phi}{c^{2}}\right) dt,
\end{eqnarray}
where we have used the weak field (\ref{h00}).

The previous equation gives us the difference between the time rates, near the gravitational field and at the infinity  
\begin{eqnarray}
d \tau = \sqrt{1-\frac{2\phi}{c^{2}}}\, dt
\simeq \left(1-\frac{\phi}{c^{2}}\right) d \tau_{\infty}.
\end{eqnarray}
Near the gravitational field of Earth, where $\phi(r)=-G_{N}M/r$, with $M$ being the mass of the Earth, the last equation leads to a difference in the frequencies given by
\begin{eqnarray}
\frac{\Delta \nu}{\nu}=\frac{gz}{c^{2}},
\end{eqnarray}
where $z$ is the hight from the Earth's surface, and $g \simeq 9.78 m/s^{2}$ is the gravitational acceleration of the Earth.

Thus, the frequency of light emitted by a source near the gravitational field will be subjected to red (blue) shifts. This gravitational frequency shifts is a direct consequence of the equivalence principle itself, and it has been firstly confirmed with a terrestrial experiment made by Pound and Rebka in 1959 \cite{weinberg, pound-rebka}. An accurate test of the gravitational redshift has been performed around 1976, using a Hydrogen maser in rocket \cite{maser in rocket, will}.

\item \textbf{Gravitational waves}:

An astonishing prediction of the general theory of relativity is the \textit{gravitational waves}. We have seen that weak deviations from flatness of spacetime would lead to the equation of motion of the weak field $h_{00}$ which has been interpreted as the gravitational potential in the Newtonian limit. In general, for the weak curvature (\ref{weak ricci}), the Einstein's field equations in vacuum take the form
\begin{eqnarray}
\label{wave}
\Box h_{\mu\nu}=0,  
\end{eqnarray}
where the d'Alembert operator is given by 
\begin{eqnarray}
\Box=\eta^{\alpha\beta}\partial_{\alpha}\partial_{\beta}=\vec{\nabla}^{2}-\frac{\partial^{2}}{c^{2}\partial t^{2}}.
\end{eqnarray}
Detailed discussion about the derivation of equation (\ref{wave}) and the gauge choices is beyond the scope of our thesis, and for more details we refer the reader to textbooks like \cite{misner}. 

Equation (\ref{wave}) is a standard wave equation which describes the \textit{propagation} of the weak field $h_{\mu\nu}$. These spacetime \textit{ripples}, or gravitational waves, have been supposed to be generated from different astrophysical sources, like the merging of two compact objects such as neutron stars and black holes \cite{hartle,misner,schutz}. The final confirmation of the detection of these waves has been announced by the LIGO\footnote{It is for Laser Interferometer Gravitational-Wave Observatory in Louisiana USA.} group in 2017 and has led to the Nobel prize in physics in the same year \cite{ligo}.
   
\end{itemize}

\newpage
\section{Palatini formalism: metric-affine gravity}
It became clear now that gravity described by general relativity is a manifestation of the curvature of spacetime. A new geometry that goes beyond Euclidean geometry had to be used as an attempt to formulate a theory of gravity in terms of curvature. Einstein has simply used the Riemannian geometry where the Levi-Civita connection is completely fixed by the metric tensor. The reason is simply that when Einstein formulated his general theory of relativity, differential geometry was generally limited to Riemannian (or metric) spaces. However, physicists and mathematicians have realized later that the general covariance of the physical laws is not restricted to those spaces, and one may consider, in addition to the metric tensor, an affine connection which is completely independent of the metric.

The simplest approach based on both metric and affine connection as an independent fundamental fields is called the Palatini formalism \cite{misner}. In turns out that this approach recovers general relativity, and it is considered then as a different formulation of the latter. 

Palatini approach to gravity starts with the following invariant action
\begin{eqnarray}
S_{\text{Pl}}[g,\Phi]=\frac{1}{2\kappa^{2}}
\int d^{4}x
\sqrt{||g||}g^{\mu\nu}R_{\mu\nu}(\Gamma)
+S_{\text{m}}[g,\Phi],
\end{eqnarray}
where in this case, the Ricci tensor which appears in the first integral is the Ricci tensor resulting from the curvature of the affine connection
\begin{eqnarray}
R_{\mu\nu}(\Gamma)=
R^{\alpha}_{\,\,\mu\alpha\nu}(\Gamma),
\end{eqnarray}
not the Ricci tensor of the metric $g_{\mu\nu}$.

Here, the action $S_{\text{m}}[g,\Phi]$ represents the couplings to matter fields, which we denote $\Phi$. For simplicity, the affine connection is taken symmetric, and then the variation of the Ricci tensor with respect to the affine connection reads
\begin{eqnarray}
\label{palatini variation of the ricci tensor}
\delta R_{\mu\nu}=
\nabla_{\lambda}(\delta \Gamma^{\lambda}_{\nu\mu})
-\nabla_{\nu}(\delta \Gamma^{\lambda}_{\lambda\mu}).
\end{eqnarray}
The total variation becomes
\begin{eqnarray}
\delta S_{\text{Pl}}=&&\frac{1}{2\kappa^{2}}
\int d^{4}x \sqrt{||g||}
\left[R_{\mu\nu}(\Gamma)-\frac{1}{2}g_{\mu\nu}g^{\alpha\beta}R_{\alpha\beta}(\Gamma)-\kappa^{2}T_{\mu\nu}  \right]\delta g^{\mu\nu} \nonumber \\
&&-\frac{1}{2\kappa^{2}}
\int d^{4}x \sqrt{||g||}
\left[\nabla_{\lambda}(\sqrt{||g||}g^{\mu\nu})
-\delta^{\mu}_{\lambda}\nabla_{\alpha}(\sqrt{||g||}g^{\alpha\nu})
\right] \delta \Gamma^{\lambda}_{\mu\nu} \nonumber \\
&&+\delta S_{\text{m}}[g,\Phi].
\end{eqnarray}
Two field equations are obtained
\begin{eqnarray}
\label{palatini field equations1}
R_{\mu\nu}(\Gamma)-\frac{1}{2}g_{\mu\nu}g^{\alpha\beta}R_{\alpha\beta}(\Gamma)=\kappa^{2}T_{\mu\nu},
\end{eqnarray}
and
\begin{eqnarray}
\label{palatini field equations2}
\nabla_{\lambda}(\sqrt{||g||}\,g^{\mu\nu})
-\delta^{\mu}_{\lambda}\nabla_{\alpha}(\sqrt{||g||}\,g^{\alpha\nu})=0.
\end{eqnarray} 
Now, by contracting the indices $\mu$ and $\lambda$ in the last equation we get
\begin{eqnarray}
\nabla_{\alpha}(\sqrt{||g||}\,g^{\alpha\nu})=0,
\end{eqnarray}
and finally equation (\ref{palatini field equations2}) becomes
\begin{eqnarray}
\label{palatini equation}
\nabla_{\lambda}(\sqrt{||g||}\,g^{\mu\nu})=0.
\end{eqnarray}
This dynamical equation must lead to a relation between the affine connection and the metric tensor. To solve this equation, we firstly expand the covariant derivative in terms of the connection, this gives
\begin{eqnarray}
\label{expanded palatini equation}
g^{\mu\nu}\partial_{\lambda}\sqrt{||g||}
+\sqrt{||g||}\,\partial_{\lambda}g^{\mu\nu}
+\sqrt{||g||}\left(\Gamma^{\mu}_{\,\,\lambda\alpha}g^{\alpha\beta} -\Gamma^{\alpha}_{\,\,\alpha\lambda}g^{\mu\nu}
+\Gamma^{\nu}_{\,\,\lambda\alpha} \right)=0,
\end{eqnarray}
Multiplying by $g_{\mu\nu}$ (contraction) we find 
\begin{eqnarray}
\label{contracted palatini equation}
2\sqrt{||g||}\,\Gamma^{\alpha}_{\,\,\alpha\lambda}=
4\partial_{\lambda}\sqrt{||g||}+\sqrt{||g||}g_{\mu\nu}\partial_{\lambda}g^{\mu\nu}.
\end{eqnarray}
The last term of the previous equation is given by
\begin{eqnarray}
\sqrt{||g||}\, g_{\mu\nu}\partial_{\lambda}g^{\mu\nu}=
-2\sqrt{||g||}\,\partial_{\lambda}\ln \sqrt{||g||}.
\end{eqnarray}
If we substitute this into equation (\ref{contracted palatini equation}), we get
\begin{eqnarray}
\Gamma^{\alpha}_{\,\,\alpha\lambda}=\partial_{\lambda}\ln \sqrt{||g||}
\end{eqnarray}
With all this together, the dynamical equation (\ref{expanded palatini equation}) reads
\begin{eqnarray}
\partial_{\lambda}g^{\mu\nu}+
\Gamma^{\mu}_{\,\,\alpha\lambda}g^{\alpha\nu}+
\Gamma^{\nu}_{\,\,\alpha\lambda}g^{\mu\alpha}=0,
\end{eqnarray}
which is equivalent to
\begin{eqnarray}
\nabla_{\lambda}g^{\mu\nu}=0.
\end{eqnarray}
Since we have $g^{\mu\nu}g_{\nu\alpha}=\delta^{\mu}_{\alpha}$, the last equation implies
\begin{eqnarray}
\label{palatini metricity}
\nabla_{\lambda}g_{\mu\nu}=0.
\end{eqnarray}
This is simply the compatibility equation, and it shows that the affine connection is reduced to the Levi-Civita connection $\Gamma^{\lambda}_{\,\,\mu\nu}(g)$, of the metric $g_{\mu\nu}$. Thus, the field equations (\ref{palatini field equations1}) are nothing but Einstein's field equations of GR. Although we have started with an action different than Einstein-Hilbert action, the GR is recovered, this means that Palatini formalism is nothing but a different formulation of GR, and the latter is derived only \textit{a posteriori}. This feature can be realized once we introduce an affine connection as an independent parameter.  

Now, the metricity equation (\ref{palatini metricity}) which has been arisen \textit{a posteriori} shows that the spacetime is again a Reimannian space. Palatini formalism is a particular case of a general type of theories called metric-affine gravity, and the metricity condition is obtained since we have simply taken a symmetric connection. In general, metric-affine theories of gravity are based on the metric tensor and an independent affine connection without metricity condition. The latter may not be obtained, but instead, the metric tensor always satisfies the \textit{nonmetricity} equation   
\begin{eqnarray}
\nabla_{\lambda}g_{\mu\nu}=Q_{\lambda\mu\nu},
\end{eqnarray}
where the nonmetricity tensor $Q_{\lambda\mu\nu}$ measures the failure of the conservation of the metric tensor under parallel displacement. 

Extended metric-affine theories of gravity have shown that $Q_{\lambda\mu\nu} \neq 0$, the case that violates the compatibility between the metric and the connection \cite{olmo: black holes in palatini}. Metric-affine gravity is a subject beyond the scope of this thesis, and for more details on this topic, we refer the reader to some references like \cite{papers on palatini,banados}.

\clearpage  
\lhead{\emph{Chapter 4}}  
  \chapter{Purely Affine gravity: metrical structure from vacuum energy and affine connection}
\vspace{-0.5 cm}
\epigraph{\textit{The essential achievement of GR, namely to overcome rigid space, is only indirectly connected with the introduction of a Riemannian metric. The directly relevant conceptual element is the displacement field $\Gamma^{\lambda}_{\,\,\mu\nu}$, which expresses the
infinitesimal displacement of vectors.}\, 

---Albert Einstein}{}

\section{Affine structure of spacetime}
The gravitational theories discussed in the last chapter are clearly based on the metric tensor as a fundamental field of gravity. In Palatini formulation, an additional degree of freedom is added as an affine connection completely independent of the metric. Thus, one raises the question of which field can be considered fundamental. If it is the metric tensor, then the theory is simply general relativity and its extensions. If not, we will be left with only the affine connection, or in the words of Schr\"{o}dinger: \textit{can one not go a step beyond Palatini and base a theory on affine connection alone?} \cite{schrodinger}.

It is worth addressing this question from a fundamental point which concerns the structure of spacetime at large and short scales. As we have seen in the first chapter, the concept of metric tensor, in both flat and curved backgrounds, is essential since it is the only \enquote{machinery} that provides us with
the measurements of distances, clock rates and angles. Spacetime at large scales has in fact this metric structure, and cosmological observations are strongly based on the notions of angles and distances. However, there is no fundamental reason for which the universe has started with this familiar geometric structure at very early times.  It is known for decades, that the existence of
singular regions in space, such as black holes and the initial singularity (big-bang) suggest a completely different and primary structure for spacetime. In fact, at short distances, where quantum
effects, translated by Heisenberg uncertainty principle, are not avoidable, the measurements of distances and clock rates break down.

Since the concepts of distances, time intervals and angles became essential at large scales, and since they are encoded in the metric tensor, it is convenient to give an origin to this later from a fundamental quantity of spacetime. In the absence of the metric tensor, the spacetime is simply an affine space. The concept of parallel transfer which we have explored in the first chapter requires an important quantity called affine connection. It is only with this connection that we can make covariant operations, like the derivation and then we are able to compare vectors and tensors in different points in space. Our fundamental quantity is then the affine connection. This quantity indirectly adds a new tensor field which is the torsion tensor. However, the space is torsion-free, if the affine connection is simply symmetric. In what follows, we will be interested only in this case; an \textit{affine torsionless connection}.    

The difficulty that we face when choosing only an affine connection is in the construction of the invariants which are necessary for the principle of least action. Since we are familiar with the operation of contraction which is performed by the metric tensor, it seems then impossible to define affine actions. However in general, invariants and volume measure do not need any process of contraction. As we have seen in the last chapter when defining the Einstein-Hilbert action, the general coordinate invariant volume element is defined only in terms of the square-root of the determinant of a rank-two tensor. This tensor was simply taken as the metric tensor itself. In affine space, the only rank-two tensor constructed by the affine connection is the Ricci tensor. It is this tensor that provides us with the simplest and relevant affine actions which will be taken as the basis towards a general affine approach to gravity. The simplest case that we shall explore in the next section goes back to Eddington and it is considered as a different formulation of GR in vacuum since it leads to Einstein's field equations with a cosmological constant. Eddington's gravity will be our guide in exploring affine gravity with scalar fields where the cosmological constant or vacuum energy plays an important role in affine space.

\section{Eddington gravity as the simplest affine theory}

Spacetime arena is supposed to be an affine space, endowed only with an affine connection $\Gamma$, and its associated curvature which leads to the Ricci tensor as given in (\ref{ricci-general})
\begin{eqnarray}
\label{ricci tensor of affine connection}
R_{\alpha\beta}(\Gamma)=
\partial_{\mu}\Gamma^{\mu}_{\,\,\alpha\beta}
-\partial_{\beta}\Gamma^{\mu}_{\,\,\alpha\mu}
+\Gamma^{\mu}_{\,\,\sigma\mu}\Gamma^{\sigma}_{\,\,\alpha\beta}
-\Gamma^{\mu}_{\,\,\sigma\beta}\Gamma^{\sigma}_{\,\,\alpha\mu}
\end{eqnarray} 
Eddington proposed an action for gravity which is based only on a symmetric affine connection and the symmetric part of the Ricci tensor (\ref{ricci tensor of affine connection}) as follows \cite{schrodinger,eddington}
\begin{equation}
\label{edd action}
S_{\text{Edd}}=\frac{2}{\Lambda}
\int d^{4}x \sqrt{||R_{(\alpha\beta)}(\Gamma)||},
\end{equation}
where $\Lambda$ is a nonzero constant and 
\begin{eqnarray}
\Gamma^{\lambda}_{\,\,\nu\mu}=
\Gamma^{\lambda}_{\,\,\mu\nu}, \quad \quad \quad
R_{(\alpha\beta)}=\frac{1}{2}(R_{\alpha\beta}+R_{\beta\alpha}).
\end{eqnarray}
Throughout this work, we will drop the symmetrization sign, and then the symmetric part of the Ricci tensor will be implicitly understood.

Here, the variation of this action will be performed with respect to the affine connection $\Gamma$. For the Ricci tensor, we have seen that it is given by relation (\ref{palatini variation of the ricci tensor}). Using relation (\ref{matrix variation}), we easily show that
\begin{eqnarray}
\delta \sqrt{||R_{\alpha\beta}||}=
\frac{1}{2}\sqrt{||R_{\alpha\beta}||}\,
(R^{-1})^{\beta\alpha}\delta R_{\alpha\beta}.
\end{eqnarray}
Then, the variation of action (\ref{edd action}) reads 
\begin{eqnarray}
\label{variation of edd action}
\delta S&&=\frac{1}{\Lambda}
\int d^{4}x
\sqrt{||R||}\,(R^{-1})^{\nu\mu} \delta R_{\mu\nu}(\Gamma) \nonumber \\
&&=\frac{1}{\Lambda}
\int d^{4}x
\sqrt{||R||}\,(R^{-1})^{\nu\mu}
\left(\nabla_{\lambda}\delta \Gamma^{\lambda}_{\nu\mu}
-\nabla_{\nu} \delta \Gamma^{\lambda}_{\lambda\mu} \right) \nonumber \\
&&=-\frac{1}{\Lambda} \int d^{4}x
\left[ \nabla_{\lambda}\left(\sqrt{||R||}(R^{-1})^{\nu\mu} \right)\delta\Gamma^{\lambda}_{\nu\mu}
-\nabla_{\nu}\left(\sqrt{||R||}(R^{-1})^{\nu\mu} \right)
\delta\Gamma^{\lambda}_{\lambda\mu} 
\right] \nonumber \\
&&+\frac{1}{\Lambda}
\int d^{4}x \left[
\nabla_{\lambda}\left(\sqrt{||R||}(R^{-1})^{\nu\mu} \delta\Gamma^{\lambda}_{\nu\mu}\right)
-\nabla_{\nu}\left(\sqrt{||R||}(R^{-1})^{\nu\mu} 
\delta\Gamma^{\lambda}_{\lambda\mu}\right)
\right]
\end{eqnarray}
Since the connection is torsionless, the last two terms are total (covariant) divergence, and then they vanish by applying the Stokes' theorem \cite{poplawski-review}. The remaining parts of integral (\ref{variation of edd action}) can be rearranged leading to the equations of motion
\begin{eqnarray}
\label{edd dynamical equation}
\frac{1}{\Lambda}\left\lbrace \nabla_{\lambda}\left(\sqrt{||R||}(R^{-1})^{\mu\nu} \right)
-\delta_{\lambda}^{\mu}
\nabla_{\rho}\left(\sqrt{||R||}(R^{-1})^{\rho\nu} \right)\right\rbrace =0.
\end{eqnarray}
This dynamical equation describes the evolution of the affine connection. The affine connection, though taken symmetric, is considered arbitrary in action (\ref{edd action}). However, the affine variational principle provides us with constraints on this connection, and then only solutions that satisfy equation (\ref{edd dynamical equation}) will describe the gravitational equations.

To proceed, let us take a look at the second term of (\ref{edd dynamical equation}). This term includes a covariant \enquote{divergence}, and then if we make $\mu=\lambda$, the dynamical equation becomes
\begin{eqnarray}
\nabla_{\rho}\left(\sqrt{||R||}(R^{-1})^{\rho\nu} \right)=0.
\end{eqnarray}
Thus, the second term of (\ref{edd dynamical equation}) vanishes, leaving only a simple equation of motion 
\begin{eqnarray}
\frac{1}{\Lambda}\nabla_{\lambda}\left(\sqrt{||R||}(R^{-1})^{\mu\nu} \right)=0.
\end{eqnarray}
This equation is solved by introducing an invertible and covariantly-constant tensor field $g_{\alpha\beta}$ such that
\begin{equation}
\label{solution to edd}
\frac{1}{\Lambda} \sqrt{||R||}(R^{-1})^{\alpha\beta}= 
\sqrt{||g||} (g^{-1})^{\alpha\beta}, \quad \text{and}\quad
\nabla_{\mu} g_{\alpha\beta}=0.
\end{equation}
The second condition, $\nabla_{\mu} g_{\alpha\beta} =0$, or the compatibility condition, which arises now dynamically, defines completely a Levi-Civita connection of the tensor $g_{\mu\nu}$ 
\begin{equation}
{}^{g}\Gamma^{\mu}_{\alpha\beta}= \frac{1}{2} g^{\mu\lambda} \left( \partial_{\alpha} g_{\beta\lambda} +
\partial_{\beta} g_{\lambda\alpha} - \partial_{\lambda} g_{\alpha\beta}\right).
\end{equation}
The tensor $g_{\mu\nu}$ plays the role of the \enquote{metric} tensor which is generated \textit{a posteriori} and not postulated \textit{a priori} as in general relativity.

Now, the gravitational field equations are described by the first identity in (\ref{solution to edd}). This identity can be written in a tensorial form as
\begin{equation}
\label{edd equations}
R_{\alpha\beta}(g) = \Lambda g_{\alpha\beta}.
\end{equation}
Since the affine connection is reduced to the Levi-Civita or the metric connection, then the tensor in the left hand side of equation (\ref{edd equations}) is nothing but the Ricci tensor of the metric $g$. Thus, equations (\ref{edd equations}) are Einstein's field equation in the presence of (only) a cosmological constant $\Lambda$. Remember that this constant must not vanish, a condition that protects action (\ref{edd action}) from going singular. The new feature of Eddington's gravity is the fact that there is no flat limit solution to the field equations, i.e, spacetimes that satisfy $R^{\lambda}_{\,\,\,\rho\mu\nu}=0$.  

Eddington's gravity then is based solely on an affine connection, and this later is forced by a dynamical equation to coincide with the Levi-Civita of a \textit{generated} metric tensor. The theory reproduces clearly the field equations of general relativity in vacuum with its metric structure. It is clear now that Eddington's gravity is a theory in vacuum, and then an extension of it is needed to include matter fields. This will be the goal of the next section.

\section{Affine gravity with matter}

In purely metric theory of gravity (general relativity), matter-gravity interaction is described by generalizing the field theory Lagrangian densities, by replacing the flat Minkowski metric $\eta_{\mu\nu}$ by the curved spacetime metric $g_{\mu\nu}$. However, coupling matter to gravity in purely affine gravity (in the absence of curved metric) is not trivial. Attempts have been made to write different models for different matter fields \cite{kijowski1,kijowski2}. The affine actions proposed in those models depend on the matter fields. Scalar field for instance is described by a Lagrangian density which is derived from Legendre transformation of its purely metric form. Kinetic and the potential terms no longer appear as a sum in this Lagrangian density. However, a classical Electrodynamics Lagrangian derived using the same transformation is found to have the form of
its purely metric Lagrangian where the metric tensor is replaced by the symmetric part of the Ricci tensor. In the limit of zero fields, those actions are undefined even in curved space when the Ricci tensor appear explicitly in their definition. This problem stems from the absence of the vacuum energy (the cosmological constant) in those
actions. In fact, in Eddington gravity, the metric tensor
is generated only if the latter is nonzero. To solve this problem, affine models of classical electrodynamics and a nonzero cosmological constant are proposed in \cite{kijowski1,poplawski}. The main result there is that there is no model of purely affine gravity that contains all matter terms of the standard model of particle physics.

Here, we will address the coupling of scalar fields in the context of pure affine gravity, and discover the new features of this gravity. But before that, it is useful to present briefly how scalar fields are coupled to gravity in the presence of metric tensors.

\subsection{Scalar field coupling in metric gravity}

\subsubsection{Minimal coupling}

Here, spacetime is endowed with a metric tensor $g_{\mu\nu}$ which describes distances and angles and invariant quantities. Among these invariants is the volume measure formed by the square root of the determinant of the metric tensor $\sqrt{-g}$.
In this theory, gravity-scalar field coupling is described by the following action
\begin{eqnarray}
S^{(1)}_{\text{Met}}=\int d^{4}x \sqrt{-g}\left[\frac{M_{Pl}^{2}}{2} R\left(g \right)-\frac{1}{2} g^{\mu\nu}\partial_{\mu}\phi\partial_{\nu}\phi-V\left(\phi\right) \right], 
\label{gr minimal action}
\end{eqnarray}
where $V\left(\phi\right)$ is the potential associated with the scalar field $\phi$.
 
The gravitational equations arise from variation of this action with respect to the metric tensor. These are given by
\begin{eqnarray}
\label{einstein equations1}
M_{Pl}^{2}G_{\mu\nu}\left(g\right)=
\partial_{\mu}\phi \partial_{\nu}\phi-\frac{1}{2}g_{\mu\nu}\left(\partial\phi\right)^{2}
-g_{\mu\nu}V\left(\phi\right),
\end{eqnarray}
where the right hand side defines the energy momentum tensor of the scalar field
\begin{eqnarray}
\label{energy momentum1}
T^{\phi}_{\mu\nu}= \partial_{\mu}\phi \partial_{\nu}\phi-\frac{1}{2}g_{\mu\nu}\left(\partial\phi\right)^{2}
-g_{\mu\nu}V\left(\phi\right),
\end{eqnarray}
which can be easily obtained from the general form (\ref{energy momentum tensor of matter}).

The dynamics of the scalar field $\phi$ is governed by the following Klein-Gordon equation derived from (\ref{gr minimal action}) by varying with respect to $\phi$
\begin{eqnarray}
\label{scalar field equation1}
\Box \phi -V^{\prime}\left(\phi\right)=0, \quad \text{where} \quad
\Box = \frac{1}{\sqrt{||g||}}
\partial_{\mu}\left(\sqrt{||g||}\partial^{\mu}\phi \right).
\end{eqnarray}

The coupling gravity-scalar field given by (\ref{gr minimal action}) is called minimal since the field $\phi$ is directly coupled to the metric $g_{\mu\nu}$. It is straightforward to construct this sort of coupling by taking the scalar field action in flat spacetime and replacing the Minkowskian metric by the curved metric tensor. 

The action (\ref{gr minimal action}) has the following two important properties:   
\begin{enumerate}
\item As in the case of flat spacetime action, kinetic terms (derivatives) of the scalar field and potentials come in summation.
\item The later property means that action (\ref{gr minimal action}) is valid for all potentials $V\left(\phi\right)$ and then gravitational equations in the standard vacuum where $\phi=\text{constant}$ (or zero) and $V\left(\phi\right)=0$ arise easily in the theory.
\end{enumerate}

\subsubsection{Nonminimal coupling and conformal transformation}

The same metrical properties of spacetime are valid here. Invariants are formed by the metric tensor $g_{\mu\nu}$ postulated \textit{a priori}.

Nonminimal coupling in metric gravity is described by a direct interaction of matter with the curvature of the spacetime. The simplest interaction is performed through the Ricci scalar, and to that end, action (\ref{gr minimal action}) is extended by adding an explicit interaction term between the scalar field $\phi$ and $R(g)$ as follows  
\begin{eqnarray}
\label{gr nonminimal action}
S_{\text{met}}^{(2)}=S_{\text{met}}^{(1)}+\int d^{4}x\sqrt{-g}\left(\frac{\xi}{2}\phi^{2}R\left(g\right)\right),
\end{eqnarray}
where $\xi$ is a dimensionless parameter.

Similarly, the following gravitational field equations are derived by varying this action with respect to the metric tensor, and they are written as
\begin{eqnarray}
\label{field equation nm gr}
M_{Pl}^{2}G_{\mu\nu}\left(g\right)=T_{\mu\nu}^{\phi}+\xi \nabla_{\mu}\nabla_{\nu}\phi^{2}-\xi \Box \phi^{2}g_{\mu\nu}-\xi \phi^{2}G_{\mu\nu}\left(g \right),
\end{eqnarray}
where $T_{\mu\nu}^{\phi}$ is the energy momentum of the scalar field given by (\ref{energy momentum1}).

Variation with respect to $\phi$ yields
\begin{eqnarray}
\Box \phi -V^{\prime}\left(\phi \right)+\xi\phi R\left(g\right)=0.
\end{eqnarray}
Like the minimal case, we enumerate here some properties of nonminimal coupling in metric gravity:
\begin{enumerate}
\item As we see from the total action (\ref{gr nonminimal action}), the nonminimal coupling $\xi \phi^{2}$ is an additive term.
\item
Due to the \textit{nonlinearity} of action (\ref{gr nonminimal action}), i.e. the presence of the second derivative of the metric tensor, the energy momentum tensor gains additional terms proportional to the derivative of the scalar field and it is given as
\begin{eqnarray}
\label{improved GR}
T_{\mu\nu}^{\text{met}}= \xi \nabla_{\mu}\nabla_{\nu}\phi^{2}-\xi \Box \phi^{2}g_{\mu\nu}-\xi \phi^{2}G_{\mu\nu}\left(g \right).
\end{eqnarray} 
\end{enumerate}
Finally we have two couplings to gravity, one minimal described by action (\ref{gr minimal action}) and a nonminimal coupling (\ref{gr nonminimal action}). It turns out that one may easily make a transition between the two actions by performing particular transformations called \textit{conformal transformation}, followed by a field and potential redefinitions. The conformal transformation is the mapping that allows the transition between two metric tensors $g_{\mu\nu}$ and $\tilde{g}_{\mu\nu}$ via the following relation
\begin{eqnarray}
\label{conformal transformation}
\tilde{g}_{\mu\nu}=\mathcal{F} g_{\mu\nu},
\end{eqnarray}
where in our case, the function $\mathcal{F}$ is given in terms of the field $\phi$ as
\begin{eqnarray}
\label{F}
\mathcal{F}(\phi)=1+\frac{\xi \phi^{2}}{M^{2}_{Pl}}.
\end{eqnarray}
Action (\ref{gr nonminimal action}) is transformed to (\ref{gr minimal action}) using (\ref{conformal transformation}) and the following redefinitions 
\begin{eqnarray}
\label{field redefinition}
d\tilde{\phi} = \sqrt{\frac{1}{\mathcal{F}\left(\phi\right)}+\frac{3\mathcal{F}^{\prime 2}\left(\phi\right)}{2 M^{2}_{Pl}\mathcal{F}^{2}\left(\phi\right) } }
\,\, d\phi, \quad \quad
\tilde{V}(\tilde{\phi})=\frac{V\left(\phi\right)}{\mathcal{F}^{2}\left(\phi\right)}.
\end{eqnarray}
These transformations lead to two possible \enquote{distinct} frames; the Einstein frame where the theory is written as (\ref{gr minimal action}), and Jordan frame in which the action takes the form (\ref{gr nonminimal action}). Classically, the two frames are considered \enquote{equivalent}. However, this equivalence breaks down when quantum fluctuations are relevant \cite{magnano,faraoni,damour1,maeda conformal transformation,gottlober,barrow2,damour2,cotsakis,bekenstein,equivalence of the frames}. We will come back to the frame ambiguity later when we study affine inflation. 

Next we will consider the purely affine theory where the metric tensor is absent and see that the last properties are no longer valid.

\subsection{Scalar field coupling in affine gravity}

\subsubsection{Minimal coupling}

Unlike metric gravity, the metric tensor is absent here, we need only an affine connection and its associated curvature. The affine connection is considered arbitrary, however, for simplicity, it can be taken symmetric $\Gamma_{\nu\mu}^{\lambda}=\Gamma_{\mu\nu}^{\lambda}$. The affine action must be based on the following quantities:
\begin{enumerate}
\item \textit{Invariant volume measure}:

A concrete theory of gravity must be described by a covariant field equations which are derived from a principle of least action. This requires an invariant measure which replaces the volume measure $\sqrt{||g||}$ of GR and other metric theories. In the absence of the metric tensor, the simplest alternative is the square root of the determinant of another possible rank-two tensor. In affine spacetime, this can be simply constructed from curvature, thus, the Ricci tensor $R_{\mu\nu}(\Gamma)$. In the presence of matter, which is taken here as a simple scalar field $\phi$, then its kinetic structure $\nabla_{\mu}\phi\nabla_{\nu}\phi$ might also play an important role in forming this invariant. In other word, the possible invariant volume measure will be considered as the square root of the determinant of the linear combination of both quantities; Ricci tensor and kinetic structure of the scalar field. For simplicity, we will be interested only in the symmetric part of the Ricci tensor, $R_{\mu\nu}=R_{(\mu\nu)}$.
\item \textit{Scalar integrand}:

The scalar field $\phi$ enters affine space through its kinetic structure, and the remaining part is its potential energy $V(\phi)$. Like any scalar function, this simply enters the action as a multiplicative term. However, attention should be given to this part, since the case $V(\phi)=0$ would lead to zero or an infinite (singular) action. To avoid this unwanted case, we must impose $V(\phi)\neq 0$ everywhere. This is a novel property which is restricted to affine gravity.      
\end{enumerate}  

With all these properties at hand, the affine action can be written as
\begin{eqnarray}
\label{min affine action}
S[\Gamma,\phi]=
\int d^{4}x\frac{\sqrt{|| M^{2}_{Pl}R_{\mu\nu}(\Gamma)-\nabla_{\mu}\phi \nabla_{\nu}\phi ||}}{V(\phi)},
\end{eqnarray}
wherein we have taken a symmetric connection, $\Gamma^{\lambda}_{\mu\nu}$, and the tensor $R_{\mu\nu}$ refers only to the symmetric part of the Ricci tensor.  

The affine gravity (AG) action in (\ref{min affine action}) is considered here as the simplest form of a pure affine theory of gravity coupled to a scalar field. As we shall see below, the equations of motion derived from this action are found to be equivalent to those of GR. This has been proposed for the first time by Kijowski where the metric tensor arises as the momentum canonically conjugate to the connection \cite{kijowski1,kijowski2}.

Unlike action (\ref{gr minimal action}) of metric gravity, the AG action (\ref{min affine action}) is singular at $V(\phi) = 0$. Thus, the scalar field must always have a nonzero potential energy. If we take $\phi = \phi_{\min}$ as the value of the scalar field for which $V(\phi)$ attains its minimum, this theory requires then $V(\phi_{min}) \neq 0$. This describes the nonzero vacuum energy.

Now, the field equations must be derived by varying action (\ref{min affine action}) with respect to the affine connection $\Gamma$. To that end, one gets
\begin{eqnarray}
\int d^{4}x
\frac{\sqrt{\left|\left| K(\Gamma,\phi)\right|\right|}}{V(\phi)}
(K^{-1})^{\alpha\beta}
\left(\nabla_{\lambda} (\delta \Gamma_{\alpha\beta}^{\lambda})
-\nabla_{\beta} (\delta \Gamma_{\alpha\lambda}^{\lambda})  \right)=0,
\end{eqnarray}
where $\nabla$ is the covariant derivative with respect to the affine connection, and the tensor $K_{\mu\nu}$ is given by
\begin{eqnarray}
\label{min tensor k(Gamma,phi)}
K_{\mu\nu}\left(\phi\right)=
M^{2}_{Pl} R_{\mu\nu}\left(\Gamma\right) - \nabla_{\mu}\phi \nabla_{\nu}\phi.
\end{eqnarray}
By integrating by parts and getting rid of the surface terms, we obtain
\begin{eqnarray}
\int d^{4}x \Bigg[ \nabla_{\nu}\left(\frac{\sqrt{\left|\left| K(\Gamma,\phi)\right|\right|}}{V(\phi)}
(K^{-1})^{\mu\nu} \delta_{\lambda}^{\kappa}\delta_{\mu}^{\sigma} \right) -\nabla_{\lambda}\left(\frac{\sqrt{\left|\left| K(\Gamma,\phi)\right|\right|}}{V(\phi)}
(K^{-1})^{\mu\nu} \delta_{\mu}^{\kappa}\delta_{\nu}^{\sigma} \right)
\Bigg] \delta \Gamma_{\kappa\sigma}^{\lambda}=0 \nonumber 
\end{eqnarray}
\begin{eqnarray}
\,
\end{eqnarray}
This leads to the dynamical equation
\begin{eqnarray}
\nabla_{\nu}\left(\frac{\sqrt{\left|\left| K(\Gamma,\phi)\right|\right|}}{V(\phi)}
(K^{-1})^{\sigma\nu} \right)\delta_{\lambda}^{\kappa}
-\nabla_{\lambda}\left(\frac{\sqrt{\left|\left| K(\Gamma,\phi)\right|\right|}}{V(\phi)}
(K^{-1})^{\kappa\sigma}\right)=0.
\end{eqnarray}
Taking the trace of the last equation, one shows that the first term vanishes, and finally this dynamical equation is equivalent to
\begin{eqnarray}
\label{min dynamical equation single field}
\nabla_{\alpha}\left\lbrace  \frac{\sqrt{\left| \right|  K_{\mu\nu}\left(\phi\right)  \left| \right|}}{V\left(\phi\right)}\left(K^{-1}\right)^{\mu\nu}
\right\rbrace =0.
\end{eqnarray}
The solution to this equation is provided by the existence of a rank-two symmetric tensor $g_{\mu\nu}$ which defines with its inverse $(g^{-1})^{\mu\nu}$, a constant scalar density satisfying 
\begin{eqnarray}
\label{min solution of the dynamical equation}
\frac{\sqrt{\left| \right|  K_{\mu\nu}\left(\phi\right)  \left| \right|}}{V\left(\phi\right)}\left(K^{-1}\right)^{\mu\nu}
=M_{Pl}^{2} \sqrt{\left| \right| g \left| \right|} \left( g^{-1}\right)^{\mu\nu}.
\end{eqnarray}

This implies that $\nabla_{\alpha}g_{\mu\nu}=0$, and then the affine connection is reduced to the Levi-Civita connection of the tensor $g_{\mu\nu}$
\begin{eqnarray}
\label{min levi-civita connection}
\Gamma^{\lambda}_{\mu\nu} \rightarrow
\Gamma^{\lambda}_{\mu\nu}(g)=
\frac{1}{2}g^{\lambda\sigma}(\partial_{\mu}g_{\nu\sigma}
+\partial_{\nu}g_{\mu\sigma}-\partial_{\sigma}g_{\mu\nu}).
\end{eqnarray}
The new tensor $g_{\mu\nu}$ with its compatibility condition that leads to its associated connection (\ref{min levi-civita connection}) plays then the role of a \textit{metric tensor}. This metric tensor is not postulated \textit{a priori} as in GR, but it arises dynamically from the affine structure. This approach provides a first argument towards the \enquote{emergence} of metrical elasticity of space which we will explore later in this paper.

Before proceeding to the scalar field dynamics, we should point out here an important point that concerns the Lorentzian signature of the generated metric. At first glance, one may notice that the metric tensor is given in terms of the affine connection and the scalar field as in (\ref{min solution of the dynamical equation}). In imposing the physical signature, the solution to this dynamical equation must be taken such that the tensor $K_{\mu\nu}(\Gamma,\phi)$ defined by (\ref{min tensor k(Gamma,phi)}), has one signature, say $(-,+,+,+)$ \cite{kijowski1}. 

Given the \textit{a posterior} metrical structure, the equations of motion now are nothing but the equality (\ref{min solution of the dynamical equation}), which is written
\begin{eqnarray}
\label{min ricci equation of motion}
M_{Pl}^{2} R_{\mu\nu}
- \nabla_{\mu}\phi \nabla_{\nu}\phi =
g_{\mu\nu} V\left(\phi \right).
\end{eqnarray}
Contracting, raising and lowering the spacetime indices in the standard way can be performed using the metric tensor. Thus, the equation of motion (\ref{min ricci equation of motion}) can be easily recast to a standard form as
\begin{eqnarray}
M_{Pl}^{2}
\left( R_{\mu\nu}-\frac{1}{2}g_{\mu\nu}R \right)=
\nabla_{\mu}\phi \nabla_{\nu}\phi
-\frac{1}{2}g_{\mu\nu}(\nabla \phi)^{2}-g_{\mu\nu}V\left(\phi\right). 
\end{eqnarray}
Now variation of the action (\ref{min affine action}) with respect to the scalar field $\phi$ leads to the following equation of motion
\begin{eqnarray}
\label{min-affine scalar field equation}
\Box \phi -V^{\prime}\left(\phi \right)=0.
\end{eqnarray}
The study made here, shows that the minimal coupling dynamics in the context of affine gravity is equivalent to that of metric gravity. The equivalence of the two formalisms has been shown for the first time in \cite{kijowski1,kijowski2}. 

As we will see latter, metric gravity and affine gravity are no longer equivalent in the case of nonminimal couplings.

\subsubsection{Nonminimal coupling}
The simplest generalization of action (\ref{min affine action}) is to introduce a \enquote{nonminimal} coupling term that enters the volume element, this is realized as follows
\begin{eqnarray}
\label{affine single0}
S\left[\Gamma,\phi \right] = \int d^{4}x \frac{\sqrt{ \left| \right| \left(M^2 + \xi \phi^2\right)R_{\mu\nu}\left(\Gamma\right) - \nabla_{\mu}\phi \nabla_{\nu}\phi \left| \right|}}{V(\phi)}, \nonumber  \\
\end{eqnarray}
where $M$ is an arbitrary constant of mass dimension.

Like (\ref{min affine action}), this action is invariant under general coordinate transformations. Additionally, the action may acquire other internal symmetries depending on the potential energy. For instance, the term inside the determinant has a $Z_{2}$ symmetry.

Following the procedure made for the minimal coupling case, one may easily derive the following dynamical equation by varying action (\ref{affine single0}) with respect to the symmetric connection $\Gamma$
\begin{eqnarray}
\label{dynamical equation single field}
\nabla_{\alpha}\left\lbrace \left(M^{2}+\xi \phi^{2} \right) \frac{\sqrt{\left| \right|  K_{\mu\nu}\left(\phi\right)  \left| \right|}}{V\left(\phi\right)}\left(K^{-1}\right)^{\mu\nu}
\right\rbrace =0,
\end{eqnarray}
where in this case, the tensor $K_{\mu\nu}$ is defined as
\begin{eqnarray}
\label{tensor k(Gamma,phi)}
K_{\mu\nu}\left(\phi\right)=
\left(M^2 + \xi \phi^2\right)R_{\mu\nu}\left(\Gamma\right) - \nabla_{\mu}\phi \nabla_{\nu}\phi.
\end{eqnarray}
Similarly, this equation is solved as 
\begin{eqnarray}
\label{solution of the dynamical equation}
\left(M^{2}+\xi \phi^{2} \right)\frac{\sqrt{\left| \right|  K_{\mu\nu}\left(\phi\right)  \left| \right|}}{V\left(\phi\right)}\left(K^{-1}\right)^{\mu\nu}
=\bar{M}^{2} \sqrt{\left| \right| g \left| \right|} \left( g^{-1}\right)^{\mu\nu},
\end{eqnarray}
where $\bar{M}$ now, is a constant of integration.

Then, the affine connection is reduced to the Levi-Civita connection of the tensor $g_{\mu\nu}$
\begin{eqnarray}
\label{levi-civita connection}
\Gamma^{\lambda}_{\mu\nu} \rightarrow
\Gamma^{\lambda}_{\mu\nu}(g)=
\frac{1}{2}g^{\lambda\sigma}(\partial_{\mu}g_{\nu\sigma}
+\partial_{\nu}g_{\mu\sigma}-\partial_{\sigma}g_{\mu\nu}).
\end{eqnarray}
Equations (\ref{solution of the dynamical equation}) are rewritten as
\begin{eqnarray}
\label{ricci equation of motion}
\left(M^{2}+\xi \phi^{2} \right)R_{\mu\nu}
- \nabla_{\mu}\phi \nabla_{\nu}\phi =
g_{\mu\nu}\left(\frac{\bar{M}^{2}}{M^{2}+\xi \phi^{2}} \right) V\left(\phi \right).
\end{eqnarray}
In a standard form, one may show that the last equations are equivalent to
\begin{eqnarray}
R_{\mu\nu}-\frac{1}{2}g_{\mu\nu}R=&&
\frac{1}{M^{2}+\xi \phi^{2}}\left[\nabla_{\mu}\phi \nabla_{\nu}\phi
-\frac{1}{2}g_{\mu\nu}(\nabla \phi)^{2}-g_{\mu\nu}V\left(\phi\right)  \right]\nonumber \\
&& +g_{\mu\nu} \frac{M^{2}-\bar{M}^{2}+\xi \phi^{2}}{\left(M^{2}+\xi \phi^{2} \right)^{2}} V\left(\phi\right).
\end{eqnarray}
For the case $\xi=0$, Einstein's field equations for minimal coupled scalar field implies that both constants $M$ and $\bar{M}$ must equal the Planck mass
\begin{eqnarray}
\bar{M}=M=M_{Pl}.
\end{eqnarray}
Finally, the last condition shows that a single scalar field $\phi$ is coupled to gravity through affine connection and its Ricci tensor via the following action \cite{affine inflation}
\begin{eqnarray}
\label{affine single}
S_{\text{AG}}\left[\Gamma,\phi\right] = \int d^{4}x \frac{\sqrt{ \left| \right| \left(M_{Pl}^2 + \xi \phi^2\right)R_{\mu\nu}\left(\Gamma\right) - \nabla_{\mu}\phi \nabla_{\nu}\phi \left| \right|}}{V(\phi)},
\end{eqnarray}
and the gravitational field equations derived from this action are written as
\begin{eqnarray}
\label{nonminimal-affine einstein equations}
R_{\mu\nu}-\frac{1}{2}g_{\mu\nu}R=&&
\frac{1}{M_{Pl}^{2}+\xi \phi^{2}}\left[\nabla_{\mu}\phi \nabla_{\nu}\phi
-\frac{1}{2}g_{\mu\nu}(\nabla \phi)^{2}-g_{\mu\nu}V\left(\phi\right)  \right]\nonumber \\
&& +g_{\mu\nu} \frac{\xi \phi^{2}}{\left(M_{Pl}^{2}+\xi \phi^{2} \right)^{2}} V\left(\phi\right)
\end{eqnarray}
Now variation of the action (\ref{affine single}) with respect to the scalar field $\phi$ leads to the following equation of motion
\begin{eqnarray}
\label{nonminimal-affine scalar field equation}
\Box \phi -V^{\prime}\left(\phi \right)+\xi \phi R\left(g\right)+\Psi\left(\phi\right)=0,
\end{eqnarray}
where the function $\Psi\left(\phi\right)$ is given by
\begin{eqnarray}
\label{Psi}
\Psi\left(\phi\right)=
\frac{\xi\phi^{2}}{M^{2}_{Pl}+\xi\phi^{2}}V^{\prime}\left(\phi\right) 
-\left(\frac{2\xi\phi}{M^{2}_{Pl}+\xi\phi^{2}}\right) g^{\mu\nu}\nabla_{\mu}\phi\nabla_{\nu}\phi.
\end{eqnarray}
In conclusion, we point out the following differences between Affine Gravity (AG) described by action (\ref{affine single}) and Metric Gravity (MG) based on action (\ref{gr nonminimal action}):
\begin{enumerate}
\item The theories are conceptionally different since they are based on different fundamental fields. In MG, matter couples to the metric, whereas this latter is absent in AG, and matter then couples to affine connection.
\item Nevertheless, the theories provide equivalent equations of motion for the minimal coupling case.
\item The theories are inequivalent in the presence of nonminimal couplings.
\end{enumerate}

\subsection{Mapping to minimal coupling in affine gravity }
The question now is how to recast the gravitational field equations (\ref{nonminimal-affine einstein equations}) to standard Einstein equations? What is the associated conformal transformation in this setup? The answer to this is that there is no need for conformal mapping to get the standard Einstein equation. In fact, one only needs to redefine the scalar field $\phi$ and its potential $V(\phi)$ as
\begin{eqnarray}
\label{affine field transformation}
d\tilde{\phi}=\frac{d\phi}{\sqrt{\mathcal{F}\left(\phi\right)}}, \quad \quad \text{and} \quad \tilde{V}[\tilde{\phi}(\phi)]=\frac{V(\phi)}{\mathcal{F}^{2}(\phi)}.
\end{eqnarray}
In terms of the new field $\tilde{\phi}$, one may easily show that the field equations (\ref{nonminimal-affine einstein equations}) and (\ref{nonminimal-affine scalar field equation}) are, respectively, written as
\begin{eqnarray}
R_{\mu\nu}-\frac{1}{2}g_{\mu\nu}R=
M_{Pl}^{-2}\left[\nabla_{\mu}\tilde{\phi} \nabla_{\nu}\tilde{\phi}
-\frac{1}{2}g_{\mu\nu}(\nabla \tilde{\phi})^{2}-g_{\mu\nu}\tilde{V}(\tilde{\phi})  \right], 
\end{eqnarray}
\begin{eqnarray}
\Box \tilde{\phi}- \tilde{V}(\tilde{\phi})=0.
\label{minimal affine field equations}
\end{eqnarray}
These equations are familiar in general relativity, they describe the dynamics of a scalar field $\tilde{\phi}$ minimally coupled to gravity via the metric tensor $g_{\mu\nu}$. In other word, both fields are coupled (through equations of motion) to the same metric which is generated dynamically in our setup. This can be seen in a standard form from the transformation of the action (\ref{affine single}) under the field redefinition (\ref{affine field transformation})
\begin{eqnarray}
\label{transformed affine action}
S_{\text{AG}}\left[\Gamma,\phi\right] \rightarrow
\int d^{4}x \frac{\sqrt{ \left| \right| M_{Pl}^2 R_{\mu\nu}\left(\Gamma\right) - \nabla_{\mu}\tilde{\phi} \nabla_{\nu}\tilde{\phi} \left| \right|}}{\tilde{V}(\tilde{\phi})}.
\end{eqnarray}
This action represents the standard minimally coupled scalar field in affine spacetime. Following the same procedure made previously, one derives the equations of motion (\ref{minimal affine field equations}).

\subsection{Multifields in affine gravity}
\label{subsec:multifields}
Coupling matter to affine gravity is not restricted to single scalar fields, in fact, affine spacetime accommodates multifields too. The general affine action which describes the scalar fields $\phi^{A}$ coupled to the affine connection, is written as \cite{short review}
\begin{eqnarray}
\label{affine multi nonminimal}
S[\Gamma,\phi^{A}] = \int d^{4}x \frac{\sqrt{ \left| \right| \mathcal{F}(\phi^{1},\dots,\phi^{N})
R_{\mu\nu} \left(\Gamma\right) -\delta_{AB} \nabla_{\mu}\phi^{A} \nabla_{\nu}\phi^{B} \left| \right|}}{V(\phi^{1},\dots,\phi^{N})}.
\end{eqnarray}
This action generalizes the affine theory of a single field (\ref{affine single}) and the dynamics of the fields may easily be obtained by following the same procedure made so far. The theory is valid for general nonzero potentials $V(\phi^{1},\dots,\phi^{N})\neq 0$, where one may impose some specific symmetries on the field space, like $SO(N)$ symmetry. In this particular cases, one may have to add an additional piece to the potentials to prevent the action from going singular at the poles of the potential function. This additional term may be simply a cosmological constant.     

The gravitational equations are derived by varying the last action with respect to the affine connection $\Gamma$. This leads to the following dynamical equation 
\begin{eqnarray}
\label{dynamical equation multifields}
\nabla_{\alpha} \left\lbrace \mathcal{F}(\phi^{1},\dots,\phi^{N})
\frac{\sqrt{ \left| \right|K(\Gamma,\phi^{A})  \left| \right|}}{V(\phi^{1},\dots,\phi^{N})} (K^{-1}(\Gamma,\phi^{A}))^{\mu\nu}   \right\rbrace =0,
\end{eqnarray}
where we have used for brevity the following tensor
\begin{eqnarray}
K_{\mu\nu}(\Gamma,\phi^{A})=
\mathcal{F}(\phi^{1},\dots,\phi^{N})
R_{\mu\nu} \left(\Gamma\right) -\delta_{AB} \nabla_{\mu}\phi^{A} \nabla_{\nu}\phi^{B}.
\end{eqnarray}
Solution to the dynamical equation (\ref{dynamical equation multifields}) requires an invertible tensor $g_{\mu\nu}$ where the connection is compatible with it, i.e, 
\begin{eqnarray}
\nabla_{\alpha} g_{\mu\nu}=0,
\end{eqnarray}
and satisfies the identity
\begin{eqnarray}
\sqrt{\left| \right| g \left| \right| }(g^{-1})^{\mu\nu}=
\mathcal{F}(\phi^{1},\dots,\phi^{N})
\frac{\sqrt{ \left| \right|K(\Gamma,\phi^{A})  \left| \right|}}{V(\phi^{1},\dots,\phi^{N})} (K^{-1}(\Gamma,\phi^{A}))^{\mu\nu}.
\end{eqnarray}
The last identity is nothing but a compact form of a gravitational field equations with matter and it is easy to put it in a tensor form as
\begin{eqnarray}
\mathcal{F}(\phi^{1},\dots,\phi^{N})
R_{\mu\nu} \left(\Gamma\right) -\delta_{AB} \nabla_{\mu}\phi^{A} \nabla_{\nu}\phi^{B}=g_{\mu\nu}\frac{V(\phi^{1},\dots,\phi^{N})}{\mathcal{F}(\phi^{1},\dots,\phi^{N})}.
\end{eqnarray}
Now the tensor $g_{\mu\nu}$ plays the role of a metric, and the connection $\Gamma$ is reduced to the Levi-Civita connection of this metric. This tensor can be used then for raising, lowering as well as contractions. To that end, one may write the last equation in terms of Einstein tensor as
\begin{eqnarray}
\label{affine multi gravitational equations}
\mathcal{F}(\phi^{1},\dots,\phi^{N})G_{\mu\nu}(g)=&&
\delta_{AB} \nabla_{\mu}\phi^{A} \nabla_{\nu}\phi^{B}
-\frac{1}{2}g^{\alpha\beta}\delta_{AB} \nabla_{\alpha}\phi^{A} \nabla_{\beta}\phi^{B}g_{\mu\nu}  \nonumber \\
&&-\frac{V(\phi^{1},\dots,\phi^{N})}{\mathcal{F}(\phi^{1},\dots,\phi^{N})}.
\end{eqnarray}
The equation of motion of a scalar field $\phi^{A}$ is obtained by varying with respect to $\phi^{B}$. This leads after simplification to the following equation
\begin{eqnarray}
\label{affine multi equations of motion}
\Box \phi^{A}-V_{,A}+\frac{1}{2}\mathcal{F}_{,A}R(g)+\Psi=0,
\end{eqnarray}
where the Comma refers to the derivative with respect to the field $\phi^{A}$, and the function $\Psi$ is given by
\begin{eqnarray}
\Psi=(1-\mathcal{F}^{-1})V_{,A}-\mathcal{F}^{-1}\mathcal{F}_{,A}g^{\alpha \beta}
\delta_{CD}\nabla_{\alpha}\phi^{C}\nabla_{\beta}\phi^{D}.
\end{eqnarray}
The action (\ref{affine multi nonminimal}) that leads to the complicated equations of motion (\ref{affine multi gravitational equations}) and (\ref{affine multi equations of motion}) can be recast to a simpler action which describes a minimally coupled multifields. This is done without altering the geometric part (connection or curvature), but only by a field redefinition of the form  
\begin{eqnarray}
\label{affine multifield redefinition}
d\phi^{A} \rightarrow
d \tilde{\phi}^{A}=\frac{M_{Pl}}{\sqrt{\mathcal{F}}} d\phi^{A}.
\end{eqnarray}
This reparametrisation must be followed by a potential rescaling as 
\begin{eqnarray}
\label{affine multifield potential rescaling}
V \rightarrow \tilde{V}=\frac{M^{4}_{Pl}}{\mathcal{F}^{2}} V(\phi^{1},\dots,\phi^{N}).
\end{eqnarray}
In this case, the action (\ref{affine multi nonminimal}) takes the following form
\begin{eqnarray}
\label{affine multi minimal}
S[\Gamma,\phi^{A}] \rightarrow \int d^{4}x \frac{\sqrt{ \left| \right| 
M^{2}_{Pl} R_{\mu\nu} \left(\Gamma\right) -\delta_{AB} \nabla_{\mu}\tilde{\phi}^{A} \nabla_{\nu}\tilde{\phi}^{B} \left| \right|}}{\tilde{V}(\tilde{\phi}^{1},\dots,\tilde{\phi}^{N})}.
\end{eqnarray}
This action represents the theory of multifields minimally coupled to gravity through affine connection. As can be easily checked by using the transformations (\ref{affine multifield redefinition}) and (\ref{affine multifield potential rescaling}), the gravitational equations (\ref{affine multi gravitational equations}) are reduced to the standard Einstein equations sourced by scalar fields $\tilde{\phi}^{A}$ and the same spacetime metric tensor $g_{\mu\nu}$. This is also the result one can obtain when performing the variation of action (\ref{affine multi minimal}) with respect to the connection and solve the obtained dynamical equations. This remarkable result is restricted to affine gravity where metrical properties are not defined \textit{a priori}, and then no conformal transformation makes sense. The absence of this latter prevents the appearance of the additional unwanted terms which are proportional to the field derivatives, and then provides us with a canonical kinetic terms of the fields. Different matter fields here which can be obtained from each other through field redefinition couple to the same and unique spacetime metric.

\subsection{Induced affine gravity}

Despite the big difference between the physics of gravity and the physics of the standard model (SM) of elementary particles, people have tried to incorporate some of the interesting phenomena of the SM into gravity. One promising attempt has been the concept of spontaneous symmetry breaking which has been the central to the Electroweak interaction. Since the latter mechanism causes some scalar fields to have nonzero vacuum expectation values, leading to generation of masses of the mediators (gauge bosons), it has been suggested then, that Newton's constant (gravitational coupling constant) could be generated in the same mechanism \cite{zee}. In fact, both weak and gravitational coupling constants appear as an inverse of mass squared. 

The mechanism is generally based on a theory of a scalar $\phi$ coupled to the spacetime scalar curvature $R\left(g\right)$ as follows
\begin{eqnarray}
\label{induced metric action}
S=\int d^{4}x \sqrt{||g||}\left[\frac{1}{2}\xi \phi^{2}R\left(g\right)-\frac{1}{2}
g^{\mu\nu} \nabla_{\mu}\phi\nabla_{\nu}\phi -V(\phi)  \right],\nonumber
\\
\end{eqnarray}
where $g_{\mu\nu}$ refers to the metric tensor of the manifold, and $\xi$ is dimensionless constant.

It is straightforward to see that Einstein-Hilbert action is obtained when the field takes a constant value $\phi=v$ (generally, it minimizes the potential) and then Newton's constant appears as
\begin{eqnarray}
\label{newton constant}
G_{N}=1/8\pi \xi v^{2},
\end{eqnarray}
where the vacuum expectation value becomes of the order of the planck mass, or $\sqrt{\xi} v \sim M_{Pl}$.

The above mechanism is called \textit{induced gravity} (IG); it is a theory of gravity based on a scalar-tensor theory and it leads to Einstein's general relativity via \textit{spontaneous symmetry breaking}.

Although it leads correctly to Einstein's general relativity, it may not reflect a complete emergence of gravity based on its \enquote{metrical} structure. In fact, classical gravity in its germinal Einstein's general relativity is a theory of the spacetime metric. This latter represents the gravitational field, and gravity then is a measure of the effects on rods and clock rates. These effects are incorporated in the metric tensor, and then it is this \enquote{metrical elasticity} which is the origin of gravity at large scales. At that end, it might be interesting if one is able to generate this metrical elasticity of space. In the metric IG based on action (\ref{induced metric action}), this metric structure is already postulated as a Lorentzian manifold, thus generation of Einstein-Hilbert action does not mean generation of \enquote{metrical elasticity}. 

In this section we will show that affine gravity may also be induced via spontaneous symmetry breaking. At the beginning, the spacetime is simply endowed with an affine connection which permits the parallel displacements without angle and distance measures. This is a consequence of the absence of the metric properties which makes the coupling, matter-geometry, non trivial. However, a scalar field, enters the setup with an explicit coupling with the curvature of this affine connection. At a non zero expectation value (vev), Newton's constant is generated, requiring that the vev is of the order of the Planck mass. This mechanism leaves a nonzero part in the potential which becomes necessary for generating the metric tensor, where at the vacuum, the obtained field equations are equivalent to GR with a cosmological constant.

Our aim in this section is to incorporate the concept of spontaneous symmetry breaking in affine spacetime and generate the gravitational sector of the affine gravity (\ref{min affine action}), and then derive the equations of motion by generating the metric tensor.  

A scalar field $\phi$ is simply coupled to the Ricci curvature tensor through the following fundamental action \cite{induced affine inflation}
\begin{eqnarray}
\label{induced affine action}
S \left[\Gamma, \phi \right]= \int d^{4}x \frac{\sqrt{\left| \right| \xi \phi^{2} R_{\mu\nu}\left(\Gamma\right)- \nabla_{\mu}\phi \nabla_{\nu} \phi \left| \right|}}{V(\phi)},
\end{eqnarray}
where $\xi$ is a dimensionless parameter.

The new coupling given by action (\ref{induced affine action}) has two important properties. Firstly, both geometry and matter field terms define the invariant volume measure, i.e, the square root of the determinant. Matter field enters this measure by its derivative (kinetic part) in a tonsorial form. Second, the potential energy enters the action separately in division, and then theory prevents zero potential, $V\left(\phi \right)\neq 0$. The second property is important for the early universe where the field $\phi$ requires a non zero potential to get all the phase of inflation done \cite{guth,linde1,albrecht,linde2,higgs inflation,bauer}. 

Here, we will assume a spontaneous symmetry breaking potential which attains its minimum at $v$
\begin{eqnarray}
\label{symmetry breaking potential}
V(\phi)=\frac{\lambda}{4} \left(\phi^{2}-v^{2}\right)^{2},
\end{eqnarray}
where $\lambda$ is some coupling constant.

Clearly, this potential tends to zero at $\phi=v$ and then leads to singular action (\ref{induced affine action}). The simplest and convenient way out to this singularity is to add a nonzero constant term $V_{0}$ and then
\begin{eqnarray}
\label{symmetry breaking potential with cc}
V(\phi)=V_{0}+\frac{\lambda}{4} \left(\phi^{2}-v^{2}\right)^{2}.
\end{eqnarray}

A nonzero constant in the potential implies and guarantees a nonzero cosmological constant even at the end of inflation (see Figure \ref{fig:ssb potentials} below). The remarkable feature is that this vacuum energy is necessary for the generation of the metric tensor in the complete absence of the scalar fields \cite{affine inflation} (see also the discussion below).
\begin{figure}[h]
\centering
    \includegraphics[width=0.6\textwidth]{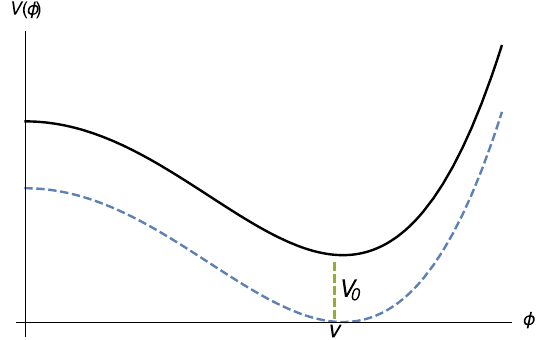}
\caption{Spontaneous symmetry breaking potentials (\ref{symmetry breaking potential}) (dashed line) and (\ref{symmetry breaking potential with cc}) (solid line). The latter never vanishes, and at the vacuum $\phi=v$, it produces the cosmological constant. This is the minimum value of the potential, and it leads to the metric tensor through the solution (\ref{density equality}) at the vacuum.}
\label{fig:ssb potentials}
\end{figure}

Now, the nonminimal coupling of the scalar field $\phi$ to the curvature in action (\ref{induced affine action}) induces the term
\begin{eqnarray}
\xi \left\langle \phi^{2} \right\rangle R_{\mu\nu}(\Gamma),
\end{eqnarray} 
which leads to the affine gravity action (\ref{min affine action}) for
\begin{eqnarray}
\xi v^{2}=M^{2}_{Pl}.
\end{eqnarray}
It has been suggested a long ago, that although gravity and the SM physics could not come into a unified picture, however, there might be a unified mechanism which provides a possible link between the two. For instance, this mechanism can be responsible for the mass scale of gravity and for a spontaneous symmetry breaking  \cite{zee}. The gravitational interactions in its pure affine picture where the metrical properties are absent, are also induced here via spontaneous symmetry breaking mechanism. This again suggests the existence of a relevant mass scale $v=M_{Pl}$. Interestingly, we will see that if this Induced Affine Gravity (IAG) tends to be equivalent to general relativity, then the gravitational constant must also be generated with the metric tensor. This will show how the affine gravity is able to induce both, Newton's constant and the metrical elasticity.

Variation of action (\ref{induced affine action}) with respect to the affine connection $\Gamma$ leads to the following dynamical equation
\begin{eqnarray}
\label{dynamical equation}
\nabla_{\mu} \left\lbrace \xi \phi^{2} \frac{ \sqrt{|| K(\Gamma,\phi)||}}{V(\phi)}
\left( K^{-1} \right)^{\alpha \beta} \right\rbrace = 0, 
\end{eqnarray}
where we have put for simplicity the following tensor
\begin{eqnarray}
\label{tensor k}
K_{\mu\nu}(\Gamma,\phi)=
\xi \phi^{2} R_{\mu\nu}(\Gamma)-\nabla_{\mu}\phi \nabla_{\nu}\phi.
\end{eqnarray}
Metrical properties will arise only after integrating the dynamical equation (\ref{dynamical equation}). Solution to this equation will be given in terms of an invertible, rank two tensor $g_{\mu\nu}$, such that
\begin{eqnarray}
\label{density equality}
M^{2}\sqrt{||g||}(g^{-1})^{\mu\nu}=
\xi \phi^{2} \frac{ \sqrt{|| K(\Gamma,\phi)||}}{V(\phi)}
\left( K^{-1} \right)^{\mu \nu}
\end{eqnarray}
and
\begin{eqnarray}
\nabla_{\alpha} g_{\mu\nu}=0,
\end{eqnarray}
where the constant $M$ is an integration constant of mass dimension.

Obviously, the connection $\Gamma$ which has been taken arbitrary in action (\ref{induced affine action}), is reduced now to the Levi-Civita connection $^{g}{\Gamma}$ of the \textit{generated} tensor $g_{\mu\nu}$, through equations of motion, thus
\begin{eqnarray}
^{g}{\Gamma}_{\mu\nu}^{\lambda}=
\frac{1}{2}g^{\lambda \sigma}
\left(\partial_{\mu} g_{\sigma\nu}+\partial_{\nu} g_{\mu\sigma}-\partial_{\sigma} g_{\mu\nu} \right).
\end{eqnarray}
The new tensor $g_{\mu\nu}$ plays the role of a metric tensor and the spacetime geometry acquires metrical structure only \textit{a posteriori}. To that end, the gravitational equations are written in a compact form (\ref{density equality}), this density equality can be put now in a tonsorial form as
\begin{eqnarray}
\xi \phi^{2} R_{\mu\nu}(g)-\nabla_{\mu}\phi \nabla_{\nu}\phi
=g_{\mu\nu}V(\phi)\left( \frac{M^{2}}{\xi \phi^{2}}\right). 
\end{eqnarray}   
The last equations can be brought to a standard form, in terms of Einstein tensor, after contracting and getting the Ricci scalar 
\begin{eqnarray}
\label{gravitational equations}
\xi \phi^{2} G_{\mu\nu}=\nabla_{\mu}\phi \nabla_{\nu}\phi
-\frac{1}{2}g_{\mu\nu}\nabla^{\lambda}\phi \nabla_{\lambda}\phi 
-g_{\mu\nu}V(\phi)\left( \frac{M^{2}}{\xi \phi^{2}}\right).  
\end{eqnarray}
These field equations are not equivalent to the ones resulting from action (\ref{induced metric action}) of metric induced gravity (see Ref.\cite{affine inflation} for similar comparison.) However, at the vacuum, $\left\langle\phi^{2} \right\rangle=v^{2}$, affine gravity (\ref{induced affine action}) is equivalent to the metric gravity (\ref{induced metric action}), where
\begin{eqnarray}
\label{planck mass}
M=\sqrt{\xi} v= M_{Pl}.
\end{eqnarray}
This can be easily checked from (\ref{density equality}) where the vacuum energy $V(v)=V_{0}$ plays an important role in generating the metric tensor.

Since there are different contributions to vacuum energy, we will assume that they are all incorporated in the piece $V_{0}$, which is associated to the observed value through \cite{affine inflation}
\begin{eqnarray}
V_{0} \sim m_{\nu}^{4},
\end{eqnarray} 
where $m_{\nu}$ is the neutrino mass.

As we have seen, induced affine gravity is realized here via spontaneous symmetry breaking. The mechanism provides the generation of both, the scale of gravity $M_{Pl}$ and the metric elasticity of space (metric tensor). The last property never holds in standard (metric) induced gravity. In Chapter 6, we will apply this setup to inflation and show that like metric induced gravity, it also provides a large tensor-to-scalar ratio. 

\subsection{Vacuum energy and the generated metric tensor}

Up to now, the transition between non-minimal and minimal coupling in affine gravity is shown without referring to any physical principle that underlies the equivalence of the theories. However, affine gravity based on the structure of the actions proposed so far, provides a good reason for that. The key point is that the affine actions are singular at $V(\phi) = 0$, which means that the scalar field must always have a non-zero potential energy. This property holds for multifields too. 

The nonzero potential of different fields may be described by a nonzero primordial part $V_{0}$ which keeps the affine action non-singular even in the absence of the fields. This turns out to be the vacuum energy. The presence of this quantity in the affine spacetime imposes (covariantly) an energy momentum tensor of vacuum $T_{\mu\nu}$ with a non-singular inverse $(T^{-1})^{\lambda \rho}$. This naturally defines a Levi-Civita connection as \cite{demir stress energy of vacuum}
\begin{eqnarray}
\label{GammaT}
{}^{T}\Gamma^{\lambda}_{\,\,\,\mu \nu} = \frac{1}{2} (T^{-1})^{\lambda \rho} \left(\partial_{\mu} T_{\nu \rho} + \partial_{\nu} T_{\rho \mu} - \partial_{\rho} T_{\mu\nu}\right)
\end{eqnarray}
with respect to which
\begin{eqnarray}
\label{compatibility}
\nabla^{T}_{\mu} T_{\alpha\beta} = 0.
\end{eqnarray}

Originally, it is this fundamental structure which provides a solution to the dynamical equations (\ref{dynamical equation single field}) and (\ref{dynamical equation multifields}). In fact, equation (\ref{dynamical equation single field}) is solved and put in the following form \cite{affine inflation}
\begin{eqnarray}
\label{affine single gravitational equations2}
(M_{Pl}^2 + \xi \phi^2) R_{\mu\nu} - \nabla_{\mu}\phi \nabla_{\nu} \phi=
\left(\frac{M_{Pl}^2}{M_{Pl}^2 + \xi \phi^2}\right)\frac{V(\phi)}{V(\phi_{min})}
T_{\mu\nu}.
\end{eqnarray}
The vacuum energy momentum tensor which is inherently contained in affine spacetime can be incorporated in its mixed form in terms of $V(\phi_{min})$ as
\begin{eqnarray}
\label{vacuum tensor}
T^{\mu}_{\nu} && \equiv V(\phi_{min}) \delta^{\mu}_{\nu} \\
&&= V(\phi_{min})  T_{\nu\alpha} (T^{-1})^{\alpha\mu}. \nonumber
\end{eqnarray}
The transition to minimal coupling is made by transforming the equations of motion (\ref{affine single gravitational equations2}) under the field redefinition (\ref{affine field transformation}). Since both vacuum energy $V(\phi_{min})$ and its energy momentum tensor $T^{\mu}_{\nu}$ are redefined, they form an invariant ratio
\begin{eqnarray}
\label{ration delta}
\frac{T^{\mu}_{\nu}}{V(\phi_{min})}=
\frac{\tilde{T}^{\mu}_{\nu}}{\tilde{V}[\tilde{\phi}(\phi_{min})]}
\equiv \delta^{\mu}_{\nu}.
\end{eqnarray}
This identity tensor which facilitates the covariant description of vacuum energy in affine spacetime reflects the metrical properties implicitly. In fact, the dimensionless metric tensor is nothing but the \enquote{unique} ratio
\begin{eqnarray}
\label{ration delta}
\frac{T_{\mu\nu}}{V(\phi_{min})}=
\frac{\tilde{T}_{\mu\nu}}{\tilde{V}[\tilde{\phi}(\phi_{min})]}
\equiv g_{\mu\nu}.
\end{eqnarray}
With this metric tensor at hand, the gravitational equations can be recast to a minimally coupled case without conformal transformation.

\clearpage  
\lhead{\emph{Chapter 5}}  
  \chapter{Standard cosmology and inflation}
\label{chapter 2}
\vspace{-0.5 cm}
\epigraph{\textit{I deal with the view now tentatively held that the whole material universe of stars and galaxies of stars is dispersing; the galaxies scattering apart so as to occupy an ever-increasing volume.}\,\\ ---Sir Arthur Eddington}{}

\section{Standard model cosmology and its shortcomings}
In this Chapter we will study some basic ideas behind relativistic cosmology. The latter is a vast subject, and much more details put it beyond the scope of this thesis. For that reason, the reader will be referred, in some cases, to some advanced and detailed references.

\subsection{Hubble law and the expansion of the universe}

Observations made in the last century have shown that our galaxy, the Milky Way, takes part of numerous similar galaxies in a large patch of space which is accessible to these observations. Local regions of this observed universe formed by stars and clusters of galaxies are subjected to different changing due to the astrophysical evolutions of these objects. However, a remarkable properties of the universe at very large distances are its \textit{homogeneity} and \textit{isotropy}. The universe looks the \enquote{same} in every point seen from every direction. This remarkable feature has been stated by Edward Milne as the \textit{cosmological principle}. From this principle, one realizes that the impression of having a center for the universe is illusory.   

\begin{figure}[h]
\centering
    \includegraphics[width=0.5\textwidth]{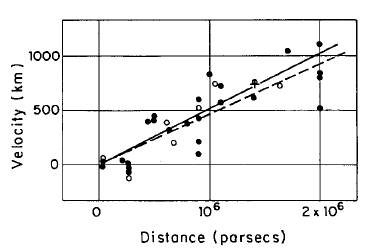}
\caption{Hubble diagram (original 1920 Hubble results) showing the linear relation (\ref{hubble law}) between distance and velocity for different galaxies \cite{harrison}.}
\label{fig:hubble}
\end{figure}

A series of observations performed in late 1920 showed that the spectra of light emitted by distant galaxies are \textit{redshifted}. This means that the frequencies of the emitted light obey a \enquote{cosmological} Doppler effect. The conclusion of these observations made finally by Edwin Hubble in 1930 demonstrated that the redshifts are the results of the recession of these distant galaxies. In other words, the observed distant galaxies are moving apart from the Milky way. A remarkable feature which has been announced by Hubble is that the redshifts increase with the distance of the galaxy from which light is emitted, thus, the farther the galaxy the faster it moves apart. This has been finally stated as the (empirical) Hubble law, which is written in terms of the velocity $v$ of any galaxy at a distance $D$    
\begin{eqnarray}
\label{hubble law}
v=H_{0}\, D,
\end{eqnarray}
where the constant of proportionality $H_{0}$ is called the \enquote{present} Hubble constant, and recent Planck observation determined its value as 
\begin{eqnarray}
\label{present hubble constant}
H_{0} =67 \, km/sec/Mpc.
\end{eqnarray}
This simply means that a galaxy 1 Mpc away, which is about three light years, recedes from us with a velocity of $67\, km/sec$. The Hubble law (\ref{hubble law}) is shown in Figure~\ref{fig:hubble} for  different distant galaxies.

The feature of the recession of the distant galaxies is applied to the whole observed universe. The cosmological principle stated above implies that every observer in a typical\footnote{A galaxy is typical if its motion is only carried along with the general cosmic flow of galaxies \cite{the first three minutes}} galaxy realizes the same effect (the recession) when observing other galaxies. The galaxies are not only receding from us but they move apart from each other. 

At the epoch, the idea of a static, unchanging universe, was the dominant belief, and it took some time for cosmologists to come out with a confident interpretation of the observations carried out by Hubble. The actual reason of the observed redshifs, or the recession of the galaxies, is the increase in the distances (size) between these galaxies. The distant galaxies are considered as comoving frames situated in an expanding space. The universe then is no longer static, but it expands carrying the galaxies along with it \cite{eddington-the expanding universe}.    

The observational facts discussed above had a theoretical reason in the framework of general relativity. The latter is a theory of the dynamics of spacetime, and it can be applied to the universe itself. Einstein's equations (\ref{final einstein equations}) relates the geometry of spacetime to the matter (and energy) contained in it, and the dynamics of the latter trivially implies a dynamical spacetime. Solutions of Einstein's field equations are after all an expanding or a contracting space. Historically, Einstein himself did not feel comfortable with these dynamical solutions, and in order to avoid them, he introduced a cosmological constant term similar to the last term in the right hand side of equation (\ref{final einstein equations}). If a specific value, which is proportional to the matter density of the universe, is given to that constant, the gravitational attraction of matter would be counterbalanced by this additional density leading to a static universe. This model of the universe has been shown to be instable by Eddington, shortly before the idea of an \textit{expanding universe} came out \cite{eddington-instability of static universe}, and Einstein has abandoned this constant\footnote{The cosmological constant is reconsidered as a possible reason for the accelerated expansion of the universe after the measurements of the luminosity of Ia supernovae in 1998 \cite{supernovae Ia}. Additionally, it has been known a long ago that this constant term is not avoidable in cosmology.} stating that it was his \textit{biggest blunder}. 

In what follows we will explore the theoretical description of the expanding universe and see how both observation and theory, have led to the \textit{Big Bang} model.

\subsection{Robertson Walker metric and the Big Bang model}
Models of the universe, or the cosmological models, are based mostly on the cosmological principle, that is to say the universe is homogeneous and isotropic on scales at least larger than 100 Mpc. This is supported by the homogeneity of Cosmic Microwave Background (CMB) radiation observed in 1965 to which we will return later.  

\begin{figure}[h]
\centering
    \includegraphics[width=0.7\textwidth]{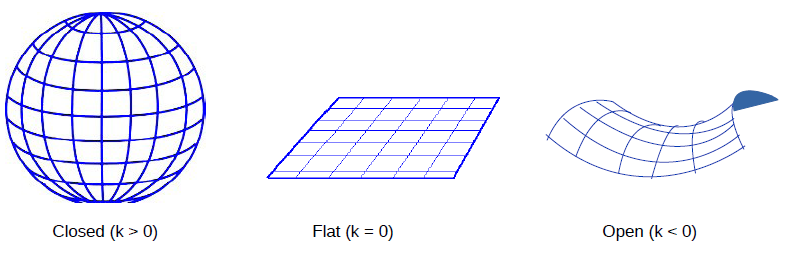}
\caption{Three possible geometries of the spatial sections of the FRW spacetime (\ref{genaral frw metric}).}
\label{fig:3geometries}
\end{figure}

The geometrical description of the cosmological models which are compatible with the cosmological principle are described by a four-dimensional spacetime metric called Friedmann-Robertson-Walker (FRW) metric and it is written as
\begin{eqnarray}
\label{genaral frw metric}
ds^{2}=
-c^{2}dt^{2}+a^{2}(t)\left[\frac{dr^{2}}{1-kr^{2}}+r^{2}d\theta^{2}
+r^{2}\sin^{2}\theta d\varphi^{2} \right].
\end{eqnarray}
The quantity $a(t)$ is called the scale factor which depends on the \textit{cosmological proper time}\footnote{In cosmology, unlike relativity, there is a preferred time parameter $t$, this is called the cosmological proper time.} $t$. The constant $k$ determines the curvature of the spatial sections of this geometry, and it falls into three categories as follows (see also Figure~\ref{fig:3geometries})
\begin{eqnarray}
&&k=0:  \text{Flat space section} \nonumber \\
&&k=+1:  \text{Positively curved space section} \nonumber \\
&&k=-1:   \text{Negatively curved space section}. \nonumber
\end{eqnarray}
 
Based on the FRW spacetime metric (\ref{genaral frw metric}), one may easily predict the recession of galaxies described by the empirical Hubble law (\ref{hubble law}). In fact, at any instant of time, the proper distance $D$ between our galaxy (at $r=0$) and another (at $r$) is given by 
\begin{eqnarray}
\label{proper distance}
D=a(t)\int_{0}^{r}\frac{dr^{\prime}}{\sqrt{1-kr^{\prime 2}}}.
\end{eqnarray}
This shows that the distance between the two galaxies changes with time and it is proportional to the scale factor.

Now, the velocity of the galaxy is given by the time derivative of $D$, thus  
\begin{eqnarray}
v=\frac{dD}{dt}=\dot{a}(t)
\int_{0}^{r}\frac{dr^{\prime}}{\sqrt{1-kr^{\prime 2}}},
\end{eqnarray}
and finally, using (\ref{proper distance}) again to get rid of the integral, we find
\begin{eqnarray}
\label{hubble law from frw}
v=\frac{\dot{a}(t)}{a(t)}\,D.
\end{eqnarray}
This is another form of Hubble law (\ref{hubble law}), and it shows how the velocity of the distant galaxy changes with distance at different times. The Hubble constant $H_{0}$ given in (\ref{present hubble constant}) represents the present value of the Hubble parameter    
\begin{eqnarray}
\label{hubble parameter}
H(t)=\frac{\dot{a}(t)}{a(t)}.
\end{eqnarray}
As we have seen here, the recession of galaxies are nothing but the consequence of the expansion of the three dimensional space section. This expansion is overall described by the time dependent scale factor. In an expanding space, light emitted from distant sources (galaxies) suffers a cosmological redshift. In fact, since light travels along a \enquote{null} geodesic, $ds^{2}=0$, one may show that if a signal of light is emitted at a time $t_{e}$ with a wavelength $\lambda_{e}$, then it will be received by an observer at $t_{0}$ with a wavelength
\begin{eqnarray}
\label{lambda redshift}
\lambda_{0}=\frac{a(t_{0})}{a(t_{e})}\,\lambda_{e}.
\end{eqnarray}
Since the scale factor increases with time, then light is redshifted, $\lambda_{0}>\lambda_{e}$, exactly as observed by Hubble (see Figure~\ref{fig:wave-redshift} below).

\begin{figure}[h]
\centering
    \includegraphics[width=0.5\textwidth]{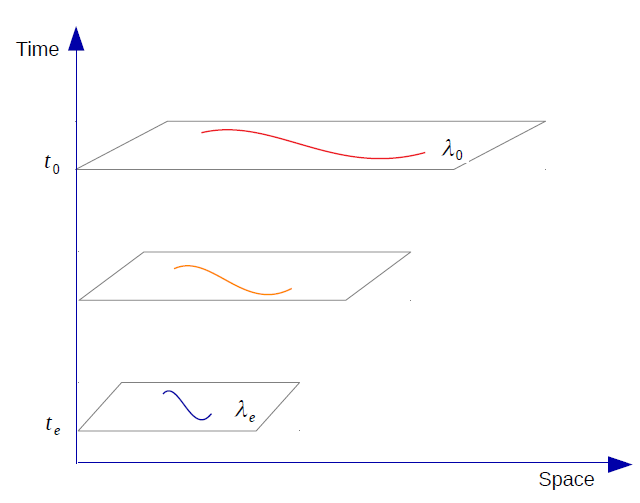}
\caption{Redshifts of light signals due to the expansion of space as given in equation (\ref{lambda redshift}).}
\label{fig:wave-redshift}
\end{figure}

The cosmological principle provides us with the possible description of the geometry of the universe, i.e, the FRW metric (\ref{genaral frw metric}), however, the dynamics of the universe, or the time evolution of the scale factor $a(t)$ remains vague. To fill this gap, we have to apply the gravitational equations where curvature of spacetime responds to the contents of the universe (matter-energy). Matter (and energy) in the universe is generally postulated as a perfect fluid with density $\rho$ and pressure $p$. In a covariant form, its energy-momentum tensor is given in terms of its (average) four-velocity $u^{\mu}=\delta^{\mu}_{0}$, by \cite{misner,schutz}       
\begin{eqnarray}
\label{fluid energy-momentum tensor}
T_{\mu\nu}=\left(\rho+\frac{p}{c^{2}}\right) u_{\mu}u_{\nu}+
\frac{p}{c^{2}}g_{\mu\nu}.
\end{eqnarray}
This allows us to adapt Einstein's equations (\ref{final einstein equations}) to the line element (\ref{genaral frw metric}) and get for the \enquote{time-time} and \enquote{space-space}, the following equations
\begin{eqnarray}
\label{freidmann equations}
&&\dot{a}^{2}+kc^{2}=\frac{8\pi G_{N}}{3}\rho \, a^{2}+\frac{\Lambda}{3} \\
&&2 a \ddot{a}+\dot{a}^{2}+kc^{2}=-\frac{8\pi G_{N}}{c^{2}}
p\, a^{2}+\frac{\Lambda}{3}.
\end{eqnarray}
These equations can be easily arranged and they lead finally to the Friedmann equations
\begin{eqnarray}
\label{first freidman equation}
\left(\frac{\dot{a}}{a}\right)^{2}=\frac{8\pi G_{N}}{3}\rho 
-\frac{kc^{2}}{a^{2}}+\frac{\Lambda}{3},
\end{eqnarray}
\begin{eqnarray}
\label{second freidman equation}
\frac{\ddot{a}}{a}=
-\frac{4\pi G_{N}}{3}\left(\rho +3\frac{p}{c^{2}} \right)+\frac{\Lambda}{3}.
\end{eqnarray}
The time evolution of the scalar factor, is given then as the solution of these equations for every type of matter (and energy). The latter which is described by the energy-momentum tensor (\ref{fluid energy-momentum tensor}), satisfies the covariant conservation law $\nabla^{\mu}T_{\mu\nu}=0$. When it is adapted to the metric (\ref{genaral frw metric}), it leads to
\begin{eqnarray}
\dot{\rho}+3\frac{\dot{a}}{a}
\left(\rho +\frac{p}{c^{2}} \right)=0.
\end{eqnarray}  
This conservation equation can easily be derived differently from the Friedmann equations (\ref{first freidman equation}) and (\ref{second freidman equation}), by taking the time derivative of the first and then using the second one.

In Table~\ref{tab:time evolution of matter and scale factor}, we summarize the time evolution of the scale factor $a(t)$ for different contents of the universe; \textit{radiation}, \textit{matter (dust)} and \textit{vacuum (cosmological constant)}.
    
\begin{table}[h]
\centering
\caption{Energy density and pressure for different perfect fluids, and time evolution of the associated scale factor $a(t)$.}
\label{tab:time evolution of matter and scale factor}
\begin{tabular}{|l|l|l|l|}
\hline
\, &Radiation &Matter (dust)& Vacuum (cosmological constant)\\
\hline
\, & \, & \, & \, \\
Equation of state & $p=\frac{1}{3}\rho$ & $p=0$& $p=-\rho$ \\
\, & \, & \, & \, \\
Energy density& $\rho \sim a^{-4}(t)$ & $\rho \sim a^{-3}(t)$ & $\rho=\text{const}$\\
\, & \, & \, & \, \\
Scale factor & $a(t)\sim t^{1/2}$ & $a(t)\sim t^{2/3}$ & $a(t)\sim
e^{\alpha t}$ \\
\hline
\end{tabular}
\end{table}    
There is a strong relation between the geometry of the universe (curvature of the space sections) and matter. This relationship can be seen from the first Friedmann equation (\ref{first freidman equation}) which can be written for the present time as 
\begin{eqnarray}
\label{fr-1st form}
\frac{kc^{2}}{a^{2}}=\frac{8\pi G_{N}}{3}\rho_{\text{tot}}
-H_{0}^{2}+\frac{\Lambda}{3},
\end{eqnarray}
where we have introduced the total energy density, including matter $\rho_{\text{m}}$ and radiation $\rho_{\text{r}}$.

It is important to define the so called \textit{critical density} as
\begin{eqnarray}
\rho_{\text{cr}}=\frac{3H_{0}^{2}}{8\pi G_{N}} \simeq 10^{-26} 
kg/m^{3},
\end{eqnarray}
and finally we get
\begin{eqnarray}
\label{curvature in terms of densities}
\frac{kc^{2}}{a^{2}}=
\frac{8\pi G_{N}}{3}
\left( \rho_{\text{m}}+\rho_{\text{r}}+ \rho_{\Lambda}
-\rho_{\text{cr}} \right),
\end{eqnarray}
where the cosmological constant contribution is given here in terms of its energy density
\begin{eqnarray}
\rho_{\Lambda}=\frac{c^{2}\Lambda}{8\pi G_{N}}.
\end{eqnarray}
The conclusion of all these is the fact that the geometry of the universe is \textit{flat} ($k=0$), \textit{negatively} curved ($k<0$) or \textit{positively} curved ($k>0$) if the total energy density of the universe (including vacuum energy) equals, less or larger than the critical density respectively. 

Now, what is the total energy density of the universe and what is its geometry? Luminous objects, such as stars and galaxies, namely \textit{baryonic}\footnote{Baryonic matter refers to all type of matter that are formed by baryons; protons and neutrons.} matter, form an average density \cite{planck}
\begin{eqnarray}
\label{density of baryons}
\rho_{\text{b}} \simeq 10^{-28} kg/m^{3}.
\end{eqnarray}
In cosmology, matter refers to the (approximately) pressureless \enquote{matter}, or \textit{dust}. It turns out that, the baryonic matter given by its energy density (\ref{density of baryons}) is not the only form of matter in the universe. In fact, there is a \enquote{strong} evidence for the existence of a (not) luminous matter in galactic haloes. This non-baryonic matter, which is called \textit{dark matter} has been postulated as an attempt to explain the flatness\footnote{Observational data show that in the outer part of the galactic haloes, the radial velocities of galaxies slowly rise or keep constant (nearly flat), which implies that there is a missing mass.} of the observed galactic rotation curves \cite{rotation-curves1,rotation-curves2}. The ratio of the energy density of dark matter to the critical density is estimated as \cite{planck}   
\begin{eqnarray}
\Omega_{dm} \simeq 0.27 \pm 0.004,
\end{eqnarray}
which is considered as a large contribution to the mass of the universe.

The final contribution to the energy density of the universe comes from thermal radiation, this is nothing but the energy density of the 2.73 K photons of the cosmic microwave background (see discussion below). This is estimated as \cite{planck}
\begin{eqnarray}
\rho_{\textit{r}} \simeq 10^{-31} kg/m^{3},
\end{eqnarray}
which is very tiny, and negligible compared with that of matter and dark matter.

The geometry of the universe has been determined from the measurements of the anisotropy of the cosmic microwave background radiation \cite{hartle}. These measurements are consistent with a flat spatial geometry, i.e, $k=0$. In this case, the contents of the universe would satisfy
\begin{eqnarray}
\rho_{\text{m}}+\rho_{\text{r}}+ \rho_{\Lambda}
=\rho_{\text{cr}}.
\end{eqnarray}
In terms of the dimensionless parameter $\Omega_{\text{i}}=\rho_{\text{i}}/\rho_{\text{cr}}$, the last identity is equivalent to
\begin{eqnarray}
\Omega_{\text{m}}
+\Omega_{\text{r}}
+\Omega_{\Lambda}=1.
\end{eqnarray}
This identity shows us that ordinary baryonic matter, radiation, as well as dark matter are not enough to \enquote{make} the universe flat!. In fact, previous estimations show that $\Omega_{\text{m}}
+\Omega_{\text{r}} \simeq 0.28$, thus a significant contribution from the cosmological constant, $\Omega_{\Lambda}\simeq 0.68$, is necessary. Up to now, this contribution has been introduced as a possible nonzero term in Einstein's equations, however, there are different possible origins for it though its physical nature and the problem of its value have not been settled yet. The energy associated to the cosmological constant has another great implication on the dynamics of the universe. In fact, if this contribution dominates the energy density of the universe as it is clear from the previous discussion, than, the second Friedmann equation (\ref{second freidman equation}) would imply an accelerating phase rather than a decelerating one as was expected for decades. The acceleration of the expansion became a real fact since its confirmation for the first time in 1998 from the measurements of the supernovae typeIa \cite{supernovae Ia}, and have led to the Nobel prize in 2011.    

Let us return now to the idea of an expanding universe, where only matter, dark matter and radiation ($\Lambda=0$) play an important role in its dynamics. If we extrapolate the history of the universe, the expansion of the universe means that its size was smaller and smaller at early times than now. The time evolution of matter $a^{-3}(t)$ and radiation $a^{-4}(t)$ clearly show that at a very early time when the scale factor was \textit{infinitely} small (goes to zero), the universe gained an \textit{infinite} energy density which led to an initial \textit{singularity}. However, this very hot and dense phase, named the \textit{hot big bang} model provides us with a good description of the early universe which is consistent with observation. As have been proposed by Gamow and his collaborators in 1948, the high energy density and high temperature would lead to thermal equilibrium between matter and radiation that filled the universe at early time \cite{gamow}. The early universe has a character similar to that of a \textit{blackbody}, and since the energy density of the latter is proportional to its temperature $T$ as $\rho \propto T^{4}$, one may easily show that the temperature of the universe drops as
\begin{eqnarray}
\label{temperature}
T \propto \frac{1}{a(t)}.
\end{eqnarray}
This relationship between the temperature and the size of the universe is the basis of the so called \textit{nucleonsynthesis} of the light elements. As we go back in time, the scale factor decreases and the universe becomes hot enough leading to high energy processes such as pair creations. For instance, electron-positron pairs would take place when a temperature $T \sim 10^{10}$ K and higher, is reached. Indeed, the energy associated to this temperature is about $k T \sim 1$ MeV ($k$ is the Boltzmann constant), which is greater than the rest mass of the electron (or positron), $m_{e}c^{2} \sim 0.5$ MeV. Different particle processes will take place, and the thermal equilibrium is maintained by primordial nuclear reactions like         
\begin{eqnarray}
e^{-}+e^{+} \rightleftharpoons \gamma +\gamma \quad \text{and} \quad
p+n \rightleftharpoons d+\gamma.
\end{eqnarray}
As the universe expands one expects the inverse processes to happen. In this case electron-positron would be annihilated leaving only photons. As the temperature goes down (with the expansion), some of the light elements like Hydrogen, Deuterium and Helium up to Lithium are synthesized. Abundances of these primordial light elements are in high agreement with their observed amounts, and this fact became one of the greatest confirmation of the hot Big Bang model \cite{gamow, turner}. 

\begin{figure}[h]
\centering
    \includegraphics[width=0.5\textwidth]{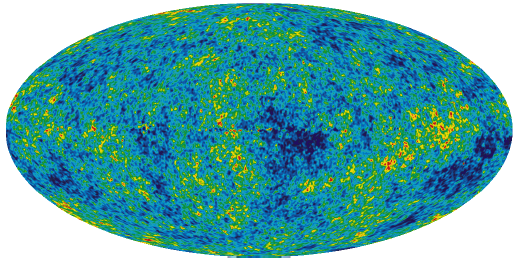}
\caption{The homogeneous cosmic microwave radiations that fill the observed sky. As we shall see later, the origin of the tiny inhomogeneities (small red regions) are due to the small perturbations that are generated during inflation \cite{planck}.}
\label{fig:cmb}
\end{figure}

When the temperature goes down to about 3000 K, it becomes possible for the nuclei and the electrons to be combined. In this \textit{recombination} era the first atoms are formed, and the universe becomes transparent\footnote{Before this, the existed photons scatter by the free electrons and then the universe was completely opaque \cite{the first three minutes}.}. It is only after this era, that photons start to propagate freely in the expanding universe, and they form the so called Cosmic Microwave Background radiation (CMB). Remarkably, this radiation which had the blackbody character in the beginning, retained the same character under the expansion of space. These radiations are subjected to redshifts and at the present epoch its temperature is dropped to 2.73 K. The CMB radiations (Figure~\ref{fig:cmb}) have been finally detected (on Earth) by Penzias and Wilson in 1965 and it was considered as the first remarkable confirmation of the the big bang model. 

\subsection{Flatness and horizon problems }

The hot Big Bang model described above is based originally on a universe which is dominated only with matter and radiation. A nonzero cosmological constant $\Lambda$ became necessary only in later times when the expansion of the universe started accelerating. A universe with $\Lambda=0$ is radiation dominated in its beginning, and this phase is followed later by a matter dominated era. Although it has been successful in describing some of the interesting (observed) phenomena, such as the origin of the CMB radiation and the primordial nucleonsynthesis of the light elements, the big bang model in the above picture fails in explaining naturally the initial conditions of the universe ! Among the shortcomings of the big bang model are the \textit{flatness} and \textit{horizon} problems.     

The former stems from the fact that the point $\Omega_{\textit{tot}}=1$ is \textit{unstable} when the universe is dominated by matter and (or) radiation. To see this fact, let us write again the Friedmann equation (\ref{fr-1st form}) as
\begin{eqnarray}
\label{flatness equation}
\frac{k}{(aH)^{2}}=\Omega_{\textit{tot}} -1.
\end{eqnarray}
Herein, the factor $(aH)^{-1}$ is called the \textit{comoving} Hubble radius, and in standard cosmology, where the universe is dominated by matter or radiation ($\Lambda=0$), it grows with time. In fact, from table \ref{tab:time evolution of matter and scale factor}, one may easily check that
\[
    \frac{1}{aH}= 
\begin{cases}
    t^{\frac{1}{3}},& \text{for matter }\\
    t^{\frac{1}{2}},              & \text{for radiation.}
\end{cases}
\]
Given this contribution, the left hand side of equation (\ref{flatness equation}) is simply increasing with time, thus, the universe becomes rapidly dominated by a nonzero curvature term. The flatness problem then can be stated as follows
\begin{center}
\textit{Why the parameter $\Omega_{tot}$ is exactly unity, but not less or larger?}
\end{center} 
 
The horizon problem is somehow related to the failure in the explaining the remarkable isotropy of the CMB radiation. In fact, only regions which have been in causal contact in the past (when radiation last scattered from matter) could have the same temperature today! It turned out however that in the standard model cosmology, different patches of the universe who are causally disconnected, have also the same temperature \cite{turner,guth-book}. This has been shown from the accurate isotropy of the CMB radiation, where $\Delta T/T \sim 10^{-5}$.

\section{Inflationary paradigm}

Inflationary scenario proposed in 1981 by Alain Guth provided a possible solutions to the horizon and flatness problems mentioned above \cite{guth}. The basic idea at the heart of inflation is that the universe, at a very early stage, has undergone a phase of a very rapid accelerated expansion that \textit{flattens} the spacial section of the universe, and makes the universe homogeneous by stretching the size of its early inhomogeneities. 

This early phase is realized generically by an exponential expansion of the form 
\begin{eqnarray}
\label{exponential expansion}
a(t)\propto e^{H t},
\end{eqnarray}
where in this case, the Hubble parameter $H$ must be given in terms of a \textit{vacuum energy} provided by a potential of a new \enquote{substance}, that dominates the energy density of the universe in its early time.

Under this rapid expansion, the Friedmann equation which has been written as (\ref{flatness equation}), takes the form
\begin{eqnarray}
\label{neglected curvature}
\Omega(t) - 1 \propto \frac{k}{e^{2H t}}.
\end{eqnarray}
The presence of the exponential term in the right hand side means that the inflationary phase drives the universe to $\Omega =1$ very rapidly, leading to a flat universe consistent with current observational data.  

\begin{figure}[h]
\centering
    \includegraphics[width=0.5\textwidth]{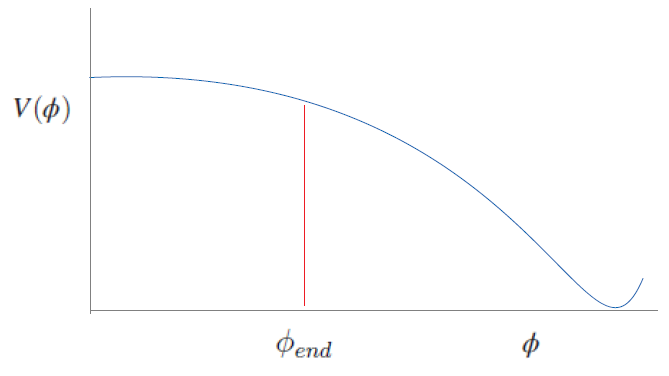}
\caption{The dominant potential energy remains nearly constant during inflation (flat potential). After inflation, the inflaton rolls down converting the potential to kinetic energy. }
\label{fig:flat-potential}
\end{figure}

This phase can be simply driven by a nearly homogeneous scalar field $\phi$. The dynamics of the latter is described in Einstein gravity by the field equations (\ref{einstein equations1}) and (\ref{scalar field equation1}). These equations can be easily adapted to the FRW universe (\ref{genaral frw metric}), and then the homogeneous field $\phi(t)$ will satisfy the following equations  
\begin{eqnarray}
\label{hubble square}
H^{2}=\frac{1}{3M^{2}_{Pl}}
\left(\frac{\dot{\phi}^{2}}{2}+V(\phi) \right),
\end{eqnarray}
and
\begin{eqnarray}
\ddot{\phi}+3H\dot{\phi}+V^{\prime}(\phi)=0,
\end{eqnarray}
where we have ignored the curvature term $k$ due to (\ref{neglected curvature}).
As stated above, in this mechanism, the energy density of the universe must be dominated by the potential energy of the \textit{inflaton} $\phi$ (see Figure~\ref{fig:flat-potential}), thus we suppose that
\begin{eqnarray}
\label{src0}
\dot{\phi}^{2} \ll V(\phi), \quad |\ddot{\phi}| \ll |3H\dot{\phi}|, \quad 
|\ddot{\phi}| \ll V^{\prime}(\phi).
\end{eqnarray}
These conditions lead to simple equations of motion
\begin{eqnarray}
\label{sr equations of motion}
H^{2} \simeq \frac{V(\phi)}{3 M^{2}_{Pl}}, \quad 
3H\dot{\phi} \simeq -V^{\prime}(\phi).
\end{eqnarray}
When applying the \textit{slow-roll conditions} (\ref{src0}), it is useful to define the so called \textit{slow-roll parameters} which are given by 
\begin{eqnarray}
\label{sr parameters}
\epsilon =\frac{M^{2}_{Pl}}{2}\left(\frac{V^{\prime}(\phi)}{V(\phi)} \right), \quad \eta = 
M^{2}_{Pl}\left(\frac{V^{\prime \prime}}{V(\phi)} \right).
\end{eqnarray}
These parameters became useful when solving for $\phi$ at the beginning and at the end of inflation. To see this clearly, let us apply the second Friedmann equation (\ref{second freidman equation}) where in this case ($\Lambda=0$) the energy density and pressure of the inflation are given by
\begin{eqnarray}
\rho= \frac{\dot{\phi}^{2}}{2}+ V(\phi), \quad \text{and}
\quad
p= \frac{\dot{\phi}^{2}}{2}- V(\phi).
\end{eqnarray}
These are the components of the energy-momentum tensor (\ref{energy momentum1}). To that end, we get
\begin{eqnarray}
\frac{\ddot{a}}{a}= -\frac{1}{3M^{2}_{Pl}}
\left(\dot{\phi}^{2}-V(\phi) \right),
\end{eqnarray}
which clearly shows that the expansion is speeding up when the potential energy dominates the right hand side.

Using equation (\ref{hubble square}), the last equation takes the following form
\begin{eqnarray}
\frac{\ddot{a}}{a}=
H^{2} \left(1- \frac{\dot{\phi}^{2}}{2M^{2}_{Pl}H^{2}} \right).
\end{eqnarray}
A simple calculation based on the (slow-roll) equations of motion (\ref{sr equations of motion}) shows that the last term in the right hand side of the previous equation is nothing but the parameter $\epsilon$
\begin{eqnarray}
\frac{\dot{\phi}^{2}}{2M^{2}_{Pl}H^{2}} \simeq
\frac{M^{2}_{Pl}}{2}\left(\frac{V^{\prime}(\phi)}{V(\phi)} \right) \equiv \epsilon, 
\end{eqnarray}
thus the second Friedmann equation takes a simple and an interesting form
\begin{eqnarray}
\ddot{a}=
aH^{2} \left(1- \epsilon \right).
\end{eqnarray}
During inflation the parameter $\epsilon$ remains smaller than unity, which guarantees that the inflaton is slowly moving, and then the inflationary phase ends when $\epsilon=1$.

As an example, inflation can be driven by a potential of the form $V(\phi)=m^{2}\phi^{2}/2$, which easily gives
\begin{eqnarray}
\epsilon= \frac{2M^{2}_{Pl}}{\phi^{2}},
\end{eqnarray}
thus inflation ends $(\epsilon=1)$ when $\phi=\phi_{end}\simeq\sqrt{2}\,M_{Pl}$. Before this $(\epsilon<1)$, $\phi > \sqrt{2}\,M_{Pl}$. 

\begin{figure}[h]
\centering
    \includegraphics[width=0.7\textwidth]{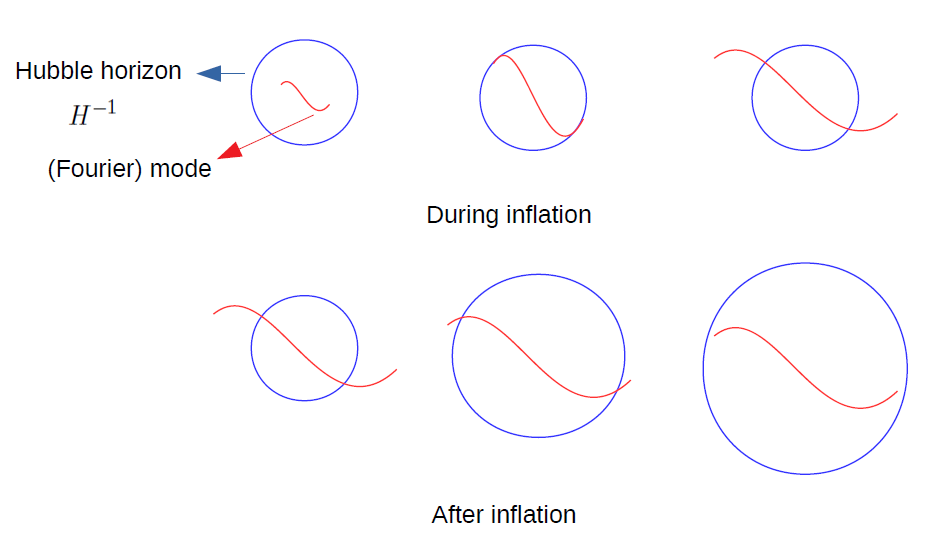}
\caption{From left to right: During inflation, the flat (nearly constant) potential renders the Hubble radius constant leaving the modes growing inside the horizon until they cross it. After inflation, the Hubble radius starts expanding (radiation era) and the modes reenters the horizon \cite{dodelson}.}
\label{fig:mode-by-mode}
\end{figure}

To solve the horizon problem, the largest scales (wavelengths) observed today might have been inside the horizon at the beginning of inflation as illustrated in Figure~\ref{fig:mode-by-mode}. For this reason,  it is useful to know how many e-folds are required for inflation. This e-foldings number of inflationary expansion will be noted $N$, and it arises from $d N=H dt$. This provides a relation between the scale factors at the beginning and at the end of inflation; $a_{end}=e^{N} \times a_{start}$. In general, the e-folds number is given as
\begin{eqnarray}
N &&\equiv \int_{t_{start}}^{t_{end}} H dt \\
&&\simeq 
-\frac{1}{M^{2}_{Pl}} \int_{\phi_{start}}^{\phi_{end}}
\frac{V(\phi)}{V^{\prime}(\phi)} d\phi.
\end{eqnarray}
For the previous given example of the mass term potential, we get 
\begin{eqnarray}
N \simeq \frac{\phi_{i}^{2}}{4M^{2}_{Pl}}-\frac{1}{2}.
\end{eqnarray}

As we see, the value of the number of the e-folds determines the initial value of the inflaton, for example at the time the observed CMB radiations are created. Determination of the exact value of $N$ is not trivial, nevertheless, one may derive its possible values as follows. From the redshift relation (\ref{lambda redshift}), the the initial scales $\lambda_{start}$ are related to the present scales $\lambda_{0}$ as  
\begin{eqnarray}
\lambda_{start}= \lambda_{0}\, \frac{a_{start}}{a_{0}},
\end{eqnarray}
where the present scales are $\lambda_{0}=H_{0}^{-1}$ (the Hubble radius).

Now, we can evaluate the remaining term in terms of the time evolution of the temperature of the universe (\ref{temperature}), thus
\begin{eqnarray}
\frac{a_{start}}{a_{0}}= \frac{a_{start}}{a_{end}}\times
\frac{a_{end}}{a_{0}} = e^{-N} \times \frac{T_{0}}{T_{end}},
\end{eqnarray}
where $a_{end}$ and $T_{end}$ refer to the scale factor and the temperature during the radiation dominated epoch (end of inflation!).
Finally, the scales evolve as follow
\begin{eqnarray}
\label{initial scale}
\lambda_{start} = H_{0}^{-1}
\left(\frac{T_{0}}{T_{end}}\right) e^{-N}.
\end{eqnarray}
Since this initial physical scale must be less than the Hubble horizon during inflation, i.e, $\lambda_{start} < H^{-1}$, then equation (\ref{initial scale}) leads to \cite{liddle-book,lyth-book}
\begin{eqnarray}
N > 60 + \log \left(\frac{T_{end}}{10^{15} \text{GeV}} \right).
\end{eqnarray}
This number, which is taken usually, $N=62$, is the number of e-foldings required to put all the regions of the observed universe in causal contact at the time of the last scattering (creation of the CMB). The energy scale, $10^{15} \text{GeV}$, which appears in the last equation refers to the scale of the Grand Unification Theory (GUT), at which inflation is supposed to occur.  

This mechanism is driven by the homogeneous background field $\phi(t)$ with a large potential energy treated as a classical source in Einstein's field equations. Another interesting feature of inflation, besides solving the flatness and horizon problems, is the generation of the tiny perturbations observed in the cosmic microwave background radiation. In fact, quantum fluctuations in $\phi$ lead to curvature perturbations, which in turn produce small fluctuations in the energy density of the early hot plasma\footnote{This is the hot plasma of relativistic particles which form the energy density of the universe in the radiation dominated era after inflation. The transition from inflation to this stage occurs after \textit{reheating}, where inflaton oscillations are followed by its decay to matter fields \cite{liddle-book,lyth-book}. Reheating is an important phase in inflation and its details go beyond the scope of this thesis.}
\begin{eqnarray}
\delta \rho = V^{\prime}(\phi)\, \delta \phi.
\end{eqnarray}
The comoving curvature perturbation is defined in terms of the inflaton fluctuation and the Hubble parameter as follows \cite{liddle-book,lyth-book}
\begin{eqnarray}
\mathcal{R}=\frac{H}{\dot{\phi}}\, \delta \phi.
\label{curvature-perturbation}
\end{eqnarray}
The \textit{power spectrum} $P_{\mathcal{R}}(k)$ of this quantity is calculated from the ensemble average of the fluctuations \cite{parker-qft} 
\begin{eqnarray}
\left\langle \mathcal{R}_{\vec{k}}
\mathcal{R}_{\vec{k^{\prime}}}
\right\rangle =
(2\pi)^{3} \delta(\vec{k}+\vec{k^{\prime}}) P_{\mathcal{R}}(k)
\end{eqnarray}
where $k$ is the momentum (in Fourier space).

For the slow-roll approximation, the power spectrum satisfies \cite{liddle-book,lyth-book}
\begin{eqnarray}
\label{scale dependence}
P_{\mathcal{R}}(k)k^{3}=\frac{1}{4\epsilon_{\ast}}\left(\frac{H_{\ast}}{M_{Pl}} \right)^{2}  \propto k^{n_{s}-1},
\end{eqnarray}
where the sign $\ast$ means the values of the parameters at the time of the horizon crossing, i.e, when the mode left the Hubble radius ($k=aH$).

The parameter $n_{s}$ which is called the spectral index, or tilt, determines the scale dependence of the perturbation, or its deviation from scale invariance ($n_{s}=1$). Calculations based on the slow-roll approximation showed that the spectral index is given in terms of the slow roll parameters (\ref{sr parameters}) as
\begin{eqnarray}
n_{s}-1= 2\eta -6\epsilon.
\end{eqnarray}
Recent Planck data strongly suggest a nearly scale invariant perturbations, precisely $n_{s} \simeq 0.965$ \cite{planck}. Measurements of the spectral index allows us then to constraint the form of the potentials that drive inflation, and come out with only models that are consistent with the measured values.

Not less important, prediction of the cosmological inflation is the production of the tensor perturbations. These tensor modes that arise from the metric fluctuation $\delta g_{\mu\nu}$ are the origin of the \textit{primordial gravitational waves}. Similarly, the power spectrum of the tensor modes, noted $P_{t}(k)$, is given in terms of the Hubble parameter at the time of the horizon crossing, but in this case
\begin{eqnarray}
P_{t}(k)k^{3} = 4\left(\frac{H_{\ast}}{M_{Pl}} \right)^{2}.
\end{eqnarray}
The amplitude of the tensor perturbations $\Delta_{t}$ is related to the amplitude of the scalar perturbation $\Delta_{s}$ by $\Delta_{t}= r \Delta_{s}$, where the parameter $r$ is called the tensor-to-scalar ratio. In the slow-roll approximation, this ratio takes the form \cite{liddle-book,lyth-book}
\begin{eqnarray}
r = 16 \epsilon.
\end{eqnarray}  
Last few years accurate cosmological data have offered a powerful discrimination between different theories, and helped in supporting or ruling out various inflationary models. The predictions of any successful model of inflation have to be consistent with the observed bounds on the $(n_{n},r)$ plan (see Figure~\ref{fig:planckresults} below).
\begin{figure}[h]
\centering
    \includegraphics[width=0.7\textwidth]{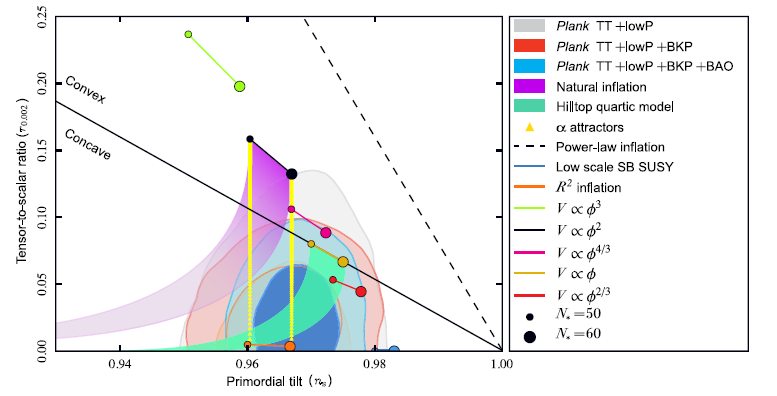}
\caption{Recent Planck results \cite{planck}. The data suggests only models with small tensor-to-scalar ratio. Some of these models are the Starobinski model \cite{starobinsky, predictions of starobinsky} and $\alpha$-attractors \cite{alpha-attractors}.}
\label{fig:planckresults}
\end{figure}

\clearpage  
\lhead{\emph{Chapter 6}}  
  \chapter{Affine inflation and frame ambiguities}
\vspace{-0.5 cm}
\epigraph{\textit{Not only is the Universe stranger than we think, it is stranger than we can think}\, --- Werner Heisenberg}{}

Standard inflation presented in the end of the previous Chapter is based on Einstein gravity coupled to a scalar field in the form (\ref{gr minimal action}). Theories of inflation driven by scalar fields coupled nonminimally to gravity have also been considered and studied in various details in metric gravity \cite{fakir,kaiser,bezrukov}. The studies are performed in both Jordan and Einstein frames where same predicted results are not guaranteed. 

In this Chapter, we came to one of the main points of this thesis. We will consider the inflationary dynamics in the context of purely affine gravity. As we have seen, in the case of scalar fields, the affine gravity approach necessitates nonvanishing
potentials, and thus, studying inflation in this context is important by itself. Throughout this chapter we will deal with the following potential 
\begin{eqnarray}
V\left(\phi\right)=V_{0}+\frac{\lambda}{4}\left(\phi^{2}-v^{2}\right)^{2},
\end{eqnarray}
where $v$ is a constant vacuum expectation value.

We will address the inflationary dynamics through three models. The first and the most important model is the standard \textit{affine inflation} where the inflaton $\phi$ is coupled nonminimally to affine gravity as in (\ref{affine single}). The second, will be a direct application of the first where the inflaton is considered as the standard model Higgs boson (Higgs affine inflation). Finally we will consider inflation in the context of induced affine gravity (\ref{induced affine action}).

\section{Affine inflation}
Here the dynamics of the inflaton $\phi$ is governed by its equations of motion (\ref{nonminimal-affine einstein equations}) and (\ref{nonminimal-affine scalar field equation}). In homogeneous flat FRW universe, the distribution of the scalar field is now described by its associated energy density and pressure, respectively, as follows
\begin{eqnarray}
\label{density and pressure}
\rho\left(\phi \right)= \frac{1}{ \mathcal{F}\left(\phi \right)} \left(\frac{\dot{\phi}^{2}}{2} +\frac{V\left(\phi \right)}{\mathcal{F\left(\phi \right)}} \right) \\
p\left(\phi \right)= \frac{1}{\mathcal{F}\left(\phi \right)} \left(\frac{\dot{\phi}^{2}}{2} -\frac{V\left(\phi \right)}{\mathcal{F}\left(\phi \right)} \right),
\end{eqnarray}
where the function $\mathcal{F}(\phi)$ is given by (\ref{F}).

As we see, the quasi-de Sitter solution which requires $p\left(\phi \right)= -\rho \left( \phi \right)$ is possible for some slowly rolling fields. The cosmological constant case is implicitly understood here for $\phi=\phi_{min}$.

In this case, the Friedman equations are derived from the gravitational field equations (\ref{nonminimal-affine einstein equations}), and are written in terms of the Hubble parameter $H$  as follows
\begin{eqnarray}
\label{hubble1}
H^{2}=\frac{1}{3 M_{Pl}^{2} \mathcal{F}\left(\phi \right)} \left(\frac{\dot{\phi}^{2}}{2} 
+\frac{V\left(\phi \right)}{\mathcal{F\left(\phi \right)}}  \right)
\end{eqnarray}
and
\begin{eqnarray}
\label{hubble2}
\dot{H}+H^{2}= -\frac{1}{3M_{Pl}^{2} \mathcal{F}\left(\phi \right)} \left(\frac{\dot{\phi}^{2}}{2} -\frac{V\left(\phi \right)}{\mathcal{F}\left(\phi \right)} \right)
\end{eqnarray}
The possible quasi-de Sitter solution (constant Hubble parameter) shows that an inflationary phase is possible in this theory. For simplicity, we will perform the calculation using the new field $\tilde{\phi}$ given by (\ref{affine field transformation}). This equation is integrated easily giving
\begin{eqnarray}
\phi(\tilde{\phi})=\frac{M_{Pl}}{\sqrt{\xi}}\sinh\left(\frac{\sqrt{\xi}}{M_{Pl}}\tilde{\phi}\right).
\end{eqnarray}
The FRW metric remains unchanged under the last redefinition, since no conformal transformation is applied. Now, we apply the slow roll conditions on $\tilde{\phi}$ as follows
\begin{eqnarray}
\label{slow roll conditions}
\frac{\dot{\tilde{\phi}}^{2}}{2} \ll \tilde{V} (\tilde{\phi}),\,\, \frac{\ddot{\tilde{\phi}}}{\dot{\tilde{\phi}}} \ll
H.
\end{eqnarray}
In this case, the new potential has the form
\begin{eqnarray}
\tilde{V}(\tilde{\phi})=\frac{\lambda}{4}\left( \frac{M^{2}_{Pl}\xi^{-1}\sinh^{2}\left(\frac{\sqrt{\xi}}{M_{Pl}}\tilde{\phi}\right)-v^{2}}
{1+\sinh^{2}\left(\frac{\sqrt{\xi}}{M_{Pl}}\tilde{\phi}\right)}\right)^{2}.
\end{eqnarray}
With these conditions, the Friedman equations (\ref{hubble1}) and (\ref{hubble2}) take the form
\begin{eqnarray}
H^{2}\simeq \frac{\tilde{V}(\tilde{\phi})}{3M^{2}_{Pl}},\,\, \text{and}\,\,3H\dot{\tilde{\phi}}\simeq
-\tilde{V}^{\prime}(\tilde{\phi}).
\end{eqnarray}
Now, the slow-roll parameters are give by
\begin{eqnarray}
\label{slow roll parameters}
\epsilon=\frac{M^{2}_{Pl}}{2}\left(\frac{\tilde{V}^{\prime}}{\tilde{V}}  \right)^{2}
\simeq 128\xi \exp \left(-4\frac{\sqrt{\xi}}{M_{Pl}}\tilde{\phi} \right) \\
\eta= M^{2}_{Pl} \left(\frac{\tilde{V}^{\prime\prime}}{\tilde{V}}  \right)
\simeq -32\xi \exp \left(-2\frac{\sqrt{\xi}}{M_{Pl}}\tilde{\phi} \right),
\end{eqnarray}
where we have taken a large field $\tilde{\phi} >M_{Pl}/\sqrt{\xi}$.

\begin{figure}[h]
\centering
    \includegraphics[width=0.5\textwidth]{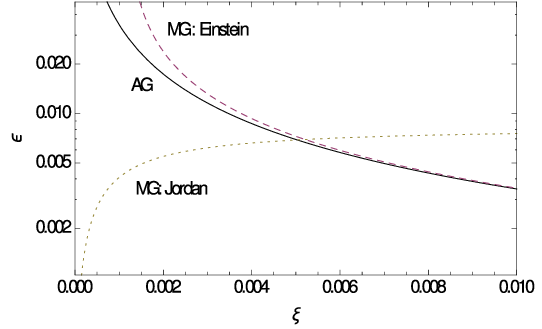}
\caption{The slow-roll parameter $\epsilon$ as a function of the coupling parameter $\xi$.}
\label{fig:slow roll parameter}
\end{figure}

The number of e-folds takes the form
\begin{align}
N&=\frac{1}{M^{2}_{Pl}}\int_{\tilde{\phi}_{f}}^{\tilde{\phi}_{i}}\frac{\tilde{V}(\tilde{\phi})}{\tilde{V}^{\prime}(\tilde{\phi})}d\tilde{\phi} \nonumber \\ &
\simeq\frac{1}{32\xi}\left[\exp\left(2 \frac{\sqrt{\xi}}{M_{Pl}}\tilde{\phi}_{i}\right)- 
\exp\left(2 \frac{\sqrt{\xi}}{M_{Pl}}\tilde{\phi}_{f}\right)\right].& \label{efoldings}
\end{align}
Inflation ends when $\tilde{\phi}=\tilde{\phi}_{f}$, or $\epsilon \simeq 1$ where the slow-roll conditions break down. The initial field $\tilde{\phi}_{i}$ is determined from the number of e-foldings $N$. Initial and final values of the inflaton are presented in Table~\ref{tab:values of inflaton}.
\begin{table}[h]
\centering
\caption{The inflaton redefinition and its initial and final values in both metric gravity and affine gravity, for large $\xi$. The field values are below Planck mass in affine gravity.}
\label{tab:values of inflaton}
\begin{tabular}{|c|c|cc|}
\hline
\, &Einstein frame (metric gravity) &Affine gravity&\\
\hline
\, & \, & \, &\\
$\xi$ & $\xi \gtrsim 6.25\times 10^{-3}$ & $\xi \gtrsim 3.12 \times 10^{-2}$& \\
\, & \, & \, &\\
$\phi(\tilde{\phi})$& $\frac{M_{Pl}}{\sqrt{\xi}} \exp \left(\sqrt{\frac{\xi}{1+6\xi}}\frac{\tilde{\phi}}{M_{Pl}} \right)$ & $\frac{M_{Pl}}{\sqrt{\xi}}\sinh\left(\frac{\sqrt{\xi}}{M_{Pl}}\tilde{\phi}\right)$ &\\
\, & \, & \, &\\
$\tilde{\phi}_{i}/M_{Pl}$ & $\sqrt{\frac{1+6\xi}{\xi}}\ln\left( \sqrt{\frac{8\xi N}{1+6\xi}}\right)$ & $\ln\left(32\xi N \right)/2\sqrt{\xi}$ &\\
\, & \, & \, &\\
$\tilde{\phi}_{f}/M_{Pl}$ &$\sqrt{\frac{1+6\xi}{16 \xi}}\ln\left( \frac{8\xi}{1+6\xi}\right)$ & $\ln\left(128\xi \right)/4\sqrt{\xi}$ &\\
\hline
\end{tabular}
\end{table}
\newpage
The slow-roll parameters are evaluated at the value $\tilde{\phi}$ when the scale of interest crossed the horizon during the inflationary phase, and they must remain smaller than unity and then deviations of the spectrum of perturbations from scale invariant spectrum are small. The slow-roll parameter $\epsilon$ is depicted in Figure~\ref{fig:slow roll parameter} as a function of $\xi$. The parameter behaves like in metric gravity only for very large $\xi$.  
The spectral index is written in its first order, $n_{s}=1-6\epsilon+2\eta$, and reads
\begin{eqnarray}
n_{s}\simeq 1-\frac{3}{4\xi N^{2}}-\frac{2}{N}.
\end{eqnarray}
In metric gravity, one may show that this quantity is given as \cite{kaiser}
\begin{align}
\label{spectral index gr}
    n_{s}\simeq 
\begin{cases}
    1-\frac{32\xi}{16\xi N-1},& \text{for } \phi_{f}^{2} \gg v^{2}\\        
    1-\frac{16\xi\left(1+\delta^{2} \right)}{8\xi N\left(1+\delta^{2}\right)+\delta^{2}}         & \text{for } \phi_{f}^{2} \simeq v^{2}
\end{cases}
\end{align}
where $\delta^{2}=\xi v^{2}/M^{2}_{Pl}$.

\begin{figure}[h]
\centering
    \includegraphics[width=0.5\textwidth]{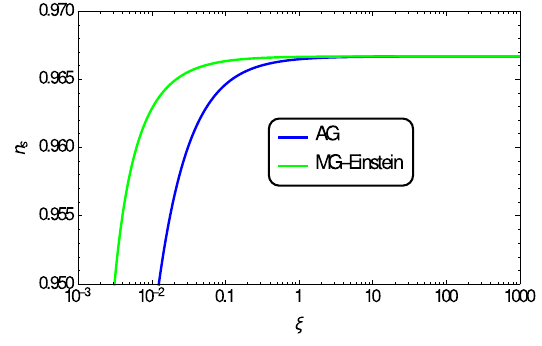}
\caption{First order spectral indices predicted by metric gravity and affine gravity.}
\label{fig:first order spectral index}
\end{figure}

Figure~\ref{fig:first order spectral index} shows the behavior of the first order spectral index for both metric gravity (MG) and affine gravity (AG) for large fields.

\begin{figure}[h]
\centering
    \includegraphics[width=0.5\textwidth]{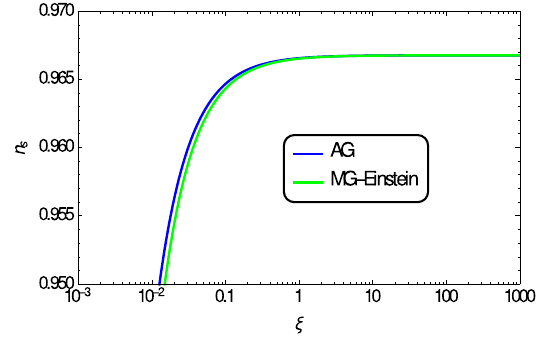}
\caption{Second order spectral indices predicted by metric gravity and affine gravity.}
\label{fig:second order spectral index}
\end{figure}

\newpage
Considering second order terms, the spectral index $n_{s}$ takes the following form \cite{liddle,stewart}
\begin{eqnarray}
\label{second order spectral index}
n_{s}= 1-6\epsilon +2\eta +\frac{1}{3}\left(44-18c\right)\epsilon^{2}+\left(4c-14 \right)\epsilon \eta 
+\frac{2}{3}\eta^{2}+\frac{1}{6}\left(13-3c \right)\zeta^{2}, 
\end{eqnarray}
where $c=4\left(\ln 2 +\gamma \right)\simeq 5.081$ and $\gamma$ is Euler's constant, and the third slow-roll parameter $\zeta^{2}$ has the following form
\begin{eqnarray}
\zeta^{2}\equiv M^{4}_{Pl}\frac{\tilde{V}^{\prime\prime\prime}\tilde{V}^{\prime}}{\tilde{V}^{2}} \simeq
\left(32 \xi\right)^{2}\exp \left(-4\frac{\sqrt{\xi}}{M_{Pl}}\tilde{\phi} \right).
\end{eqnarray}
Since the slow-roll parameters of affine inflation decay exponentially, deviations from the first order spectral index is very tiny. This is not the case for metric gravity as it is illustrated in Figure~\ref{fig:second order spectral index}. Finally, the tensor-to-scalar ratio $r\simeq 16\epsilon$ reads
\begin{eqnarray}
r\simeq \frac{2}{\xi N^{2}},
\end{eqnarray}
which takes a very small values, $r \lesssim 1.7 \times 10^{-5}$, for the bound $\xi \gtrsim 3.12 \times 10^{-2}$ (see figure~\ref{fig:first order spectral index}).

Recent data, provides a power spectrum of the primordial perturbations of the order \cite{planck}
\begin{eqnarray}
\frac{H^{2}}{8\pi^{2} \epsilon M^{2}_{Pl}} \simeq 2.4 \times 10^{-9},
\end{eqnarray}
which allows us to put a constraint on the following ratio
\begin{eqnarray}
\label{ratio lambda xi}
\lambda / \xi \simeq 2.66 \times 10^{-11}.
\end{eqnarray}
This ratio will be important later when we address Higgs affine inflation, where the measured self coupling $\lambda$ will require a large nonminimal coupling $\xi$. 

\newpage
This study shows that like metric gravity, slow-roll inflation arises naturally in the context of affine gravity, and leads to observed quantities that fit the recent data.

\section{Other affine inflationary models}

\subsection{Higgs affine inflation}
\label{higgs inflation}

Like any scalar field, the SM Higgs boson may drive the cosmic inflation. In this case, the predictions must be in agreement with the SM measured parameters such as the Higgs mass and the self coupling parameter. However, for a Higgs boson minimally coupled to metric gravity (GR), the observed power spectrum requires an extremely small quartic coupling $\lambda \simeq \mathcal{O}(10^{-13})$. Nevertheless, it has been shown that, this constraint can be relaxed by adding a nonminimal coupling term, Higgs-curvature, to the action. Then, the SM quartic coupling $\lambda \simeq \mathcal{O}(10^{-1})$ is attained for large nonminimal coupling parameter $\xi \simeq 10^{4}$. The non-minimal coupling then motivates the SM Higgs inflation, where the predictions are in agreement with recent Planck results \cite{bezrukov, planck}. Our aim here is to study \enquote{Higgs affine inflation}, where the SM Higgs boson is supposed to be coupled to affine gravity rather than metric gravity.

Here, the mechanism is similar to that of the previous section, where $\phi\equiv h$ being the SM Higgs boson \cite{short review}. In this case, and from equation (\ref{ratio lambda xi}), the SM quartic coupling $\lambda \simeq 0.13$ implies
\begin{eqnarray}
\xi \simeq 4.8 \times 10^{9}.
\end{eqnarray}
The affine nonminimal coupling is then larger than its value in metric gravity. This leads to an extremely small tensor to scalar ratio
\begin{eqnarray}
r\simeq \mathcal{O}\left(10^{-13} \right).
\end{eqnarray}
As we see, the tensor contribution is tiny and negligible. Recent observations suggest a very small upper bound for tensor perturbations, the tensor to scalar ratio is of the order $r < 0.08 $. Future observations are expected to provide us with a precise bounds, since then, one may decide whether Higgs affine inflation could be considered as a good model for the early universe. In Table~\ref{tab:2} we summarize the results obtained here and compare them with Higgs inflation in metric gravity.

\begin{table}[h]
\centering
\caption{Higgs affine inflation suggests a strong Higgs-curvature coupling $\xi$, and a negligible tensor-to-scalar ratio.}
\label{tab:2}
\begin{tabular}{|c|c|cc|}
\hline
Parameters &Higgs Inflation (metric gravity) &Higgs Affine Inflation&\\
\hline
\, & \, & \, &\\
$\xi$ & $10^{4}$ & $10^{9}$& \\
\, & \, & \, &\\
$n_{s}$& $0.97$ & $0.97$ &\\
\, & \, & \, &\\
$r$ & $0.0032$ & $\mathcal{O}(10^{-13})$ &\\
\hline
\end{tabular}
\end{table}

\subsection{Induced affine inflation}
\label{induced affine inflation}
Induced affine inflation is the inflationary dynamics based on induced affine gravity action (\ref{induced affine action}). A detailed study of this model has been done in Ref. \cite{induced affine inflation}. 

It has been shown that for ordinary inflation where the fields start with values $\phi_{\text{start}} \ll v$, the scale factor follows a power law
\begin{eqnarray}
a\left(t\right) \propto t^{1/8\xi}.
\end{eqnarray}
In such theories, the spectrum of density perturbation is sensible to the value of the power $p$ \cite{abbott, lyth}. 

In metric gravity, the conformal transformation which leads to different power law would clearly provide a significant difference between the density perturbations which are calculated in two conformal frames. However, field redefinition in affine gravity does not alter the physics, but it enters only as a new variable leading to a unique observable spectral index and tensor-to-scalar ratio. These are given in terms of the coupling $\xi$ as 
\begin{eqnarray}
n_{s}-1=-\frac{16\xi}{1-8\xi},\quad \quad r=128 \xi.
\end{eqnarray}
Recent Planck bound, $r<0.12$ implies $\xi < 10^{-3}$. This clearly drags the spectral index $n_{s}$ up to its required bound. Thus, the induced gravity inflation, in both metrical and affine gravity setups, cannot satisfy the recent Planck bounds on $r$ and $n_{s}$ simultaneously. The reason is that induced gravity inflation supports only large tensor-to-scalar ratio, a feature which is not specific to induced affine gravity; it already happens in the metric induced gravity (see Figure~\ref{fig: tilt} below).

\begin{figure}[h]
\centering
    \includegraphics[width=0.5\textwidth]{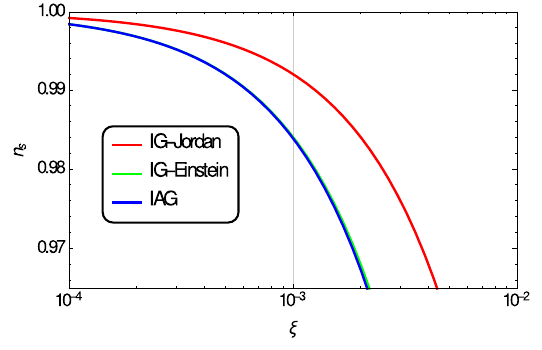}
\caption{Spectral indices predicted by metric induced gravity (IG) and induced affine gravity (IAG). The tensor-to-scalar bound, $r<0.12$, drags the spectral index to larger values for both theories.}
\label{fig: tilt}
\end{figure}

We conclude this chapter by addressing briefly a serious problem one faces when studying inflation using nonminimal coupling. As we have seen from the inflationary predictions summarized in Figures \ref{fig:first order spectral index}, \ref{fig:second order spectral index} and \ref{fig: tilt}, metric gravity suffers from Einstein-Jordan ambiguities where the observable quantities are frame dependent. 

The ambiguity in metrical gravity is traced back to the conformal transformation (\ref{conformal transformation}) that maps one frame to the other. Since this transformation is nothing but a field and metric redefinitions, one expects then physics to be identical in both frames. As we have stated in chapter 4, this is true only at the classical level. The problem arises when we consider the quantum fluctuations of the fields. When passing from Jordan to Einstein frame, the mixing between the inflaton and metric fluctuation is not avoided. An important quantity which is not invariant under conformal transformation is the curvature perturbation (\ref{curvature-perturbation}). This undoubtedly has an effects on the form of the spectral indices, and then leads to different results in different frames.  Attempts have been made to overcome this ambiguities, and to come out with a unique description of inflation and other cosmological scenarios, but the debate has not settled down \cite{kaiser-frame independent,fakir-frame independent,
sasaki,sasaki-power law,karam}.

The advantage of pure affine gravity, which we have considered throughout this thesis, is that it provides us with a unique \textit{geometric} frame (a unique metric). In this case, the inflaton dynamics is described in one and the same frame with metric tensor $g_{\mu\nu}$. This is clearly seen from the fact that 
nonminimal coupling actions are transformed to minimal actions by making only a field redefinition. The uniqueness of the metric tensor ensures then the invariance of the intrinsic curvature perturbations and the observable parameters such as the spectral index.

\clearpage  
\lhead{\emph{Chapter 7}}  
  \chapter{Higher dimensional affine gravity}
\vspace{-0.5 cm}
\epigraph{\textit{But the creative principle resides in mathematics. In a certain sense, therefore, I hold it true that pure thought can grasp reality, as the ancients dreamed.}\,\\---Albert Einstein}{}

In this chapter we will consider affine gravity, particularly Eddington's gravity, in a higher dimensional space. The latter will be considered as the product of two spaces. Some of the results will be based on geometric operations such as exterior derivatives of differential forms, and quantities like connection and curvature forms, which are directly given without details. In this case, the reader may be referred to some text books on differential geometry such as \cite{chern,Lichnerowicz} and others.

\section{Immersed space}

The spacetime is described by a four dimensional space $M_{4}$ which is immersed in an affine eight dimensional space $M_{8}$ which is the product of two identical four dimensional real manifolds $W_{4}$ \cite{Clerc}
\begin{equation}
M_{8}=W_{4}\times W_{4}. \label{stru}
\end{equation}
The following index notation will be used \cite{Lichnerowicz,Yano}: 
\newline
For Latin indices: $i,j,..=1,...8$ and for Greek indices: $\alpha,\beta,..=1,...4$. We also introduce on the indices the operation $\ast$ such that $i^{\ast}=i\pm4$, (then $(i^{\ast})^{\ast}=i$).

This means that Latin indices take both Greek indices, $\alpha$ and $\alpha^{\ast}$ via the operation $\ast$, i.e, $i= \alpha, \alpha^{\ast}=1,...8$. 
\newline
One may show that the above construction confers to the large space $M_{8}$ a hypercomplex structure \cite{Clerc,Crum,AB}.
\newline
The hypercomplex coordinates, noted $X^{\alpha}=x^{\alpha}+Ix^{\alpha^{\ast}}$ are elements of the Hypercomplex Ring $\mathbf{H}$, such that $I^{2}=1$ and $x^{\alpha},x^{\alpha^{\ast}}$ are real coordinates in $W_{4}\times W_{4}$.

The spacetime will be defined as the diagonal submanifold $M_{4}$ where
\cite{Clerc,AB}
\begin{equation}
x^{\alpha^{\ast}}=0.
\end{equation}
The real elements $x^{\alpha},x^{\alpha^{\ast}}$ define the associated diagonal coordinates.

This construction is similar to that of complex manifolds, and one similarly defines the almost hypercomplex structure on the tangent space of $M_{8}$ by the operator $J$ such that \cite{Lichnerowicz}
\begin{equation}
J\left(  \frac{\partial}{\partial x^{\alpha}}\right) = \frac{\partial}{\partial x^{\alpha^{\ast}}}, \quad J\left(  \frac{\partial}{\partial x^{\alpha^{\ast}}}\right) = \frac{\partial}{\partial x^{\alpha}}. 
\end{equation} 
Thus, this operator satisfies $J^{2}=id$, with $id$ refers to the identity operator on the tangent space of $M_{8}$. This operator is defined in the real basis $M_{8}$ by a tensor with components \cite{Crum,Lichnerowicz}
\begin{equation}
\label{realmat}
J^{i}_{j}=
\left( {\begin{array}{cc}
0 & \mathcal{I}_{4} \\
\mathcal{I}_{4} & 0
\end{array} } \right),
\end{equation}
where $\mathcal{I}_{4}$ is the $4\times4$ unit matrix.
Thus, the operator $J$ has the components
\begin{equation}
\label{comp}
J^{\alpha}_{\beta} = J^{\alpha^{\ast}}_{\beta^{\ast}} =0, \quad
J^{\alpha}_{\beta^{\ast}} = J^{\alpha^{\ast}}_{\beta}= \delta^{\alpha}_{\beta},
\end{equation}
and it corresponds to the multiplication by $I$ (remember that $I^{2}=1$). To see this clearly, we define a hypercomplex basis by the hypercomplex vectors
\begin{equation}
\frac{\partial}{\partial X^{\alpha}} = \frac{1}{2}\left(\frac{\partial}{\partial x^{\alpha}}+I\frac{\partial}{\partial x^{\alpha^{\ast}}}  \right), \quad \frac{\partial}{\partial X^{\alpha^{\ast}}} = \frac{1}{2}\left(\frac{\partial}{\partial x^{\alpha}}-I\frac{\partial}{\partial x^{\alpha^{\ast}}}  \right), 
\end{equation}
such that
\begin{equation}
J\left(  \frac{\partial}{\partial X^{\alpha}}\right) = I\frac{\partial}{\partial X^{\alpha}}, \quad J\left(  \frac{\partial}{\partial X^{\alpha^{\ast}}}\right) = -I\frac{\partial}{\partial X^{\alpha^{\ast}}}. 
\end{equation} 
Now, the operator $J$ has a representation in the hypercomplex basis, which is given by the matrix
\begin{equation}
J^{i}_{j}=
\left( {\begin{array}{cc}
I\mathcal{I}_{4} & 0 \\
0 & -I\mathcal{I}_{4}
\end{array} } \right).
\end{equation}  
Here, the real representation of the linear group $GL(4,\mathbf{H})$ can be described by the subgroup of $GL(8,\mathbf{R})$ defined by the matrices which commute with (\ref{realmat}).  

The connection form $\omega^{i}_{j}$ is given in the frame of $M_{8}$ by the matrix \cite{Lichnerowicz}
\begin{equation}
\omega^{i}_{j}=\left( {\begin{array}{cc}
\omega^{\alpha}_{\beta} & \omega^{\alpha}_{\beta^{\ast}} \\
\omega^{\alpha^{\ast}}_{\beta} & \omega^{\alpha^{\ast}}_{\beta^{\ast}}
\end{array} } \right).
\end{equation}   
In the natural diagonal frame of $M_{8}$, the affine connection satisfies \cite{Clerc, Crum}
\begin{equation}
\omega^{\alpha}_{\beta} = \omega^{\alpha^{\ast}}_{\beta^{\ast}}, \quad \omega^{\alpha}_{\beta^{\ast}} = \omega^{\alpha^{\ast}}_{\beta}.
\end{equation}
Generally, the connection form is written in terms of its components $\Gamma^{i}_{jk}$ as
\begin{equation}
\omega^{i}_{j} = \Gamma^{i}_{jk} dx^{k}.
\end{equation}
To that end, the affine connections in the natural diagonal frame bundle of $M_{8}$ satisfy the conditions 
\begin{equation}
\label{condt}
\Gamma^{i}_{jk}=\Gamma_{j^{\ast}k}^{i^{\ast}},\quad \Gamma
_{jk}^{i}=\Gamma_{j^{\ast}k^{\ast}}^{i}.
\end{equation}
These conditions can be derived from the relation $\nabla J=0$, where $\nabla$ is the covariant derivative with respect to the connection $\Gamma_{jk}^{i}$, and $J$ is the operator of the almost hypercomplex structure given above by its components (\ref{comp}).

Now, let us turn to the restriction in the spacetime $M_{4}$, where the above conditions induce for all diagonal frame of $M_{4}$ the equations
\begin{equation}
\Gamma_{\beta\gamma}^{\alpha}=\Gamma_{\beta^{\ast}\gamma}^{\alpha^{\ast}
}=\Gamma_{\beta\gamma^{\ast}}^{\alpha^{\ast}}=\Gamma_{\beta^{\ast}\gamma
^{\ast}}^{\alpha},\quad \Gamma_{\beta\gamma}^{\alpha^{\ast}}
=\Gamma_{\beta^{\ast}\gamma}^{\alpha}=\Gamma_{\beta\gamma^{\ast}}^{\alpha
}=\Gamma_{\beta^{\ast}\gamma^{\ast}}^{\alpha^{\ast}}.\label{D7}
\end{equation}
With this structure at hand, one may show that the coefficients $\Gamma_{jk}^{i}$ with even number of
asterisks transform like connections, while those with odd number of
asterisks transform like tensors in all natural diagonal frame of
$V_{4}$ \cite{Lichnerowicz,Yano}. This allows us to define an affine connection $\mathcal{L}_{\gamma\beta}^{\alpha}$ and a tensor $\Lambda
_{\beta\gamma}^{\alpha}$ as follows
\begin{eqnarray}
\Gamma_{\beta\gamma}^{\alpha}=\Gamma_{\beta^{\ast}\gamma}^{\alpha^{\ast}
}=\Gamma_{\beta\gamma^{\ast}}^{\alpha^{\ast}}=\Gamma_{\beta^{\ast}\gamma
^{\ast}}^{\alpha}=\mathcal{L}_{\gamma\beta}^{\alpha},\\
\Gamma_{\beta\gamma}^{\alpha^{\ast}}=\Gamma_{\beta^{\ast}\gamma}^{\alpha
}=\Gamma_{\beta\gamma^{\ast}}^{\alpha}=\Gamma_{\beta^{\ast}\gamma^{\ast}
}^{\alpha^{\ast}}=\Lambda_{\beta\gamma}^{\alpha}, \label{D8}
\end{eqnarray}
where the affine connection $\mathcal{L}_{\beta\gamma}^{\alpha}$ has no symmetric (antisymmetric) character.

We proceed by defining the curvature form induced in $M_{4}$ as
\begin{equation}
\widehat{\Omega}_{j}^{i}
=\frac{1}{2}\widehat{R}_{j\lambda\mu}^{i}dx^{\lambda} \wedge dx^{\mu},
\end{equation}
where the hat denotes the restriction in $M_{4}$ ($x^{\mu^{\ast}}=0$) and $R_{j\lambda\mu}^{i}$ are the components of the Riemann tensor.

Thus, the induced Riemann tensor in $M_{4}$ takes the form
\begin{equation}
\widehat{R}_{j\lambda\mu}^{i}=\partial_{\lambda}\Gamma_{j\mu}
^{i}-\partial_{\mu}\Gamma_{j\lambda}^{i}+\Gamma_{\rho\lambda}^{i}
\Gamma_{j\mu}^{\rho}+\Gamma_{j\mu}^{\rho^{\ast}}\Gamma
_{\rho^{\ast}\lambda}^{i}-\Gamma_{\rho\mu}^{i}\Gamma_{j\lambda}^{\rho}
-\Gamma_{\rho^{\ast}\mu}^{i}\Gamma_{j\lambda}^{\rho^{\ast}}.
\label{D19}
\end{equation}
In this case, we can construct the two independent Ricci-type tensors as
\begin{equation}
\mathcal{P}_{\alpha\beta}=\widehat{R}_{\beta\lambda\alpha}^{\lambda}, \quad
\mathcal{Q}_{\alpha\beta}=\widehat{R}_{\alpha^{\ast}\lambda\beta}^{\lambda},
\end{equation} 
which are given explicitly as follows \cite{Clerc,AB}
\begin{eqnarray}
\mathcal{P}_{\alpha\beta}=\partial
_{\lambda}\mathcal{L}_{\alpha\beta}^{\lambda}-\partial_{\alpha}\mathcal{L}
_{\lambda\beta}^{\lambda}+\mathcal{L}_{\lambda\rho}^{\lambda}\mathcal{L}
_{\alpha\beta}^{\rho}-\mathcal{L}_{\alpha\rho}^{\lambda}\mathcal{L}
_{\lambda\beta}^{\rho}+\Lambda_{\rho\lambda}^{\lambda}\Lambda_{\beta\alpha}^{\rho}
-\Lambda_{\rho\alpha}^{\lambda}\Lambda_{\beta\lambda}^{\rho},\label{P}\\
\mathcal{Q}_{\alpha\beta}
=\partial_{\lambda}\Lambda_{\alpha\beta}^{\lambda}-\partial_{\beta}
\Lambda_{\alpha\lambda}^{\lambda}+\mathcal{L}_{\lambda\rho}^{\lambda}
\Lambda_{\alpha\beta}^{\rho}-\mathcal{L}_{\beta\rho}^{\lambda}\Lambda
_{\alpha\lambda}^{\rho}+\Lambda_{\rho\lambda}^{\lambda}\mathcal{L}
_{\beta\alpha}^{\rho}-\Lambda_{\rho\beta}^{\lambda}\mathcal{L}_{\lambda\alpha
}^{\rho}. \label{Q}
\end{eqnarray}
The first motivation that led to this mathematical construction was the generalization of Einstein-Schr\"odinger theory \cite{Clerc, Crum} as an attempt to unify gravity and classical electrodynamics, where the spacetime $M_{4}$ was supposed to be endowed with a metric structure. Another interesting application  of the formalism has been done to describe a dynamical dark energy \cite{AB,AB2}.

Although the formalism is mathematically complicated, however, it may lead to a possible modification of gravity. Here, we will be interested only in the extensions of Eddington's purely affine gravity. These extensions will arise from the Lagrangian densities which are constructed from the Ricci-type tensors (\ref{P}) and (\ref{Q}).      
      
\section{Eddington's gravity}

Here, we will focus on the simplest extension of Eddington's gravity, where the action is constructed from the symmetric part of the Ricci tensor (\ref{P}). For simplicity, the affine connection $\mathcal{L}$ will be taken symmetric. In this case, we have \cite{azri-immersed}
\begin{equation}
S=\int d^{4}x \sqrt{|| \mathcal{P}_{(\alpha\beta)}||}. \label{action1}
\end{equation} 
The variation of the Ricci tensor is given by
\begin{equation}
\delta {\mathcal{P}}_{\alpha\beta} = \nabla_{\mu}\Big( \delta\mathcal{L}^{\mu}_{\beta\alpha}\Big) -
\nabla_{\beta}\Big(\delta\mathcal{L}^{\mu}_{\mu\alpha}\Big),
\end{equation}
where we have omitted the sign of symmetry, however, it must be implicitly understood. 

Following the same procedure made so far in deriving the field equations, the variational principle applied to action (\ref{action1}) leads to the dynamical equation
\begin{equation}
\nabla_{\mu} \left[ \sqrt{{\texttt{Det}}\left[{\mathcal{P}}\right]}
\left({\mathcal{P}}^{-1}\right)^{\alpha\beta}\right]=0, \label{motion1}
\end{equation} 
which is solved as
\begin{equation}
\sqrt{{\texttt{Det}}\left[{\mathcal{P}}\right]}
\left({\mathcal{P}}^{-1}\right)^{\alpha\beta}  = \lambda
\sqrt{g} g^{\alpha\beta}, \label{imersed-density identity}
\end{equation}
where $\lambda$ is a constant and $g_{\alpha\beta}$ is an invertible rank two tensor which satisfies
\begin{eqnarray}
\nabla_{\gamma} g_{\alpha\beta} =0.
\end{eqnarray}
This condition forces the affine connection to coincide with the Levi-Civita connection of the tensor $g_{\alpha\beta}$ which will play the role of the metric tensor. Thus
\begin{eqnarray}
\mathcal{L}^{\mu}_{\alpha\beta}= \frac{1}{2} g^{\mu\lambda} \left( \partial_{\alpha} g_{\beta\lambda} +
\partial_{\beta} g_{\lambda\alpha} - \partial_{\lambda} g_{\alpha\beta}\right),
\end{eqnarray}
and the density equality (\ref{imersed-density identity}) becomes
\begin{equation}
{\mathcal{P}}_{\alpha\beta} = \lambda g_{\alpha\beta}. \label{Eq1}
\end{equation}
Finally, using equation (\ref{P}), the gravitational field equations (\ref{Eq1}) take the form
\begin{equation}
{\mathcal{R}}_{\alpha\beta} 
= \lambda g_{\alpha\beta}+\Lambda_{\rho (\alpha}^{\lambda}\Lambda_{\beta)\lambda}^{\rho
}-\Lambda_{\rho\lambda}^{\lambda}\Lambda_{(\beta\alpha)
}^{\rho}. \label{field1}
\end{equation}
This is nothing but Einstein's equations with a \enquote{generated} energy-momentum tensor of matter which is given by
\begin{equation}
T_{\alpha\beta}=\left(\delta^{\nu}_{\beta}\delta^{\mu}_{\alpha}-\frac{1}{2}g_{\alpha\beta}g^{\mu\nu}\right)
\left(\Lambda_{\rho (\mu}^{\lambda}\Lambda_{\nu)\lambda}^{\rho
}-\Lambda_{\rho\lambda}^{\lambda}\Lambda_{(\nu\mu)
}^{\rho}\right). \label{matter0}
\end{equation}
The setup described here shows that matter can also be generated dynamically when spacetime is considered as a subspace of a higher dimensional space. In this case, the metric tensor, the cosmological constant as well as the energy momentum tensor of matter appear dynamically.

The second possible extension of Eddington's gravity using the same formalism, is to take the second Ricci tensor $\mathcal{Q}_{\alpha\beta}$ given by (\ref{Q}) in addition to $\mathcal{P}_{\alpha\beta}$. However, a dynamical equation like (\ref{motion1}) is not guaranteed, and in this case, a \enquote{current}-like term would appear leading to a nonmetricity equation. For more details, the reader is referred to Ref \cite{azri-immersed}.

\section{Separate Einstein-Eddington spaces}

In this section we will be interested in Eddington's affine gravity in the so called separate space. This is a higher dimensional space which is supposed to have a product structure. The aim of this section is to derive the gravitational equations that arise in separate Einstein's space, a space with only a cosmological constant.

Given a $2N$-dimensional space which admits a locally product structure, i.e, the existence of a separating coordinate system $x^{j}$ such that in any intersection of two neighbourhoods $x^{k}$ and $x^{k\prime}$ we have \cite{tachibana,Yano}
\begin{equation}
x^{\mu\prime}=x^{\mu\prime}\left(x^{\mu} \right), \quad \quad
x^{\mu^{\ast}\prime}=x^{\mu^{\ast}\prime}\left(x^{\mu^{\ast}} \right),
\end{equation}
where the Greek indices are given as $\mu=1,...,N$ and $\mu^{\ast}=N+1,...,2N$.

This means that the higher space appears as the product of two spaces $\mathcal{M}$ and $\mathcal{M}^{\ast}$ defined by their coordinate systems $x^{\mu}$ and $x^{\mu^{\ast}}$ respectively.

Additionaly, if the space is endowed with a metric tensor \textit{a priori}, then we define the separate Einstein's spaces as the product spaces which are described by their Ricci tensors $\mathcal{R}_{ij}$ which are splited into \cite{tachibana,Yano}
\begin{equation}
\label{ricci-ricci}
\mathcal{R}_{\mu\nu}= (a+b) g_{\mu\nu},\quad \mathcal{R}_{\mu^{\ast}\nu^{\ast}}= (a-b) g_{\mu^{\ast}\nu^{\ast}}.
\end{equation}
Here, $a$ and $b$ are constants.

It is clear that these spaces have constant curvature. We call these
spaces, the maximally symmetric spaces. The curvatures are given by two nonzero cosmological terms $a+b$ and $a-b$ respectively.

Next, we will provide a derivation of the equations (\ref{ricci-ricci}) in the context of Eddington gravity, using only an affine connection. 

\subsection{Gravitational equations in the separate space}

Herein, the $2N$-dimensional product space is endowed with a symmetric affine connection given by its components $\Gamma_{ij}^{k}$, such that $i,j=1,...2N$.

The curvature tensor, noted $\mathcal{R}_{ijk}^{l}$, has a standard form in terms of the affine connection
\begin{equation}
\mathcal{R}_{ijk}^{l}=\partial_{i}\Gamma_{jk}^{l}-\partial_{j}\Gamma_{ik}^{l}+\Gamma_{im}^{l}\Gamma_{jk}^{m}-\Gamma_{jm}^{l}\Gamma_{ik}^{m}.
\end{equation}
The Ricci tensor $\mathcal{R}_{ij}$ arises as
\begin{equation}
\mathcal{R}_{ij}=\mathcal{R}_{ikj}^{k}.
\end{equation}
We define the $2N$ dimensional Eddington's action as follows \cite{azri-separate}
\begin{equation}
\label{2n action}
S= 2 \int d^{2N} x \sqrt{||\mathcal{R}_{ij}||}.
\end{equation}
Here, the Lagrangian density is defined by
\begin{equation}
\label{lagran1}
\mathcal{L}= 2\sqrt{||\mathcal{R}_{ij}||},
\end{equation}
where we have taken only the symmetric part of the Ricci tensor, additionally, the affine connection $\Gamma$ is taken symmetric.

Following \cite{kijowski1,kijowski2}, we construct the canonical momentum conjugate to the connection $\Gamma$ as follows 
\begin{eqnarray}
\label{pi}
\pi^{ij}= \frac{\partial \mathcal{L}}{\partial \mathcal{R}_{ij}},
\end{eqnarray}
which will be at the heart of the metrical structure.

Using the the Lagrangian density (\ref{lagran1}), the last equation becomes
\begin{equation}
\label{pi1}
\sqrt{||\mathcal{R}_{ij}||}\mathcal{R}^{ij}=\pi^{ij},
\end{equation}
where $\mathcal{R}^{ij}$ is the inverse of the Ricci tensor.

In the following, we will apply Euler-Lagrange equations where the field configuration is the affine connection, then
\begin{equation}
\label{euler}
\partial_{l}\left( \frac{\partial\mathcal{L}}{\partial\left(  \partial_{l}\Gamma^{i}_{jk}\right)} \right) - \frac{\partial\mathcal{L}}{\partial\Gamma^{i}_{jk}}=0.
\end{equation}
This leads to the dynamical equation \cite{kijowski1,kijowski2}
\begin{equation}
\label{dyn}
\nabla_{k} \pi^{ij}=0.
\end{equation}
where the operator $\nabla$ is the covariant derivative associated to the affine connection $\Gamma$.

A possible $2N$ dimensional solution of equation (\ref{dyn}) is given as follows
\begin{equation}
\label{sol1}
\pi^{ij}=\sqrt{||a\mathcal{G}_{ij}+b\mathcal{F}_{ij} ||} \left( a\mathcal{G}+b\mathcal{F}\right)^{ij},
\end{equation}
where $a, b$ are constants, and the $2N$ tensors $\mathcal{G}_{ij}$ and $\mathcal{F}_{ij}$ have the components 
\begin{equation}
\mathcal{G}_{ij}=
\left( {\begin{array}{cc}
g_{\mu\nu} & 0 \\
0 & g_{\mu^{\ast}\nu^{\ast}}
\end{array} } \right), \quad \quad
\mathcal{F}_{ij}=
\left( {\begin{array}{cc}
g_{\mu\nu} & 0 \\
0 & -g_{\mu^{\ast}\nu^{\ast}}
\end{array} } \right).
\end{equation}
The tensors defined above will be important in defining the so called \textit{projective} operators which map the higher dimensional space into the separate spaces $\mathcal{M}$ and $\mathcal{M}^{\ast}$.

Now, let us turn to equation (\ref{pi1}) which finally takes the form
\begin{equation}
\label{field1}
\mathcal{R}_{ij}=\left( a\mathcal{G}_{ij}+b\mathcal{F}_{ij} \right).
\end{equation}
Additionally, the dynamical equation (\ref{dyn}) is written in the separate spaces $\mathcal{M}$ and $\mathcal{M^{\ast}}$ as follows
\begin{equation}
\label{dyn2}
\nabla_{\kappa}g_{\mu\nu}=0, \quad \text{and} \quad \nabla_{\kappa^{\ast}}g_{\mu^{\ast}\nu^{\ast}}=0.
\end{equation}

The generated metric tensors $g_{\mu\nu}$ and $g_{\mu^{\ast}\nu^{\ast}}$ lead to the following \enquote{separate} Levi-Civita connections
\begin{eqnarray}
&&\Gamma^{\mu}_{\alpha\beta}= \frac{1}{2} g^{\mu\lambda} \left( \partial_{\alpha} g_{\beta\lambda} +
\partial_{\beta} g_{\lambda\alpha} - \partial_{\lambda} g_{\alpha\beta}\right), \\
&&\Gamma^{\mu^{\ast}}_{\alpha^{\ast}\beta^{\ast}}= \frac{1}{2} g^{\mu^{\ast}\lambda^{\ast}} \left( \partial_{\alpha^{\ast}} g_{\beta^{\ast}\lambda^{\ast}} +
\partial_{\beta^{\ast}} g_{\lambda^{\ast}\alpha^{\ast}} - \partial_{\lambda^{\ast}} g_{\alpha^{\ast}\beta^{\ast}}\right)
\end{eqnarray}
in the separate spaces $\mathcal{M}$ and $\mathcal{M}^{\ast}$ respectively.

Mapping the vectors and tensors from the $2N$ dimensional space into the $N$ separate spaces is made via the projection operators
which are defind as \cite{tachibana, Yano}
\begin{equation}
\label{proj}
\mathcal{P}_{ij}= \frac{1}{2} \left( \mathcal{G}_{ij}+\mathcal{F}_{ij}\right), \quad \text{and} \quad \mathcal{Q}_{ij}= \frac{1}{2} \left( \mathcal{G}_{ij}-\mathcal{F}_{ij}\right),
\end{equation}
where for every vector $v^{i}$ with components $\left(v^{\mu}, v^{\mu^{\ast}} \right)$ we have
\begin{equation}
\mathcal{P}_{i}^{k}v^{i}=\left(v^{\mu},0 \right), \quad \text{and} \quad
\mathcal{Q}_{i}^{k}v^{i}=\left(0,v^{\mu^{\ast}} \right),
\end{equation}
with $\mathcal{P}_{i}^{k}=\mathcal{G}^{kl} \mathcal{P}_{li}$ and $\mathcal{Q}_{i}^{k}=\mathcal{G}^{kl} \mathcal{Q}_{li}$.

The separability of the higher dimensional space allows us to write the field equation (\ref{field1}) in two independent and separate field equations in the spaces $\mathcal{M}$ and $\mathcal{M}^{\ast}$ respectively. These equations are given as
\begin{eqnarray}
\label{field2}
\mathcal{R}_{\mu\nu}=\left( a+b\right) g_{\mu\nu},
\end{eqnarray}
and
\begin{eqnarray}
\label{field3}
\mathcal{R}_{\mu^{\ast}\nu^{\ast}}=\left( a-b\right)g_{\mu^{\ast}\nu^{\ast}}.
\end{eqnarray}

These equations govern the dynamics of the so called Einstein's spaces which have a constant curvature. Detailed studies of these spaces in the context of metric theory are give in Ref. \cite{tachibana,Yano}.

The derivation presented in this section is different from the one given in the referred works. It is based only on affine spaces endowed with an affine connection and its associated curvature. The metric tensors arise \textit{a posteriori} as in Eddington gravity, and finally the theory is reduced to separate spaces with two cosmological constants $a+b$ and $a-b$ respectively.

Next, we will present a possible application of this formalism. We will focus on the cosmological constant in the separate spaces and show how this constant vanishes in one of the spaces due to \textit{projective symmetry}.

\subsection{Zero cosmological constant from {\it projective} symmetry}

As we have seen so far, the cosmological constant is at the heart of the affine approach to gravity. In this sense, a nonzero cosmological constant facilitates the generation of the metrical structure and drives the affine models to metrical gravity. In what follows, we will discuss a mechanism that allows us to render the cosmological constant to zero \textit{a posteriori}. Although, this is generally not possible, however, the structure of the separate spaces discussed above provides us with a particular cases, where one of the spaces may be free of the cosmological term.  

Previously, we have shown that Einstein's space may describe two maximally symmetric spaces (universes) with nonzero cosmological constants given by
\begin{equation}
\Lambda=a+b \quad \text{and} \quad \Lambda^{\ast}=a-b,
\end{equation}
where $a$ and $b$ are nonzero constants.

Generating the metric tensors forbids a zero cosmological constant in both spaces. However, the \textit{symmetric} conditions $b=-a$, or $b=a$ render one of the cosmological constants to zero. In the first case, we have
\begin{equation}
\label{bi1}
\mathcal{R}_{\mu\nu}=0, \quad
\mathcal{R}_{\mu^{\ast}\nu^{\ast}}=2a g_{\mu^{\ast}\nu^{\ast}},
\end{equation}
where space $\mathcal{M}$ becomes empty. 

The other symmetric case ($b=a$) leads to
\begin{equation}
\mathcal{R}_{\mu\nu}=2a g_{\mu\nu} \quad \text{and} \quad \mathcal{R}_{\mu^{\ast}\nu^{\ast}}=0.
\end{equation}

The two cases $b=a$ and $b=-a$ correspond to the projection of the action (\ref{2n action}) on the spaces $\mathcal{M}$ and $\mathcal{M}^{\ast}$ respectively. This can be simply shown by using the projection tensors (\ref{proj}), thus
\begin{equation}
\mathcal{R}_{ik}\mathcal{P}^{k}_{j}= \left(\mathcal{R}_{\mu\nu}, 0\right) \quad \text{and} \quad \mathcal{R}_{ik}\mathcal{Q}^{k}_{j}=\left(0, \mathcal{R}_{\mu^{\ast}\nu^{\ast}} \right).
\end{equation}
A zero cosmological constant arises then in one of the spaces due to the projection on the separate spaces. The result of this projective symmetry can be translated as follows; while one of the universes is sensitive to a possible large vacuum energy due to the cosmological term, the other one becomes completely empty.   

In \cite{linde-double universe}, Linde has proposed the \textit{antipodal} symmetry in a two interacting universes and has shown that the effective cosmological constant vanishes in both spaces when applying that symmetry.

Clearly, our setup is not able to solve the cosmological constant problem, since it does not explain the tiny value of the vacuum energy that arises in the other space. Nevertheless, the separate spaces with zero and nonzero vacuum energy may, after all, describe two states of \textit{one} universe. The large vacuum energy at the early state is driven to zero at the final stage. We will return to this case in the following discussion when we introduce scalar fields in the setup.

Now, in the presence of a simple scalar field $\phi \left( x^{i}\right)$, where $i= 1,...,8$, the affine Lagrangian density takes the following form
\begin{equation}
\label{lagran2}
\mathcal{L}=2\frac{\sqrt{||\mathcal{R}_{ij}(\Gamma)-\partial_{i}\phi \partial_{j}\phi||}}{V(\phi)},
\end{equation}
where $V\left( \phi \right)$ is a potential energy, and for brevity, we will take $8\pi G=1$.

The canonical momentum (\ref{pi}) that corresponds to the above Lagrangian becomes
\begin{equation}
\label{pi2}
\pi^{ij}= \frac{\sqrt{||\mathcal{R}_{ij}-\partial_{i}\phi \partial_{j}\phi||}}{V(\phi)}
\left( \mathcal{R}-\partial\phi.\partial\phi\right)^{ij}.
\end{equation}
In this case, Euler-Lagrange equations (\ref{euler}) imply a dynamical equation similar to (\ref{dyn}), which finally allows us to write the field equations
\begin{equation}
\mathcal{R}_{ij}=\left( a\mathcal{G}_{ij}+b\mathcal{F}_{ij} \right)V\left( \phi\right) + \partial_{i}\phi \partial_{j}\phi.
\end{equation}
In the philosophy of the separate spaces presented above, the last equation is written in two forms
\begin{eqnarray}
\label{matter1}
\mathcal{R}_{\mu\nu}=\left( a+b\right)V\left( \phi\right)g_{\mu\nu} + \partial_{\mu}\phi \partial_{\nu}\phi,
\end{eqnarray}
\begin{eqnarray}
\label{matter2}
\mathcal{R}_{\mu^{\ast}\nu^{\ast}}=\left( a-b\right)V\left( \phi\right)g_{\mu^{\ast}\nu^{\ast}} + \partial_{\mu^{\ast}}\phi \partial_{\nu^{\ast}}\phi.
\end{eqnarray}
In vacuum, we have seen that the two universes are completely separate. This is however not the case in the presence of matter. 

Now, the dynamics of the scalar field is described by its equation of motion derived from the variation with respect to $\phi$. Again, this would lead to two equations of motion
\begin{eqnarray}
\Box_{g} \phi-\left( a+b\right)V^{\prime}\left( \phi\right)=0 \quad \text{and} \quad \Box_{g^{\ast}}\phi-\left( a-b\right)V^{\prime}\left( \phi\right)=0,
\end{eqnarray}
where the operators $\Box_{g}$ and $\Box_{g^{\ast}}$ are defined in the spaces $\mathcal{M}$ and $\mathcal{M^{\ast}}$ respectively.

In order to study the cosmological evolution of the scalar field, we will adapt the previous equations of motion to the flat Friedmann-Robertson-Walker metrics
\begin{equation}
d s^{2}=-dt^{2}+a^{2}\left( t\right) d \overrightarrow{x}^{2},\quad
d s_{\ast}^{2}=-d t_{\ast}^{2}+a^{2}_{\ast} \left( t_{\ast}\right)
d \overrightarrow{x_{\ast}}^{2},
\end{equation}
where the asterisks ($\ast$) refer to the coordinates of space $\mathcal{M}^{\ast}$.

The second Friedmann equation arises as follows
\begin{eqnarray}
\label{fr1}
\frac{\overset{\cdot\cdot}{a}}{a}=-\frac{4\pi G}{3}
\left[ 2\dot{\phi}^{2}-2\left(a+b \right)V\left(\phi \right)\right],
\\
\label{fr2}
\frac{\overset{\cdot\cdot}{a_{\ast}}}{a_{\ast}}=-\frac{4\pi G}{3}
\left[ 2\dot{\phi}^{2}-2\left(a-b \right)V\left(\phi \right)\right],
\end{eqnarray}
where the time derivative in the last equation is with respect to $t_{\ast}$.

Applying the projective symmetry discussed above, the Friedmann equations (\ref{fr1}) and (\ref{fr2}) become
\begin{eqnarray}
&&\label{frr1}
\frac{\overset{\cdot\cdot}{a}}{a}=-\frac{4\pi G}{3}
\left[ 2\dot{\phi}^{2}-4aV\left(\phi \right)\right],
\\
&&\frac{\overset{\cdot\cdot}{a_{\ast}}}{a_{\ast}}=-\frac{4\pi G}{3}
\left( 2\dot{\phi}^{2}\right).
\label{frr2}
\end{eqnarray}
Here, we have taken the case $b=a$ (equivalent to $b=-a$).

In cosmology, the early accelerated phase of the universe (the initial state here) is governed by the so called \text{\it gravitational} mass density $\rho+3p$, where $\rho$ and $p$ are the density and pressure of the inflaton respectively. The gravitational mass density forms the quantity in the right hand side of equation (\ref{frr1}). In fact, from the energy momentum tensor of the scalar field (for $b=a$)
\begin{equation}
T_{\mu\nu}=\partial_{\mu}\phi\partial_{\nu}\phi-\frac{1}{2}g_{\mu\nu}\left[\left(\partial\phi\right)^{2}+4aV\left(\phi \right) \right],
\end{equation}
we easily find the energy density and pressure of the scalar field as
\begin{equation}
\rho=\frac{1}{2}\dot{\phi}^{2}+2aV\left(\phi \right),\quad
p=\frac{1}{2}\dot{\phi}^{2}-2aV\left(\phi \right),
\end{equation}
and finally, the gravitational mass density takes the form\begin{equation}
\rho_{grav}=\rho+3p = 2\dot{\phi}^{2}-4aV\left(\phi \right).
\end{equation}
On the other hand, the right hand side of (\ref{frr2}) is governed by the \textit{inertial mass density} of the field. This is given by 
\begin{equation}
\dot{\phi}^{2}= \rho_{\ast}+p_{\ast}=\rho_{iner}.
\end{equation}
Thus, in the presence of matter, the role of the projective symmetry is to eliminate the effects of the gravitational mass density in the final state, and the dynamics of the universe in this case is governed by only its inertial mass density. In the case of the vacuum energy (cosmological constant) where $p=-\rho$, the inertial mass density $\rho+p$ vanishes, which is consistent with (\ref{bi1}).

\clearpage  
\lhead{\emph{Chapter 8}}  
  \chapter{Concluding remarks}
\vspace{-0.5 cm}
\epigraph{\textit{The important thing is not to stop questioning}\, --- Albert Einstein}{}

Purely metrical structure of spacetime where gravity is described by general relativity is essential for the very large scales of the universe. However this structure may not be required in the very beginning. At early stages, the spacetime is purely affine in a sense that it does not accommodate notions of angles and lengths. These notions arise a posteriori with the metric structure when this latter is generated. The absence of the metric tensor leaves spacetime with a very simple structure, the affine structure, in which the affine theory of gravity is viable.

As we have mentioned in this thesis, purely affine gravity is not a new theory, it goes back to previous classic works of Einstein, Eddington and Schr\"{o}dinger as an attempts to a unified picture of gravity and electrodynamics \cite{schrodinger}. The failure of the purpose of unification has led people to abandoning the affine approach by considering it as a pure mathematical construction that lacks physical interpretations. Other affine approach to gravity has been proposed later as a different formulation of general relativity where the metric tensor appears as a momentum canonical conjugate to the affine
connection, and the derived field equations are equivalent to those of GR with scalar and
possibly gauge fields \cite{kijowski1,kijowski2}. In the recent few years, attempts have been made to consider
general and different approaches to pure affine gravity, in vacuum and in the presence of matter and even in higher dimensions \cite{demir-eddington,kemal,liebscher,oscar,poplawski}.

In this thesis, we have studied this affine gravity in the presence of scalar fields. At the first step where the field is minimally coupled, we have seen that the theory is defined only for nonzero potential, this led out to a nonzero vacuum energy in the theory. We have argued that this nonzero vacuum is the origin of the metric tensor from which Einstein's equations are written. Transition to nonminimal coupling is investigated, where the coupling is made through the Ricci tensor. It turned out that, unlike the first case, the nonminimal coupling in affine gravity differs from general relativity. The differences rely on both, the improved energy-momentum tensor and the modified equation of the field. We have seen that the improved energy momentum tensor depend on the potential of the scalar field rather than derivatives of the field $\phi$ as in general relativity. This is a consequence of the linearity of the Ricci tensor in first derivative of the affine connection. We have shown that the transformation from nonminimal to minimal coupling is simply obtained through the scalar field redefinition. This shows that there is only one frame in which affine gravity is formulated. This is arguably clear since there is only one generated metric tensor. This means that Jordan and Einstein frames of general relativity are not present in affine gravity.

The main goal of this thesis is to show that affine spacetime though difficult to accommodate all matter fields, it serves a viable framework for studying the early universe. In fact, in the inflationary regime and before reheating phase, only scalar fields that drive the rapid expansion are required. We have shed light on two particular examples. The first is the standard \textbf{affine inflation} where a non-SM scalar field is coupled nonminimally to affine gravity and drives cosmological inflation. In this model, the scalar spectral index as well as the tensor-to-scalar ratio are in agreement with the recent Planck results for some values of the nonminimal coupling parameter \cite{affine inflation}. The second model is based on a new approach to induced gravity. In this \textbf{induced affine gravity}, it has been shown that both gravity scale and the metric tensor gain an emergent character \cite{induced affine inflation}. As in (metric) induced gravity, Planck mass arises spontaneously in terms of the vacuum expectation value of a non-SM heavy scalar. Additionally, the metric tensor appears dynamically from nonzero vacuum energy which is left after symmetry breaking. \textbf{Induced affine inflation}, however, results in a relatively large tensor-to-scalar ratio, a feature which is generic of the models in which gravity is induced by the vacuum expectation value of a scalar field. Last but not least, we tackled the conformal frame ambiguities. It has been shown that since affine gravity stands on a unique \enquote{generated} metric tensor, the familiar Jordan and Einstein (conformal) frames are absent, the case which makes the affine inflation predictions unique and frame ambiguity-free.  

We have to mention here that up to now affine gravity is considered as an incomplete theory. In fact, a program should be pursued of incorporating all the SM matter fields in order to complete the affine picture of matter-gravity interactions. Speculatively speaking, the SM matter fields may also be generated dynamically at the end of inflation where the inflaton energy is converted to SM particles and the universe becomes radiation dominated. In this case, a reheating process in the context of affine gravity must be studied. The final point concerns the quantum correction to the affine actions which have been proposed throughout this thesis. Since these actions are not polynomials in the fields then one might go beyond the standard techniques of field theory when performing the covariant quantization. An alternative way is to transform these actions into polynomials which lead to the same field equations of motion and go through the quantization in its standard form, however, in this case one may lose the aim of affine gravity by proposing different forms of the action \cite{martenilli}.

\clearpage  
\lhead{\emph{References}}  









\addtocontents{toc}{\vspace{2em}} 

\clearpage  



\addtocontents{toc}{\vspace{2em}}  
\backmatter


\end{document}